\documentclass[11pt]{article}
\usepackage{epsfig}
\author{O. \"Ogetbil}
\title{Stable de Sitter Vacua in 4 Dimensional Supergravity Originating from 5 Dimensions}
\usepackage{amssymb}
\usepackage{amsmath}
\usepackage{verbatim}
\usepackage[plainpages=false,hypertexnames=false,pdfpagelabels,pdfauthor={O. Ogetbil},pdftitle={Stable de Sitter Vacua in 4 Dimensional Supergravity Originating from 5 Dimensions},pdfsubject={N=2 Supergravity},pdfkeywords={supergravity, de Sitter, N=2, ground states, vacua, solutions},pdfproducer={Latex with hyperref},pdfcreator={pdflatex},colorlinks=false,pdfborder={0 0 0}]{hyperref}
\numberwithin{equation}{section}

\newsavebox{\uuunit}
\sbox{\uuunit}
    {\setlength{\unitlength}{0.825em}
     \begin{picture}(0.6,0.7)
        \thinlines
        \put(0,0){\line(1,0){0.5}}
        \put(0.15,0){\line(0,1){0.7}}
        \put(0.35,0){\line(0,1){0.8}}
       \multiput(0.3,0.8)(-0.04,-0.02){10}{\rule{0.5pt}{0.5pt}}
     \end {picture}}
\newcommand {\unity}{\mathord{\!\usebox{\uuunit}}}
\newcommand{\sdsum}{{\scriptstyle \circledS}}
\begin{document}
%\maketitle
\pagenumbering{alph}
\vspace*{1.8cm}
\begin{center}
{\bf{\large Stable de Sitter Vacua in 4 Dimensional Supergravity Originating from 5 Dimensions}}  \\

\vspace{1cm}
\begin{large}

 O. \"Ogetbil\footnote{oogetbil@phys.psu.edu}
\end{large}
\\
%\end{center}
\vspace{.35cm}

\vspace{.3cm}  \emph{Physics Department \\
Pennsylvania State University\\
University Park, PA 16802, USA} \\

\vspace{1cm}
{\bf Abstract}
\end{center}

%\begin{abstract}
 The five dimensional stable de Sitter ground states in $\mathcal{N}=2$ supergravity obtained by gauging $SO(1,1)$ symmetry of the real symmetric scalar manifold (in particular a generic Jordan family manifold of the vector multiplets) simultaneously with a subgroup $R_s$ of the $R$-symmetry group descend to four dimensional de Sitter ground states under certain conditions. First, the holomorphic section in four dimensions has to be chosen carefully by using the symplectic freedom in four dimensions; and second, a group contraction is necessary to bring the potential into a desired form. Under these conditions, stable de Sitter vacua can be obtained in dimensionally reduced theories (from 5D to 4D) if the semi-direct product of $SO(1,1)$ with $\mathbb{R}^{(1,1)}$ together with a simultaneous $R_s$ is gauged. We review the stable de Sitter vacua in four dimensions found in earlier literature for $\mathcal{N}=2$ Yang-Mills Einstein supergravity with $SO(2,1)\times R_s$ gauge group in a symplectic basis that comes naturally after dimensional reduction. Although this particular gauge group does not descend directly from five dimensions, we show that, its contraction does. Hence, two different theories overlap in certain limits. Examples of stable de Sitter vacua are given for the cases: (i) $R_s=U(1)_R$, (ii) $R_s=SU(2)_R$, (iii) $\mathcal{N}=2$ Yang-Mills/Einstein Supergravity theory coupled to a universal hypermultiplet. We conclude with a discussion regarding the extension of our results to supergravity theories with more general homogeneous scalar manifolds.
%\end{abstract}

\thispagestyle{empty}
\newpage
\pagenumbering{arabic}

\tableofcontents\setcounter{page}{1}
\newpage

\section{Introduction}
Supergravity theories are local gauge theories of supersymmetry and were first formulated in 70's \cite{FNF76, DZ76, Nie81}\footnote{For a review about gauged supergravity theories of various dimensions that have been studied extensively since then, see \cite{SS89}.}. There are two ways of studying supergravity in a certain dimension. One can either construct it directly from field content and symmetries (both local and global) that the action must have; or one can obtain them from higher dimensions by dimensional reduction. Supergravity theories that are obtained purely by dimensional reduction from 10 or 11 dimensional supergravity are low energy effective limits of some superstring theory/M-theory. In such cases, their scalar manifold is the moduli space of the compactification. For certain extended supergravity theories, gauging a symmetry of the action may yield a potential term $V(\phi)$ of scalar fields. The ground states of the resulting theory are determined by the critical points (say $\phi_0$) of the potential term.

Scalar fields play a fundamental role in the description of cosmological models. In fact, the assumption that the energy-momentum tensor is dominated by scalar potential energy density $V(\phi)$ has been the starting point of many inflationary models\footnote{For a general review and further references on inflationary cosmology, see \cite{Lin07}.} \cite{Lin90, LL00}. If the value of the potential at its critical point is positive ($V'\arrowvert_{\phi_0}=0,\,V(\phi_0) >0$)\footnote{$V'\equiv \partial V / \partial \phi$}, the case with zero kinetic energy ($\dot{\phi}=0$) corresponds to de Sitter space with a positive cosmological constant. The current accelerated expansion of the universe \cite{Per97, Rie98} can be explained either by a positive vacuum energy $V(\phi_0)$, or a scalar field in a slow-roll regime $\dot{\phi}^2/2 \ll V(\phi)$ on a near de Sitter background (quintessence) \cite{Wei87, Wet87, RP87}.

There are two possible ways of explaining the positive vacuum energy in terms of scalar potentials $V(\phi)$. The observed cosmological constant may correspond to the minimum of a scalar potential, in which case the universe will continue to accelerate forever. However the de Sitter regime might be transient, i.e. it might correspond to a local maximum or a saddle point of the scalar potential. Models with slow-roll inflation $(|V''|\ll |V|)$ and fast-roll inflation $(|V''|\sim |V|)$ have been considered in \cite{KLPS01}. In such cases the scalar potential either vanishes as the field rolls to $\phi \rightarrow \infty$ and the universe reaches a Minkowski stage; or the scalar field rolls to the minimum of the potential with $V(\phi)<0$ (or $V(\phi)\rightarrow -\infty$, such that the potential does not have a minimum at all) and the universe may eventually collapse.
\begin{figure}
 \centering
 \includegraphics[width=419pt,viewport=14 14 550 285]{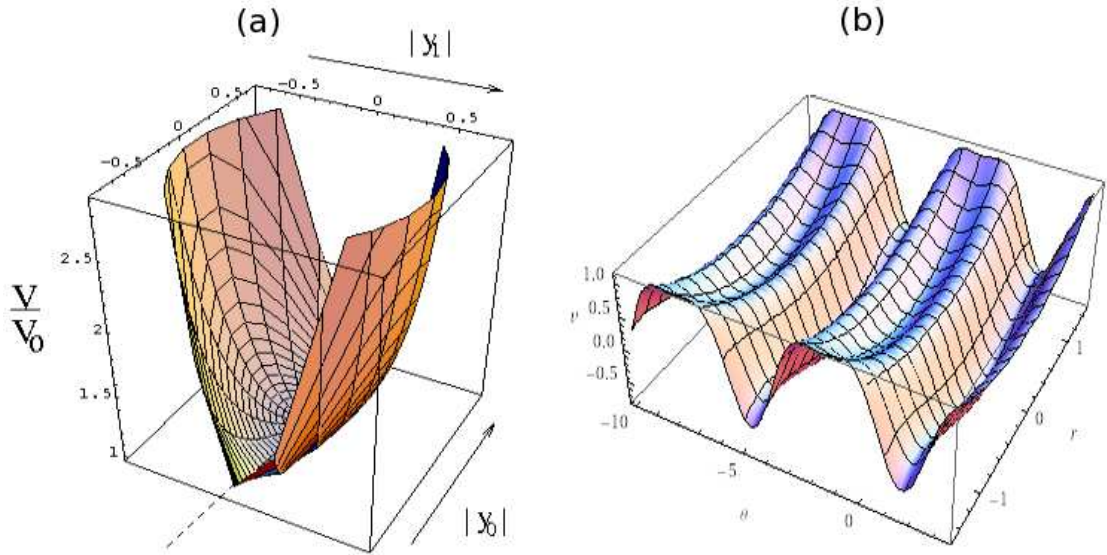}
 % pot_extrema.eps: 1179666x1179666 pixel, 300dpi, 9987.84x9987.84 cm, bb=14 14 550 285
 \caption[Examples of de Sitter extrema]{Examples of de Sitter extrema in supergravity theories: (a) Stable minima with a flat direction. The potential belongs to the $4D,\mathcal{N}=2$ supergravity coupled to 3 vector multiplets, considered in \cite{FTP02}. Figure taken from \cite{FTP03}. (b) Saddle point, where the scalar rolls into a Minkowski minimum on one side and anti-de Sitter minimum on the other. This is $4D,\mathcal{N}=2$ supergravity coupled to 1 hypermultiplet, considered in \cite{BM03}. The potential includes instanton corrections.}
 \label{fig_pot}
\end{figure}

The evidence of a small positive cosmological constant attracted interest in finding stable de Sitter ground state solutions in supersymmetric theories. In the context of supersymmetric theories, anti-de Sitter ground states emerge naturally in contrast to de Sitter ground states. This is due to the fact that the de Sitter superalgebras usually have non-compact $R$-symmetry subalgebras, which leads to existence of ghosts if the supersymmetry is to be fully preserved. Nevertheless exact supersymmetry is not observed in nature and supersymmetry must be a broken symmetry. There are two main approaches to study de Sitter ground state solutions of supersymmetric theories. One can start from a fundamental theory (a superstring or M-theory), study compactifications on various internal manifolds and, with the combined effects of the warped geometries of the internal manifold and tree-level corrections to the $4D$ K\"{a}hler potential, obtain a potential in four dimensions that admits de Sitter critical points \cite{KKLT03, BHM02, MSS02, BKQ03, BCK04, Buc04, Gra05, VZ06, ISS06, Sil07, Cov08}. On the other hand one can search for such potentials in the extended gauged supergravity theories directly \cite{Hul01a, Hul01b, GH01, KLPS01, dRWP03, dRWPT03, FTP02, FTP03, BM03, KP04, RRW06}. Figure \ref{fig_pot} shows two examples obtained from $4D,\mathcal{N}=2$ supergravity theories \cite{FTP02,FTP03,BM03}. A novel result of these studies is that the mass squared of the scalar fields is on the order of the cosmological constant, i.e. the value of the scalar potential at its extremum, 
\begin{equation}
 m_\phi^2 \sim \Lambda.
\end{equation}
Any quantum corrections to the scalar masses will be related to the cosmological constant $\Lambda= 3 H_0^2 \sim 10^{-120} M_{\rm Planck}^4$ and will be very small \cite{KLPS01}.

In this paper, we will take the second approach and start with studying five dimensional gauged supergravity theories \cite{GST84, GRW85, GRW86, PPvN85} that received renewed attention more recently due their role within the AdS/CFT correspondences in string theory \cite{Mal98, GKP98, Wit98, AGMOO99}, Randall-Sundrum (RS) braneworld scenario \cite{RS99a, RS99b, KL00} and M-/superstring theory compactifications on Calabi-Yau manifolds with fluxes \cite{CCDF95, FKS95, FK96, FGK97}. It is believed that the $5D, \mathcal{N}=8$ gauged supergravity \cite{GRW85, GRW86, PPvN85} is a consistent nonlinear truncation of the lowest lying Kaluza-Klein modes of type IIB supergravity on $AdS_5\times S^5$ \cite{GM85, KRvN85, NV00, CLPST00, Tra01}. Moreover, certain brane world scenarios based on M-theory compactifications have $5D, \mathcal{N}=2$ gauged supergravity as their effective field theories \cite{HW95, HW96, LOSW99a, LOSW99b, ELPP98}.

We adopt the convention introduced in \cite{GZ99} to classify the gaugings of $\mathcal{N}=2$ supergravity theories in 5 and 4 dimensions. The ungauged $\mathcal{N}=2$ supergravity coupled to vector- and/or hypermultiplets is referred as (ungauged) Maxwell-Einstein supergravity theories (MESGT). In the absence of hypermultiplets, these theories have a global symmetry group of the form $G\times SU(2)_R$\footnote{The global symmetry group is $G\times SU(2,1)$ if a universal hypermultiplet is coupled to the theory where $SU(2,1)$ is the isometry group of the hyperscalar manifold. Note that $SU(2)_R\subset SU(2,1)$.} in 5 dimensions, where $G$ is generally the isometry group of the scalar manifold of the vector multiplets\footnote{For the non-generic Jordan family, which will be defined in the next section, a parabolic subgroup of $G$ is the symmetry of the whole Lagrangian \cite{dWP92}.} and $SU(2)_R$ is the automorphism group of the underlying supersymmetry algebra, which is also commonly referred as the ``$R$-symmetry group''. Theories obtained by gauging a $U(1)_R$ subgroup of $SU(2)_R$ by coupling a linear combination of vector fields to the fermions \cite{GST84}, which are the only fields that transform nontrivially under $SU(2)_R$, are called gauged Maxwell-Einstein supergravity theories (gauged MESGT). On the other hand, if only a subgroup $K$ of the symmetry group $G$ of the action is being gauged, the theory is referred as a Yang-Mills/Einstein supergravity theory (YMESGT). Note that the theories which include tensor fields fall into this category. A theory with a gauge group $K\times U(1)_R$ is called gauged Yang-Mills/Einstein supergravity theory (gauged YMESGT).

Pure $5D, \mathcal{N}=2$ supergravity was constructed in \cite{Cre80, CN80}, coupling to vector multiplets was done in \cite{GST83b, GST84} and tensor fields were added to the theory in \cite{GZ99}. Coupling of hypers to these theories was done in \cite{CD00}. Vacua of $U(1)_R$ gauged $5D, \mathcal{N}=2$ MESGT's and YMESGT's without hypers and tensors were studied in \cite{GST84}. Vacua of the generic Jordan family models, which will be defined in the next section, with Abelian gaugings and tensors have been investigated in \cite{GZ00}, the full $R$-symmetry group gauging was done in \cite{GZ01} and a study for vacua of some other gauged theories were carried out in \cite{CS05}. We will give two examples from the literature \cite{GZ00, CS05, Oge06} of the stable de Sitter vacua of $5D, \mathcal{N}=2$ supergravity theories coupled to vector, tensor multiplets and a universal hypermultiplet. Then, following the dimensional reduction process of \cite{GST83b, GMZ05b}, we will look for de Sitter ground states in four dimensions. The analysis in $5D$ is somewhat easier than in $4D$, mainly because in $4D$, the $U$-duality is an on-shell symmetry, whereas in $5D$, it is a symmetry of the Lagrangian. Moreover $5D$ theories have real geometry while the geometry in $4D$ is complex. Therefore, whereas our study in $5D$ in an earlier work \cite{Oge06} covered all possible ground states, in $4D$, motivated by experimental observations, we will concentrate only on de Sitter solutions.

The organization of this paper is as follows. In section \ref{sec2}, we start with reviewing the field content of the $5D, \mathcal{N}=2$ supergravity. The potential terms arising from non-compact $SO(1,1)$ real scalar manifold isometry gauging and a subgroup $R_s$ of the $R$-symmtery group $SU(2)_R$ will be given. It will turn out that an $SO(1,1)\times R_s$ gauged YMESGT has stable de Sitter ground states in 5 dimensions. Section \ref{sec3} takes the story down to 4 dimensions. The symplectic freedom related to the de Roo-Wagemans rotations will be used to find de Sitter ground states. In fact, the stable five dimensional de Sitter ground states we will demonstrate in section \ref{sec2} and those found in \cite{FTP02,FTP03} coincide in certain limits. This relation is revealed by introducing contractions on the gauge groups. Most of the calculations of this section uses the symmetric generic Jordan family as the scalar manifold, although in the last subsection we discuss about extending our results to the more general homogeneous scalar manifolds. Section \ref{sec4} collects the summary of all of our the results and proposes future directions. In the first appendix, one can find the bosonic part of the four and five dimensional Lagrangians, the elements of very special geometry and the derivation of the potential terms from more fundamental quantities. In appendix \ref{appendixhypers}, we list the Killing vectors and their corresponding prepotentials of the hyperscalar manifold isometries that will be used to carry out the hyper-gaugings throughout the paper. Appendix \ref{app_trf} gives the quadratic coordinate transformations between the parametrization we use in the paper and Calabi-Vesentini coordinates that was used in \cite{FTP02, FTP03}. Certain scalar potential terms are given in the last appendix in their full form due to their lengthiness. They will be referred within the text in section \ref{sec3}. The contents of this paper constitute part of the author's PhD thesis \cite{Oge08a}.

\section{5 Dimensional \texorpdfstring{$\mathcal{N} = 2$}{N=2} Supergravity Theories\label{sec2}}
\subsection{The Basics and the Scalar Potential Terms}
\setcounter{equation}{0}
The field content of the ungauged (before tensor- or hypermultiplet coupling) $\mathcal{N}=2$ MESGT is
\begin{equation}
\{e^{\hat{m}}_{\hat{\mu}},\Psi^i_{\hat{\mu}},A^I_{\hat{\mu}},\lambda^{i\tilde{a}},\varphi^{\tilde{x}}\}
\end{equation}
where
\begin{eqnarray}
\begin{array}{rcl}
i&=&1,2,\\
I&=&1,2,...,\tilde{n}+1,\\
\tilde{a}&=&2,3,...,\tilde{n}+1,\\
\tilde{x}&=&2,3,...,\tilde{n}+1.\nonumber
\end{array}
\end{eqnarray}
The 'graviphoton' is combined with the $\tilde{n}$ vector fields of the $\tilde{n}$ vector multiplets into a single ($\tilde{n}$+1)-plet of vector fields $A^I_{\hat{\mu}}$ labelled by the index $I$. The indices $\tilde{a},\tilde{b},...$ and $\tilde{x},\tilde{y},...$ are the flat and the curved indices, respectively, of the $\tilde{n}$-dimensional target manifold $\mathcal{M}^5_{VS}$ of the real scalar fields, which we will define below.\\[2pt]
The bosonic part of the Lagrangian is given in the appendix \ref{appendixveryspecial}. The global symmetries of these theories are of the form $SU(2)_R \times G_{(5)}$, where $SU(2)_R$ is the $R$-symmetry group of the $\mathcal{N}=2$ Poincare superalgebra and $G_{(5)}$ is the subgroup of the group of isometries of the scalar manifold that extends to the symmetries of the full action. Gauging a subgroup $K_{(5)}$ of $G_{(5)}$ requires dualization of some of the vector fields to self-dual tensor fields if they are transforming in a non-trivial representation of $K_{(5)}$. More formally, the field content, when $2 n_T$ of the vector fields are dualized to tensor fields, becomes
\begin{equation}
\{e^{\hat{m}}_{\hat{\mu}},\Psi^i_{\hat{\mu}},A^I_{\hat{\mu}},B^M_{{\hat{\mu}}{\hat{\nu}}},\lambda^{i\tilde{a}},\varphi^{\tilde{x}}\}
\end{equation}
where now
\begin{eqnarray}
\begin{array}{rcl}
i&=&1,2,\\
I&=&1,2,...,n_V +1,\\
M&=&1,2,...,2n_T,\\
\tilde{I}&=&1,2,...,\tilde{n}+1,\\
\tilde{a}&=&2,3,...,\tilde{n}+1,\\
\tilde{x}&=&2,3,...,\tilde{n}+1,
\end{array}\nonumber
\end{eqnarray}
with $\tilde{n}=n_V+2n_T$. Tensor multiplets come in pairs with four spin-$1/2$ fermions (i.e. two $SU(2)_R$ doublets) and two scalars. Tensor coupling generally introduces a scalar potential of the form \cite{GZ99}:

\begin{equation}
P_{(5)}^{(T)}=\frac{3\sqrt{6}}{16}h^I\Lambda^{MN}_I h_M h_N.\label{pt}
\end{equation}
Here $\Lambda^{MN}_I$ are the transformation matrices of the tensor fields and $h_{\tilde{I}}, h^{\tilde{I}}$ are elements of the ``very special'' geometry of the scalar manifold $\mathcal{M}^5_{VS}$ that has the metric $\stackrel{o}{a}_{\tilde{I}\tilde{J}}$ which is used to raise and lower the indices $\tilde{I},\tilde{J},...\,$.  

When the full $R$-symmetry group $SU(2)_R$ is being gauged the potential gets the contribution
\begin{equation}
 P_{(5)}^{(R)}=-4 C^{ij\tilde{K}}\delta_{ij} h_{\tilde{K}},\label{su2r}
\end{equation}
where $i,j$ are adjoint indices of $SU(2)$. If instead, the $U(1)_R$ subgroup is being gauged, the contribution to the potential becomes
\begin{equation}
P_{(5)}^{(R)}=-4C^{IJ\tilde{K}}V_I V_J h_{\tilde{K}}.\label{u1r}
\end{equation}
The expressions that lead to the derivation of the above potential terms can be found in appendix \ref{appendixveryspecial}.

We will look at the cases, where the scalar manifold $\mathcal{M}^5_{VS}$ is a symmetric space. Such spaces are further divided in two categories, depending whether they are associated with a Jordan algebra or not. The spaces that are associated with Jordan algebras are of the form $\mathcal{M}^5_{VS}=\frac{Str_0 (J)}{Aut(J)}$, where $Str_0 (J)$ and $Aut(J)$ are the reduced structure group and the automorphism group, respectively, of a real, unital Jordan algebra $J$, of degree three \cite{GST83b, GST86}. More specifically,
\begin{itemize}
\item Generic Jordan Family:
\begin{equation}\begin{array}{rlr}
J=\mathbb{R}\oplus \Sigma_{\tilde{n}}:&\displaystyle \mathcal{M}^5_{VS}=\frac{SO(\tilde{n}-1,1)\times SO(1,1)}{SO(\tilde{n}-1)},&\tilde{n}\geq 1.\nonumber
\end{array}\end{equation}
\item Magical Jordan Family:
\begin{equation}\begin{array}{rll}
J_3^{\mathbb{R}}:\quad&\displaystyle \mathcal{M}^5_{VS}=\frac{SL(3,\mathbb{R})}{SO(3)},&\quad\tilde{n}=5,\\[9pt]
J_3^{\mathbb{C}}:\quad&\displaystyle \mathcal{M}^5_{VS}=\frac{SL(3,\mathbb{C})}{SU(3)},&\quad\tilde{n}=8,\\[9pt]
J_3^{\mathbb{H}}:\quad&\displaystyle \mathcal{M}^5_{VS}=\frac{SU^{*}(6)}{Usp(6)},&\quad\tilde{n}=14,\\[9pt]
J_3^{\mathbb{O}}:\quad&\displaystyle \mathcal{M}^5_{VS}=\frac{E_{6(-26)}}{F_4},&\quad\tilde{n}=26.\label{3families}
\end{array}\end{equation}
\item Generic non-Jordan Family:
\begin{equation}
\mathcal{M}^5_{VS}=\frac{SO(1,\tilde{n})}{SO(\tilde{n})},\quad \tilde{n}\geq 1.\nonumber
\end{equation}
\end{itemize}

In addition to the supergravity multiplet, $n_V$ vector multiplets and $2 n_T$ tensor multiplets one can couple hypermultiplets into the theory. A universal hypermultiplet
\begin{equation}
\{\zeta^a,q^X\}
\end{equation}
contains a spin-1/2 fermion doublet $A=1,2$ and four real scalars $X=1,...,4$. The total manifold of the scalars $\phi=(\varphi,q)$ then becomes
\begin{equation}
\mathcal{M}^5_{scalar} = \mathcal{M}^5_{VS} \otimes \mathcal{M}_{Q}\nonumber
\end{equation}
with $dim_{\mathbb{R}}\mathcal{M}^5_{VS}=n_V+2n_T$ and $dim_{\mathbb{Q}}\mathcal{M}_{Q}=1$. The quaternionic hyperscalar manifold $\mathcal{M}_{Q}$ of the scalars of a single hypermultiplet has the isometry group $SU(2,1)$. Gauging a subgroup of this group introduces an extra term in the scalar potential \cite{CD00}
\begin{equation}
P_{(5)}^{(H)}=2\mathcal{N}_{iA}\mathcal{N}^{iA}
\end{equation}
where $\mathcal{N}^{iA}=\frac{\sqrt{6}}{4} h^I K^X_I f^{iA}_X$ with $f^{iA}_X$ being the quaternionic vielbeins, $f^{iA}_X f_{YiA}=g_{XY}$, and $g_{XY}$ is the metric of the quaternionic-K\"{a}hler hypermultiplet scalar manifold \cite{CDKP01}
\begin{equation}
ds^2=\frac{dV^2}{2V^2}+\frac{1}{2V^2} (d\sigma +2\theta d\tau - 2 \tau d\theta)^2 + \frac{2}{V} (d\tau^2+d\theta^2),\label{hypermetric}
\end{equation}
and $K^X_I$ being the Killing vectors given in appendix \ref{appendixhypers} together with their corresponding prepotentials. The determinant of the metric is $1/V^6$ and it is positive definite and well behaved everywhere except $V=0$. But since in the Calabi-Yau derivation V corresponds to the volume of the Calabi-Yau manifold \cite{LOSW99b}, we restrict ourselves to the positive branch $V>0$.

When the $R$-symmetry is gauged in a theory that contains hypers, the potential $P_{(5)}^{(R)}$ gets some modification due to the fact that the fermions in the hypermultiplet are doublets under the $R$-symmetry group $SU(2)_R$. It becomes
\begin{equation}
P_{(5)}^{(R)}=-4 C^{IJ\tilde{K}} \vec{P}_I\cdot\vec{P}_J h_{\tilde{K}}\label{pr_h}
\end{equation}
where $\vec{P}_I$ are the prepotentials corresponding to the Killing vectors $K^X_I$.

The total scalar potential, which includes terms from tensor coupling, $R$-symmetry gauging and hyper coupling, is given by
\begin{equation}
\begin{array}{rcl}
P_{(5)}\equiv e^{-1}\mathcal{L}_{pot}&=&- g^2 P_{(5)}^{(T)} - g_R^2 P_{(5)}^{(R)} - g_H^2 P_{(5)}^{(H)}\\
&\equiv&- g^2 P_{(5)}\\[5pt]
&=&- g^2 (P_{(5)}^{(T)} + \lambda P_{(5)}^{(R)} + \kappa P_{(5)}^{(H)}),\label{totalpot}
\end{array}
\end{equation}
where $\lambda=g_R^2/g^2$, $\kappa=g_H^2/g^2$; $g_R$, $g_H$ and $g$ are coupling constants, which need not to be all independent. Any point on the scalar manifold where the first derivatives of the total scalar potential with respect to all scalars vanish will be a solution to the corresponding model.\\[5pt]

%\underline{\textbf{Supersymmetry of the solutions:}}
\paragraph{Supersymmetry of the solutions:}

Demanding supersymmetric variations of the fermi-ons vanish at the critical points of the theory, the conditions that need to be satisfied are found as \cite{GZ00, CDKP01}
\begin{equation}
 \langle W^{\tilde{a}}\rangle = \langle P^{\tilde{a}}\rangle = \langle \mathcal{N}_{iA}\rangle=0
\end{equation}
where $W^{\tilde{a}}$ and $P^{\tilde{a}}$ are defined in (\ref{wapadef}). Any ground state that does not satisfy all of these conditions are not supersymmetric. One can see that any supersymmetric solution must be of the form
\begin{equation}
 P_{(5)}\arrowvert_{\phi^c} = -4 \lambda \text{ }\vec{P} \cdot \vec{P}(\phi^c)
\end{equation}
which is negative semi-definite. Hence we know from beginning that any de Sitter type ground state of the theories we will consider will have broken supersymmetry. The para\-met\-ri\-za\-tion of the Killing vectors of the hyperscalar manifold, which is outlined in appendix \ref{appendixhypers}, yields $K_I^X\arrowvert_{q^c}\neq 0$, for non-compact generators. Here, the point $q^c=\{V=1,\sigma=\theta=\tau=0\}$ is the base point of the hyperscalar manifold, i.e. the compact Killing vectors of the hyper-isometry generate the isotropy group of this point. This point will be used as the hyper-coordinate candidate of the critical points. As a consequenece $\langle \mathcal{N}_{iA}\rangle\neq 0$; and hence theories including non-compact hyper-gauging will not have supersymmetric critical points either.
\\[10pt] 

\subsection{Gauging a compact symmetry group of the hyper-isometry}

The total potential is of the form $P_{(5)}=P_{(5)}^{(T)} +\lambda P_{(5)}^{(R)}+\kappa P_{(5)}^{(H)}$. The most general way of doing simultaneous $U(1)_R$ gauging together with $U(1)_H$ gauging of the hypermultiplet isometry ($\lambda = \kappa$) is done by selecting a linear combination of compact Killing vectors from (\ref{Tvex}). One can easily see that at the base point $q^c = \{V=1, \sigma=\theta=\tau=0\}$ of the hyperscalar manifold all these compact generators vanish. Therefore one has $\mathcal{N}^{iA}=0$ and as a consequence \cite{CS05}
\begin{equation}
 P_{(5)}^{(H)}\arrowvert_{q^c}=\frac{\partial P_{(5)}^{(H)}}{\partial \varphi}\arrowvert_{q^c}=\frac{\partial P_{(5)}^{(H)}}{\partial q}\arrowvert_{q^c}= 0.\label{u1saddle}
\end{equation}
$P_{(5)}^{(T)}$ is a function of the real scalars $\varphi^{\tilde{x}}$ only. On the other hand, $P_{(5)}^{(R)}$ of (\ref{pr_h}) is of the form $P_{(5)}^{(R)}\sim f(\varphi) g(q)$, where $g(q)=\vec{P}_I \cdot \vec{P}_J (q) \delta^{IJ}$ for the generic family. $g(q)$ has an extremum point at the base point of the hyperscalar manifold (i.e. $\frac{dg}{dq}\arrowvert_{q^c}=0$). This leads to
\begin{equation}
\frac{\partial P_{(5)}^{(T)}}{\partial q}\arrowvert_{q^c}= \frac{\partial P_{(5)}^{(R)}}{\partial q}\arrowvert_{q^c}=\frac{\partial P_{(5)}}{\partial q}\arrowvert_{q^c}=\frac{\partial^2 P_{(5)}}{\partial \varphi \partial q}\arrowvert_{q^c}=0\nonumber
\end{equation}
and hence the Hessian is in block diagonal form. The fact that $g(q) \geq 0$ makes it impossible to convert the non-minimum critical points that correspond to the upper block of the Hessian ($\frac{\partial^2 P_{(5)}}{(\partial \varphi)^2}$) to minimum points of the potential or change its sign at the critical point. Therefore a $U(1)_H$ gauging will not change the nature of an existing critical point. One can arrive at the same result by gauging a $SU(2)_H$ subgroup of the isometry group $SU(2,1)$ of the hyperscalar manifold.

Equation (\ref{u1saddle}) does not hold for non-compact generators of the hyper isometry. Indeed gauging a non-compact hyper-symmtery generally leads to stable and unstable de Sitter ground states in 5 dimensions as was shown in a previous work \cite{Oge06}. However this topic will not be covered in this paper. Instead we will concentrate on studying de Sitter ground states that result from gauging a non-compact symmetry of the real scalar manifold of the vector multiplets.

\subsection{Two Models with Stable de Sitter Ground States\label{GJF}}
The real scalar manifolds of the two models we will discuss belong to the generic Jordan family\footnote{It is possible to embed these models into magical Jordan family theories, provided that there is a sufficient number of vector fields to perform the respective gaugings \cite{Oge06}.}. These two models will play an important role in the four dimensional stable dS vacua calculations in section \ref{sec3}.

The theory being considered is $\mathcal{N}=2$ supergravity coupled to $\tilde{n}$ Abelian vector multiplets and with real scalar manifold $\mathcal{M}^5_{VS}=SO(\tilde{n}-1,1)\times SO(1,1)/SO(\tilde{n}-1),\tilde{n}\ge 1$. The cubic polynomial can be written in the form \cite{GZ00}
\begin{equation}
N(h)=\frac{3\sqrt{3}}{2}h^1[(h^2)^2 - (h^3)^2 - ... - (h^{\tilde{n}+1})^2].
\end{equation}
The non-zero $C_{\tilde{I}\tilde{J}\tilde{K}}$'s are
\begin{equation}
C_{122}=\frac{\sqrt{3}}{2},\qquad C_{133}=C_{144}=...=C_{1,\tilde{n}+1,\tilde{n}+1}=-\frac{\sqrt{3}}{2}\nonumber
\end{equation}
and their permutations. The constraint $N=1$ can be solved by
\begin{equation}
  h^1=\frac{1}{\sqrt{3}||\varphi||^2},\qquad h^{a}=\sqrt{\frac{2}{3}}\varphi^a\label{hiconditions}
\end{equation}
with $a,b=2,3,..,\tilde{n}+1$ and $||\varphi||^2 =\varphi^a\eta_{ab}\varphi^{b}$, where $\eta_{ab}=(+--...-)$. The scalar field metric metrics $g_{\tilde{x}\tilde{y}}$ and vector field metric $\stackrel{o}{a}_{\tilde{I}\tilde{J}}$ that appear in the kinetic terms in the Lagrangian are positive definite in the region $||\varphi||^2 >0$. In order to have theories that have a physical meaning, our investigation is restricted to this region. As a consequence one must have $\varphi^2\neq 0$.\\[3pt]

The isometry group of the real scalar manifold $\mathcal{M}^5_{VS}$ is $G_{(5)} = SO(\tilde{n}-1,1)\times SO(1,1)$. The gauging of an $SO(1,1)$ or an $SO(2)$ subgroup of $SO(\tilde{n}-1,1)$ will lead to dualization of vectors to tensor fields and this gives a scalar potential term. In the generic Jordan family there are no vector fields that are nontrivially charged when the gauge group is non-Abelian, and hence gauging a non-Abelian subgroup of $G_{(5)}$ will not give a scalar potential term. It is also possible to gauge the $R$-symmetry group $SU(2)_R$ or its subgroup $U(1)_R$.
\paragraph{Gauging $SO(1,1)$ Symmetry:}

The $SO(1,1)$ subgroup of the isometry group of the scalar manifold acts nontrivially on the vector fields $A^2_{\hat{\mu}}$ and $A^3_{\hat{\mu}}$. Hence these vector fields must be dualized to antisymmetric tensor fields. The index $\tilde{I}$ is decomposed as
\begin{equation}
 \tilde{I}=(I,M)\nonumber
\end{equation}
 with $I,J,K=1,4,5,...,\tilde{n}+1$ and $M,N,P=2,3$. The fact that the only nonzero $C_{IMN}$ are $C_{1MN}$ for the theory at hand requires $A_{\hat{\mu}}^1$ to be the $SO(1,1)$ gauge field because of $\Lambda^M_{IN}\sim \Omega^{MP} C_{IPN}$ (c.f equation (\ref{pt})). All the other $A_{\hat{\mu}}^I$ with $I\neq 1$ are spectator vector fields with respect to the $SO(1,1)$ gauging. The potential term (\ref{pt}) that comes from the tensor coupling is found to be (taking $\Omega^{23}=-\Omega^{32}=-1$)
\begin{equation}
 P_{(5)}^{(T)}=\frac{1}{8} \frac{\left[ (\varphi^2)^2 - (\varphi^3)^2\right]}{||\varphi||^6} . \label{so11pt}
\end{equation}
For the function $W_{\tilde{x}}$ that enters the supersymmetry transformation laws of the fermions, one obtains
\begin{equation}
 \begin{array}{rcl}
  W_4=W_5=&...&=W_{\tilde{n}+1}=0,\\
W_2&=&-\frac{\varphi^3}{4||\varphi||^4},\\
W_3&=&\frac{\varphi^2}{4||\varphi||^4}.
 \end{array}
\end{equation}
Since $W_3$ can never vanish, there can be no $\mathcal{N}=2$ supersymmetric critical point.

Taking the derivative of the total potential $P_{(5)}=P_{(5)}^{(T)}$ with respect to $\varphi^{\tilde{x}}$, one finds
\begin{equation}
 \begin{array}{rcll}
 \partial_{\varphi^2} P_{(5)} &=& B \varphi^2,\\
\partial_{\varphi^3} P_{(5)} &=& - B \varphi^3,\\
\partial_{\varphi^b} P_{(5)} &=& -B \varphi^b +\frac{\varphi^b}{4||\varphi||^6},\quad&b=4,...,\tilde{n}+1
\end{array}\nonumber
\end{equation}
where
\begin{equation}
 B=-\frac{3}{4}\frac{(\varphi^2)^2-(\varphi^3)^2}{||\varphi||^8}+\frac{1}{4||\varphi||^6}<0 .\label{so11B}
\end{equation}
Since $\partial_{\varphi^2} P_{(5)}$ cannot be brought to zero the potential $P_{(5)}=P_{(5)}^{(T)}$ alone does not have any critical points. However, one can gauge $R$-symmetry to get additional potential terms.

\subsubsection{\texorpdfstring{$SO(1,1)\times SU(2)_R$}{SO(1,1) x SU(2)R} Symmetry Gauging}
For such a gauging one needs at least $\tilde{n}\geq 5$. Choosing $A_{\hat{\mu}}^4,A_{\hat{\mu}}^5,A_{\hat{\mu}}^6$ as the $SU(2)_R$ gauge fields one finds
\begin{equation}
 P_{(5)}=P_{(5)}^{(T)}+\lambda P_{(5)}^{(R)}\nonumber
\end{equation}
with
\begin{equation}
 P_{(5)}^{(R)}=6||\varphi||^2
\end{equation}
and $P_{(5)}^{(T)}$ given in (\ref{so11pt}). Taking the derivative of the total potential with respect to $\varphi^{\tilde{x}}$ one finds
\begin{equation}
\begin{array}{rcl}
 \partial_{\varphi^2} P_{(5)}&=& (B+12\lambda)\varphi^2\\
 \partial_{\varphi^3} P_{(5)}&=& -(B+12\lambda)\varphi^3\\
 \partial_{\varphi^b} P_{(5)}&=& -(B+12\lambda)\varphi^b +\frac{\varphi^b}{4||\varphi||^6},\qquad b=4,...,\tilde{n}+1
\end{array}
\end{equation}
with $B$ defined in (\ref{so11B}). Setting the first equation to zero means
\begin{equation}
 B=-12\lambda\label{so11B2}
\end{equation}
since $\varphi^2\neq 0$. The last equation then implies $\varphi^b_c=0$. From (\ref{so11B2}) we find
\begin{equation}
 \frac{1}{||\varphi_c||^6}=24\lambda.
\end{equation}
 The value of $||\varphi_c||^2=(\varphi_c^2)^2-(\varphi_c^3)^2$ is fixed by $\lambda$ but not $\varphi_c^2$ and $\varphi_c^3$ individually. The value of the potential at these critical points is
\begin{equation}
 P_{(5)}\arrowvert_{\varphi^c}=\frac{3}{8||\varphi_c||^4}
\end{equation}
and therefore it corresponds to a one parameter family of de Sitter ground states. The stability of the critical points is checked by calculating the eigenvalues of the Hessian of the potential, which are easily found as
\begin{equation}
 \{0,\frac{3\left[(\varphi_c^2)^2+(\varphi_c^3)^2\right]}{||\varphi_c||^8},\underbrace{\frac{1}{4||\varphi_c||^6},...,\frac{1}{4||\varphi_c||^6}}_{(\tilde{n}-2) \text{ times}}\}.\nonumber
\end{equation}
The eigenvalues are all non-negative, thus the one parameter family of de Sitter critical points is found to be stable \cite{Oge06}.

\subsubsection{\label{so11u1rhypers} \texorpdfstring{$SO(1,1)\times U(1)_R$}{SO(1,1) x U(1)R} Symmetry Gauging}

\begin{figure}
 \centering
 \includegraphics[viewport=14 14 216 126]{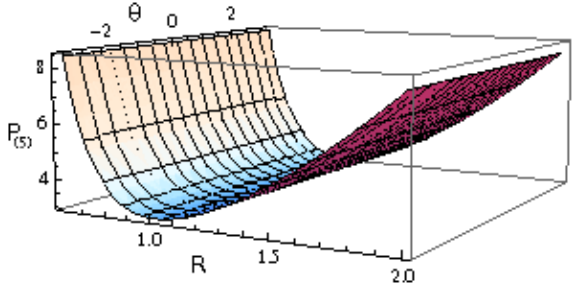}
 % stable.eps: 0x0 pixel, 300dpi, 0.00x0.00 cm, bb=14 14 216 126
\caption[The extrema of the potential $P_{(5)}$ due to $SO(1,1)\times U(1)_R$ gauging]{The extrema of the potential $P_{(5)}(R,\theta)$ due to $SO(1,1)\times U(1)_R$ gauging, evaluated at $\varphi^4=0$; $V_1=0$ and $\lambda=1$; with parametrization $\varphi^2 = R\, \cosh\theta,\,\varphi^3 = R\, \sinh\theta$. The zero eigenvalue of the Hessian corresponds to the flat direction of the potential at its minima.}\label{5dplot}
\end{figure}

The calculation in \cite{GZ00} for $\tilde{n}=3$ was later generalized to arbitrary $\tilde{n}\geq 3$ in \cite{CS05}. Let us briefly quote their results. A linear combination $A_{\hat{\mu}} [U(1)_R]=V_I A^I_{\hat{\mu}}$ of the vector fields is taken as the $U(1)_R$ gauge field. The scalar potential is now
\begin{equation}
  P_{(5)}=P_{(5)}^{(T)}+\lambda P_{(5)}^{(R)}\nonumber
\end{equation}
where
\begin{equation}
 P_{(5)}^{(R)}=-4\sqrt{2} V_1 V_i \varphi^i ||\varphi||^{-2}+2 |V|^2 ||\varphi||^2\label{u1rforso11}
\end{equation}
with $i=4,...,\tilde{n}+1$ and $|V|^2=V_i V_i$. Demanding $\partial_{\varphi^{\tilde{x}}} P_{(5)}=0$, one obtains the following conditions
\begin{equation}
 \begin{array}{rcl}
  \frac{\varphi^i_c}{||\varphi_c||^4}&=&16\sqrt{2}\lambda V_1 V_i\\
\frac{1}{||\varphi_c||^6}&=&-\frac{1}{2}(16\sqrt{2}\lambda V_1 |V|)^2 +8\lambda |V|^2\label{so11u1rconds}
 \end{array}
\end{equation}
with the constraints
\begin{equation}\begin{array}{rcl}
|V|^2&>&0\\
32\lambda (V_1)^2 &<&1.\label{so11u1rconstraints}
\end{array}\end{equation}
Given a set of $V_I$ subject to (\ref{so11u1rconstraints}), we see that $||\varphi||^2$ and $\varphi^i$ (and thus $(\varphi^2)^2-(\varphi^3)^2$) are completely determined by (\ref{so11u1rconds}) but $\varphi^2$ and $\varphi^3$ are otherwise undetermined. The value of the potential at these one parameter family of critical points becomes
\begin{equation}
 P_{(5)}\arrowvert_{\varphi^c}=3\lambda||\varphi||^2 |V|^2 (1-32\lambda(V_1)^2)
\end{equation}
and this corresponds to de Sitter vacua. The stability is checked by calculating the eigenvalues of the Hessian of the potential at the critical point. We can use the $SO(1,1)$ invariance together with the $SO(\tilde{n}-2)$ of the ${\varphi}^i$ to take for any critical point $\varphi_c = (\varphi^2,0,\varphi^4,0,...,0)$. With these choices the Hessian becomes block diagonal at the critical point. $\varphi^3$ is a zero mode and the sector $\varphi^5,...,\varphi^{\tilde{n}+1}$ consists of a unit matrix times $\frac{1}{4}||\varphi||^{-6}$. The only non-diagonal part of the Hessian is
\begin{equation}
 \partial_{\tilde{x}}\partial_{\tilde{y}}P_{(5)}\arrowvert_{\tilde{x},\tilde{y}=2,4}=\gamma\text{\scriptsize  $ \left(\begin{array}{cc}
                                                                                                                                                        (\varphi^2)^2 [6(\varphi^2)^2 + 5(\varphi^4)^2]&-\varphi^2 [8(\varphi^2)^2 \varphi^4 + 3 (\varphi^4)^3]\\
-\varphi^2 [8(\varphi^2)^2 \varphi^4 + 3 (\varphi^4)^3]&\frac{1}{4}[2(\varphi^2)^4+37 (\varphi^2)^2 (\varphi^4)^2 +5 (\varphi^4)^4]
                                                                                                                                                       \end{array}\right)$}\nonumber
\end{equation}
 with $\gamma = ||\varphi||^{-8} [2(\varphi^2)^2-(\varphi^4)^2]^{-1}$. The determinant and the trace of this part of the Hessian are
\begin{equation}\begin{array}{rcl}
 \text{det } \partial\partial P_{(5)} &=&\displaystyle  \frac{12 (\varphi^2)^6 -12 (\varphi^2)^4 (\varphi^4)^2 + 11 (\varphi^2)^2 (\varphi^4)^4}{4 ||\varphi||^{14} [2(\varphi^2)^2-(\varphi^4)^2]^2}\\[9pt]
\text{tr } \partial\partial P_{(5)} &=&\displaystyle  \frac{26(\varphi^2)^4 + 57 (\varphi^2)^2 (\varphi^4)^2 +5 (\varphi^4)^4}{4 ||\varphi||^8 [2(\varphi^2)^2-(\varphi^4)^2] }\nonumber
\end{array}\end{equation}
which are both positive because of $(\varphi^2)^2 > (\varphi^4)^2$ and therefore the family of critical points is found to be stable. We note that, although the above quantities are both positive, they are slightly different than the ones found in \cite{CS05}, where the authors fixed the coupling constants with $\lambda=1$. Figure \ref{5dplot} shows the plot of the potential (\ref{u1rforso11}) for the special case $\tilde{n}=3$, $V_1=0$ and $\lambda=1$. 

\begin{comment}At this point we want to emphasize that the stable dS vacua found by gauging $SO(1,1)\times SU(2)_R$ or $SO(1,1)\times U(1)_R$ will play an important role in the four dimensional stable dS vacua calculations in section \ref{sec3}.\end{comment}

\section{4 Dimensional \texorpdfstring{$\mathcal{N} = 2$}{N=2} Supergravity Theories\label{sec3}}

Having discussed two possible gaugings that result in de Sitter ground states from the scalar potentials of $\mathcal{N}=2$ supergravity theories with symmetric scalar manifolds in 5 dimensions, we now move on to de Sitter ground states of the 4 dimensional $\mathcal{N}=2$ supergravity theories obtained by dimensional reduction. The details of the dimensional reduction process can be found in \cite{GST83b, GMZ05b}. Here we quote the necessary tools for the calculation of the scalar potentials. The bosonic sectors of the Lagrangians before and after the dimensional reduction are given in appendix \ref{appendixveryspecial}.

Before we begin, let us see what kind of ground states we would get just by considering ordinary dimensional reduction. The dimensionally reduced potential derived from (\ref{totalpot}), in the absence of hypers\footnote{Adding hypers results in an additional $P_{(4)}^{(H)}$ in the dimensionally reduced potential (\ref{POT4__}), which is given in (\ref{phafterred}). The two terms of $P_{(4)}^{(H)}$ have the same powers of $\sigma$ and $A^I$ as the first and third terms above and can be absorbed in them by proper field redefinitions and
hence it will not change our result.} reads (\ref{potin4}, \ref{prin4})
\begin{equation}
P_{(4)}=e^{-\sigma}P_{(5)}^{(T)}+\lambda e^{-\sigma}P_{(5)}^{(R)}+\frac{3}{4}e^{-3\sigma}\stackrel{o}{a}_{\tilde{I}\tilde{J}}(A^I M^{\tilde{I}}_{I\tilde{K}}h^{\tilde{K}})(A^J M^{\tilde{J}}_{J\tilde{L}}h^{\tilde{L}}),\label{POT4__}
\end {equation}
where $M^{\tilde{I}}_{I\tilde{K}}$ are the $K_{(5)}$-transformation matrices defined in (\ref{mijk}). The scalars of the above potential are $\varphi^{\tilde{x}}$, $A^{I}$ and $\sigma$. Taking the $\sigma$-derivative of the potential, setting it equal to zero and plugging the result back into the potential gives us the value of the potential at the critical point $\phi^c$ as
\begin{equation}
P_{(4)}\arrowvert_{\phi^c}=-\frac{3}{2}e^{-3\sigma}\stackrel{o}{a}_{\tilde{I}\tilde{J}}(A^I M^{\tilde{I}}_{I\tilde{K}}h^{\tilde{K}})(A^J M^{\tilde{J}}_{J\tilde{L}}h^{\tilde{L}}).\label{d4potcr}
\end{equation}
The derivative of the potential with respect to any $A^I$ must vanish at the critical point. Hence we arrive at
\begin{equation}
A^I\frac{\partial P_{(4)}}{\partial A^I}\arrowvert_{\phi^c}=\frac{3}{2}e^{-3\sigma}A^I A^J\stackrel{o}{a}_{\tilde{I}\tilde{J}}M^{\tilde{I}}_{I\tilde{K}}M^{\tilde{J}}_{J\tilde{L}}h^{\tilde{K}}h^{\tilde{L}}=0.
\end{equation}
So if a critical point exists the potential vanishes there (c.f. equation (\ref{d4potcr})), and there is no possibility for an (anti-)de Sitter ground state. Since cosmological observations imply that the universe has a very small positive cosmological constant, we must find a way around this problem.

It was shown in \cite{GMZ05b} that the dimensionally reduced $5D$ Yang-Mills-Einstein supergravity theories coupled to tensor multiplets result in $4D$ theories that have gauge groups of the form $K_{(4)}=K_{(5)}\ltimes \mathcal{H}^{n_T +1}$, where $\mathcal{H}^{n_T +1}$ is a Heisenberg group of dimension $n_T +1$ and $\ltimes$ denotes semi-direct product. On the other hand, stable de Sitter vacua were found for $4D, \mathcal{N}=2$ theories in \cite{FTP02}, where the authors showed that the three necessary ingredients to obtain stable de Sitter vacua are non-Abelian, non-compact gauge groups, $SO(2,1)$ in particular; Fayet-Iliopoulos (FI) terms that are possible only for $SU(2)$ or $U(1)$ factors, which can be identified by the $SU(2)_R$ or $U(1)_R$ gaugings; and the de Roo-Wagemans (dRW) rotation. 
The last ingredient uses additional symmetries in 4 dimensions, where the isometry group is larger than in 5 dimensions. In order to make use of these symmetries we first need to review the structure of the complex geometry of 4 dimensional $\mathcal{N}=2$ supergravity theories. Once this is achieved it will be easier to see the 5 dimensional origins of de Sitter ground states that we will show how to obtain in 4 dimensions.

\subsection{The Geometry}
The scalar manifold of the theory we studied in the last section, when reduced to 4 dimensions, is the special K\"{a}hler manifold \cite{GST83a, GST83b}
\begin{equation}
 \mathcal{M}_{VS}^4 =  \mathcal{ST}[2,n-1]=\frac{SU(1,1)}{U(1)}\times\frac{SO(2,n-1)}{SO(2)\times SO(n-1)}.\label{skmanin4}
\end{equation}
In 4 dimensions, there are $n=\tilde{n}+1$ vector multiplets and $n$ complex scalars. The $(n+1)$ field strengths $\mathcal{F}^{A\mu\nu}$ and their magnetic duals $\mathcal{G}_{A\mu\nu}$ transform in the $(2,n+1)$ representation of the ${U}$-duality group $\mathcal{U}=SU(1,1)\times SO(2,n-1)$. The models with stable de Sitter vacua that we will discuss in this section originate from the 5 dimensional YMESGT's with gauge groups $SO(1,1) \times U(1)_R$ or $SO(1,1) \times SU(2)_R$. The $SO(1,1)$ factor, as we will show, will become a subgroup of $SO(2,n-1)$ in 4 dimensions. This is similar to the models with stable de Sitter vacua found in \cite{FTP02} where the full $SO(2,1)$ is gauged. Note that the $SU(1,1)_G$ symmetry of the pure $5D,\, \mathcal{N}=2$ supergravity reduced to 4 dimensions is not the $SU(1,1)=SO(2,1)$ factor in the 4 dimensional $U$-duality group $\mathcal{U}$. It is rather a diagonal subgroup of $SU(1,1)$ times an $SO(2,1)$ subgroup of $SO(2,n-1)$ under which the following decompositions occur \cite{GMZ05a}:

\begin{footnotesize}
\begin{equation}
 \begin{array}{ccccl}
 SO(2,1)\times SO(2,n-1)&\supset&SO(2,1)\times SO(2,1)\times SO(n-2)&\supset&SO(2,1)_G\times SO(n-2)\\
(2,n+1)&=&(2,3,1)\oplus(2,1,n-2)&=&(4,1)\oplus(2,1)\oplus(2,n-2)\nonumber
 \end{array}
\end{equation}
\end{footnotesize}

Note that the four dimensional graviphoton transforms in the spin-$3/2$ representation of $SO(2,1)_G$ along with some linear combination of the other vectors in the theory and due to the mixing, one can say that it does not descend directly from the five dimensional graviphoton. Instead, it is a linear combination of the vector that comes from the dimensional reduction of the f\"{u}nfbein and the vector that is obtained by the dimensional reduction of the five dimensional graviphoton. We will address this issue in subsection \ref{anewbasis}.

The scalars can be used to define the complex coordinates \cite{GST83b, GMZ05b}
\begin{equation}
 z^{\tilde{I}} = \frac{1}{\sqrt{3}} A^{\tilde{I}} +\frac{i e^\sigma}{\sqrt{2}} h^{\tilde{I}}.\label{zidefined}
\end{equation}
These $n$ complex coordinates can be interpreted as the inhomogeneous coordinates of the $(n+1)$-dimensional complex vector $(\tilde{I}=1,...,n)$
\begin{equation}
 X^A=\left(\begin{array}{c}
            X^0\\
	    X^{\tilde{I}}
           \end{array}
\right)=\left(\begin{array}{c}
            1\\
	    z^{\tilde{I}}
           \end{array}
\right).
\end{equation}
One can introduce the prepotential\footnote{Note that the prepotential given here differs by a factor $\sqrt{6}$ from that of \cite{GMZ05b}.}
\begin{equation}
F(X^A)=-\frac{1}{3\sqrt{3}} C_{\tilde{I}\tilde{J}\tilde{K}} \frac{X^{\tilde{I}} X^{\tilde{J}} X^{\tilde{K}}}{X^0}
\end{equation}
to write the holomorphic (symplectic) section
\begin{equation}\begin{array}{rcl}
 \Omega_0 &=& \left(\begin{array}{c}
                   X^A\\F_B
                  \end{array}
	\right)=\left(\begin{array}{c}
                   X^A\\\partial_B F
                  \end{array}
	\right)\\&=&\left(\begin{array}{c}
                   X^0\\X^I\\X^M\\F_0\\F_I\\F_M
                  \end{array}
	\right)=\left(\begin{array}{c}
                   1\\z^I\\z^M\\ 
\frac{1}{3\sqrt{3}} [C_{IJK} z^I z^J z^K + 3 C_{IMN} z^I z^M z^N]\\
-\frac{1}{\sqrt{3}} [C_{IJK} z^J z^K + C_{IMN} z^M z^N]\\
-\frac{2}{\sqrt{3}}C_{MNI} z^N z^I
                  \end{array}
	\right)
\end{array}\label{om0}\end{equation}
with $\tilde{I}=(I,M)$. The reason why the above manifold is called a \textit{special K\"{a}hler manifold} is that one can write a K\"{a}hler potential in terms of the holomorphic section $\Omega$
\begin{equation}
 \mathcal{K}=-\log\left(i\langle\Omega|\bar{\Omega}\rangle\right)=-\log\left[i\left(\bar{X}^A F_A - \bar{F}_A X^A\right)\right].
\end{equation}
The K\"{a}hler potential is used to form the K\"{a}hler metric on the scalar manifold $\mathcal{M}_{VS}^4$ of the four-dimensional theory as
\begin{equation}
 g_{\tilde{I}\bar{\tilde{J}}}\equiv \partial_{\tilde{I}} \partial_{\bar{\tilde{J}}}  \mathcal{K}.\label{gibarj}
\end{equation}
It is also possible to introduce the covariantly holomorphic section\cite{FS89, Str90, CDF90, CDGP90}
\begin{equation}
 V=\left(\begin{array}{c}
          L^A\\M_B
         \end{array}
\right)\equiv e^{ \mathcal{K}/2} \Omega = e^{ \mathcal{K}/2}\left(\begin{array}{c}
          X^A\\F_B
         \end{array}
\right)
\end{equation}
which obeys
\begin{equation}
 \nabla_{\bar{\tilde{I}}} V = \left(\partial_{\bar{\tilde{I}}} - \frac{1}{2}\partial_{\bar{\tilde{I}}} \mathcal{K}\right) V=0.
\end{equation}
By defining 
\begin{equation}
 U_{\tilde{I}}=\nabla_{\tilde{I}}V=\left(\partial_{\tilde{I}}+\frac{1}{2}\partial_{\tilde{I}}\mathcal{K}\right)V\equiv\left(\begin{array}{c}
          f^A_{\tilde{I}}\\h_{B|\tilde{I}}
         \end{array}
\right)
\end{equation}
the \textit{period matrix} is introduced via relations
\begin{equation}
 \bar{M}_A=\bar{\mathcal{N}}_{AB}\bar{L}^B\qquad ; \qquad h_{A|\tilde{I}}=\bar{\mathcal{N}}_{AB} f^B_{\tilde{I}}
\end{equation}
which can be solved by introducing two $(n+1)\times(n+1)$ vectors
\begin{equation}
 f^A_{\tilde{C}}=\left(\begin{array}{c}
                        f^A_{\tilde{I}}\\\bar{L}^A
                       \end{array}
\right)\qquad ; \qquad h_{A|\tilde{C}}=\left(\begin{array}{c}
                        h_{A|\tilde{I}}\\\bar{M}_A
                       \end{array}
\right)
\end{equation}
and setting
\begin{equation}
 \bar{\mathcal{N}}_{AB} = h_{A|\tilde{C}}\cdot \left(f^{-1}\right)^{\tilde{C}}_{\phantom{\tilde{C}}B}.
\end{equation}
Whenever the prepotential $F$ exists, the period matrix has the form \cite{dWP84, Cre84, dWLP85}
\begin{equation}
 \mathcal{N}_{AB} = \bar{F}_{AB} + 2i \frac{\text{Im}(F_{AC}) \text{Im}(F_{BD}) L^C L^D}{\text{Im}(F_{CD}) L^C L^D}
\end{equation}
where $F_{AB}=\partial_A \partial_B F$.

A symplectic rotation $C$ of the holomorphic section obeys $C^T \omega C=\omega$ for
\begin{equation} \omega=\left(\begin{array}{cc}
                                                                               0&{\unity_{n+1}}\\-{\unity_{n+1}}&0
                                                                              \end{array}
\right).\nonumber
\end{equation}
\subsection{Gauge Group Representation and dRW Angles}
The special K\"{a}hler manifold (\ref{skmanin4}) of vector multiplets has the isometry group $G_{(4)}=SU(1,1)\times SO(2,n-1)$. If we are to gauge a subgroup $K_{(4)}\subset G_{(4)}$, then the symplectic representation $R$ of $G_{(4)}$, under which the electric field strengths and their magnetic duals transform must be decomposed as
\begin{equation}
 \begin{array}{rcl}
  G_{(4)}&\supset&K_{(4)},\\
R&=&\text{adj.} + \text{adj.} + \text{singlets} + \text{singlets} .
 \end{array}
\end{equation}
The electric and magnetic field strengths are in the doublet representation of $SU(1,1)$ and in \textbf{$n+1$} vector representation of $SO(2,n-1)$. The non-compact non-Abelian gauge group $K_{(4)}=SO(2,1)$ which is a necessary ingredient to obtain stable de Sitter vacua in $4D,\, \mathcal{N}=2$ supergravity is embedded in $SO(2,n-1)$. The $SO(2,1)$ generators $t_A$ form an adjoint representation. The symplectic embedding of this representation into the fundamental representation of $Sp(2(n+1),\mathbb{R})$ is given by
\begin{equation}
 T_A = \left(\begin{array}{cc}
{t}_{A} & 0\\
0& -t^T_A
\end{array}
\right)\, \in \, Sp\left(2(n+1),\mathbb{R}\right),\quad A=0,2,3,
\end{equation}
and the corresponding algebra $[T_A,T_B]=f^C_{AB}\, T_C$ is
\begin{equation}
 [T_0, T_{2}] = T_{3},\qquad [T_{2}, T_{3}] = -T_{0},\qquad [T_{3}, T_{0}] = -T_{2}.\label{so21_alg}
\end{equation}
Here, $f^C_{AB}$ are the structure constants of the algebra.

In addition to the $SO(2,1)$, one can gauge a $U(1)_R$ (or $SU(2)_R$) $R$-symmetry group for theories with $n>2$ (or $n>4$) vector multiplets using the remaining vectors (or a linear combination of them) as gauge fields. The de Roo - Wagemans (dRW) angles, as first introduced for $\mathcal{N}=4$ supergravity \cite{dRW85, Wag90} and later used in $\mathcal{N}=2$ supergravity as an ingredient to obtain de Sitter vacua \cite{FTP02, FTP03}, parametrize the relative embedding of the $R$-symmetry group within $Sp\left(2(n+1),\mathbb{R}\right)$. They mix the electric and magnetic components of the symplectic section prior to the gauging by a ``non-perturbative'' rotation. The dRW rotation matrix has to be chosen in such a way that it commutes with $SO(2,1)$ symmetry gauging. For example, we will use the following dRW matrix for the models where we gauge a $SO(2,1)\times U(1)_R$ symmetry \cite{FTP02, FTP03}:
\begin{equation}
 \mathcal{R} = \left(\begin{array}{cccc}
                      \unity_n&0&0&0\\
		      0&\cos\theta&0&\sin\theta\\
		      0&0&\unity_n&0\\
		      0&\sin\theta&0&\cos\theta
                     \end{array}\label{dRWmatrix}
\right).
\end{equation}
The holomorphic section and the covariantly holomorphic section are rotated via
\begin{equation}\begin{array}{rcl}
 \Omega \rightarrow \Omega_R &=& \mathcal{R}\, \Omega,\\
V \rightarrow V_R &=& \mathcal{R}\, V.
\end{array}\end{equation}

\subsection{Symplectic Rotation}

The symplectic section (\ref{om0}) is written in the most natural way when one comes from 5 down to 4 dimensions. But it has shortcomings. The translations $z^M\rightarrow z^M+b^M$ act on the symplectic section in such a way that the electric components mix with magnetic ones so that the transformation matrix is not block-diagonal, which is not suitable if symmetries are to be gauged in the standard way. In this section we will give two inequivalent examples of symplectic rotations that will bring $\Omega_0$ in bases where this problem does not occur. 
\subsubsection[GMZ Rotation]{G\"{u}nayd\i n-McReynolds-Zagermann (GMZ) Rotation\label{gmzrot}}
We start with observing how $\Omega_0$ varies under the infinitesimal translation $z^M \rightarrow z^M + b^M$ \cite{GMZ05b}:
\begin{equation}
\Omega_0=\left(
 \begin{array}{c}
 X^0\\ X^I\\ X^M\\ F_0\\ F_I\\F_M
\end{array}\right)\rightarrow
\left(
 \begin{array}{c}
 X^0\\ X^I\\ X^M\\ F_0\\ F_I\\F_M
\end{array}\right) +
\left(
 \begin{array}{c}
 0\\ 0\\ b^M X^0\\ -b^M F_M\\ -\frac{2}{\sqrt{3}}b^M C_{IMN} X^N\\-\frac{2}{\sqrt{3}}b^N C_{IMN} X^I
\end{array}\right)\label{omatrf}
\end{equation}
In the original basis a combined infinitesimal translation and infinitesimal $K$ transformation with parameter $\alpha^I$ is generated by
\begin{equation}
 \mathcal{O}  = \unity_{2n+2} +  \left(  \begin{array}{cc}
 B & 0 \\
 C & -B^{T}
 \end{array}  \right)  ,
 \end{equation}
  with
\begin{equation}
B =  \left( \begin{array}{ccc}
0 & 0 & 0 \\
0 & \alpha^I f_{IJ}^{K} & 0\\
b^{M} & 0 & \alpha^I \Lambda_{IN}^{M}
 \end{array}   \right)   , \qquad C =
   \left(    \begin{array}{ccc}
   0&0&0\\
   0 & 0 & B_{IM}\\
   0 & B_{MI} & 0 \end{array}
      \right)   ,
\end{equation}
where
\begin{equation}
B_{IM} := \frac{-2}{ \sqrt{3}} C_{IMN}b^{N}.
\end{equation}

By having a closer look at (\ref{omatrf}) we see that $(X^0, F_I, X^M)$ transform among themselves, as do $(F_0, X^I, F_M)$. In order to make the translations block diagonal we exchange $F_0$ with $X^0$ and $F_M$ with $X^M$. The symplectic rotation 
\begin{equation}
  \left(  \begin{array}{c}
  X^{A} \\
  F_{B}
  \end{array}   \right)   \rightarrow    \left(  \begin{array}{c}
  \check{X}^{A} \\
  \check{F}_{B}
  \end{array}   \right)   \equiv    \mathcal{S}   \left(  \begin{array}{c}
  X^{A} \\
  F_{B}
  \end{array}   \right) , \qquad  \left(  \begin{array}{c}
  F_{\mu\nu}^{A} \\
  G_{\mu\nu B}
  \end{array}   \right)   \rightarrow    \left(  \begin{array}{c}
  \check{F}_{\mu\nu}^{A} \\
  \check{G}_{\mu\nu B}
  \end{array}   \right)   \equiv  \mathcal{S}   \left(  \begin{array}{c}
  F_{\mu\nu}^{A} \\
  G_{\mu\nu B}
  \end{array}   \right),\nonumber
\end{equation}
\begin{equation}
\left(  \begin{array}{c}
  L^{A} \\
  M_{B}
  \end{array}   \right)   \rightarrow    \left(  \begin{array}{c}
  \check{L}^{A} \\
  \check{M}_{B}
  \end{array}   \right)   \equiv    \mathcal{S}   \left(  \begin{array}{c}
  L^{A} \\
  M_{B}
  \end{array}   \right)
\end{equation}
that achieves this is \cite{GMZ05b}
\begin{equation}
\mathcal{S} =  \left(  \begin{array}{cccccc}
0 & 0 & 0 & 1 & 0 & 0 \\
0 & \delta^{J}{}_{I} & 0 &0& 0& 0\\
0 & 0 & 0 & 0 & 0 & D^{MN}\\
-1 & 0 & 0 & 0 & 0 & 0 \\
0 & 0& 0& 0& \delta_{I}{}^{J}& 0 \\
0 & 0 & D_{MN}  &0&0&0
\end{array}
\right) .\label{GMZ_Sdef}
\end{equation}
where $D_{MN}=-\sqrt{2}\, \Omega_{MN}$ and $D_{MN} D^{NP} = \delta^P_M$.\\[2pt]

%\underline{\textbf{The Potential Terms:}}
\paragraph{The Potential Terms:}

The holomorphic Killing vectors 
\begin{equation}
K_{A}^{\ensuremath{\tilde{I}}} = i g^{\ensuremath{\tilde{I}}\bar{\ensuremath{\tilde{J}}}} \partial_{\bar{\ensuremath{\tilde{J}}}} P_{A} ,\label{kill4}
\end{equation}
that are determined in terms of the Killing prepotentials \cite{ABCDFFM96, dWP84,BW83, DFF91, dWLP85}
\begin{equation}
P_{A}=e^{\mathcal{K}} ( \check{F}_{B} f_{AC}^{B} \bar{\check{X}}^{C} + \bar{\check{F}}_{B}f_{AC}^{B}\check{X}^{C}) \label{prekill4}
\end{equation}
can be used to show that the potential in the canonical form
\begin{equation}
 V=e^{K} (\check{X}^{A}\bar{K}_{A}^{\ensuremath{\tilde{I}}}) g_{\bar{\ensuremath{\tilde{I}}}\ensuremath{\tilde{J}}}(\bar{\check{X}}^{B}
 K_{B}^{\ensuremath{\tilde{J}}})\end{equation}
is indeed equal to $P_{(4)}^{(T)}$ of (\ref{potin4}) \cite{GMZ05b}. Here, $f^A_{BC}$'s are the structure constants of the gauge group.

Now we turn to the calculation of the potential $P_{(4)}^{(R)}$ rising from the $R$-symmetry gauging. For gauge groups with $U(1)$ or $SO(3)=SU(2)$ factors there is a superrenormalizable term, known as a Fayet-Iliopoulos (FI) term\cite{FI74, Wei00} that can be added to the Lagrangian. The variation of this term under a supersymmetry transformation is a total derivative and it yields a supersymmetric term in the action. FI terms are used in effective field theories for standard model building or cosmology quite often. It has been recently emphasized that these terms in $\mathcal{N}=1$ or $\mathcal{N}=2, D=4$ supersymmetric models are related to $R$-symmetry gauging \cite{BDKP04,Pro04}. Here we will verify this statement by reformulating an already known $P_{(4)}^{(R)}$ potential, coming from 5 dimensions, in terms of complex geometry elements and comparing the expressions. The potential term we will consider is given by \cite{ABCDFFM96}
\begin{equation}
 V'=(U^{(AB)} - 3 \bar{\check{L}}^{(A} \check{L}^{B)}) \mathcal{P}_A^x \mathcal{P}_B^x\label{v'}
\end{equation}
where $U^{AB}$ is defined as
\begin{equation}
U^{AB} \, \equiv \,  f^A_{\tilde{I}} \, f^B_{\bar{\tilde{J}}} \,
g^{\tilde{I}\bar{\tilde{J}}} =
-{\frac{1}{2}} \, \left ( {\rm Im}{\cal N} \right )^{-1 \vert
AB} \, -\, \bar{\check{L}}^A \check{L}^B.
\end{equation}
The negative definite term in (\ref{v'}) is the gravitino mass contribution, while the $U^{AB}$ term is the gaugino shift contribution. $\mathcal{P}_A^x$ are called the \textit{triholomorphic moment map}s for the gauge group action on quaternionic scalars with $x$ being an $SU(2)$ index. When a hypermultiplet is coupled to the theory, the potential (\ref{v'}) carries contact interactions between the real- and hyperscalars. In this case the triholomorphic moment maps $\mathcal{P}_A^x$ that describe the action of the $R$-symmetry gauge group on the quaternionic scalars are associated to the Killing prepotentials of the isometries of the hyperscalar manifold \cite{ABCDFFM96, Mar06}.  This is analogous to the 5 dimensional theory (c.f. appendices \ref{appendixveryspecial} and \ref{appendixhypers}). An FI term can be assigned to the moment maps if (and only if \cite{Pro04}) hypers are absent from the theory. For such models the triholomorphic moment maps satisfy the equivariance condition \cite{ABCDFFM96, FTP02, Mar06}
\begin{equation}
 -\epsilon^{xyz} \mathcal{P}_A^y \mathcal{P}_B^z = f^C_{AB} \mathcal{P}_C^x .
\end{equation}
In the $SU(2)_R$ case one can set $f^x_{yz} = e \epsilon_{xyz}$, where $e$ is some number, and this condition is satisfied via
\begin{equation}
 \mathcal{P}_A^x = \left\{ \begin{array}{ll}
                           -\, \delta^x_y&\textrm{for } A=3+y\\
			    0		&\textrm{otherwise},
                          \end{array}\right.\label{mmsu2}
\end{equation}
whereas in the $U(1)_R$ case, for each generator one can set an FI term
\begin{equation}
 \mathcal{P}_A^x = \left\{ \begin{array}{ll}
                           e\, \delta^x_3&A\textrm{: index for the $U(1)_R$ gauge vector}\\
			    0		&\textrm{otherwise}.
                          \end{array}\right.
\end{equation}
\\
\textbf{\underline{\textit{Example:}}}\, Let us now calculate the $V'$ potential for a specific model with $n=4$ vector multiplets where the $U(1)_R$ gauge field is a linear combination of $A_\mu^1$ and $A_\mu^4$. This is indeed the model we discussed in subsection \ref{so11u1rhypers} before the dimensional reduction. Using (\ref{u1rforso11}) and (\ref{prin4}) one can write the $U(1)_R$ potential in 4 dimensions as

\begin{equation}
 P_{(4)}^{(R)}=e^{-\sigma}P_{(5)}^{(R)}=e^{-\sigma}\left(-4\sqrt{2} V_1 V_4 \varphi^4 ||\varphi||^{-2} + 2 (V_4)^2 ||\varphi||^2\right).\label{pr14}
\end{equation}
On the other hand, the moment map for this type of gauging can be written as
\begin{equation}
 \mathcal{P}^x_A = \delta^{x3} \left(e_1 \delta_{A1} + e_4 \delta_{A4}\right)
\end{equation}
where $e_1$ and $e_4$ parametrize the linear combination of the gauge fields. Then the potential (\ref{v'}) becomes
\begin{equation}
 V'_{n=4} = e_1^2 (U^{(11)} - 3 \bar{\check{L}}^{(1} \check{L}^{1)}) + 2e_1 e_4  (U^{(14)} - 3 \bar{\check{L}}^{(1} \check{L}^{4)}) + e_4^2 (U^{(44)} - 3 \bar{\check{L}}^{(4} \check{L}^{4)})
\end{equation}
and after some calculation one can find
\begin{equation}
 \begin{array}{rcl}
  U^{(11)} - 3 \bar{\check{L}}^{(1} \check{L}^{1)}&=&0,\\
U^{(14)} - 3 \bar{\check{L}}^{(1} \check{L}^{4)}&=&-\frac{\text{Im}z^4}{(\text{Im}z^2)^2-(\text{Im}z^3)^2-(\text{Im}z^4)^2},\\
U^{(44)} - 3 \bar{\check{L}}^{(4} \check{L}^{4)}&=&\frac{1}{2\text{Im}z^1}.\label{gmzus}
 \end{array}
\end{equation}
By using (\ref{zidefined}) and (\ref{hiconditions}) on (\ref{gmzus}), we conclude that $V'_{n=4} = P_{(4)}^{(R)}$ if we identify $e_1 = \pm (\frac{8}{3})^{\frac{1}{4}} V_1$ together with $e_4 = \pm (\frac{8}{3})^{\frac{1}{4}} V_4$.

One can arrive at a similar conclusion by gauging the full $SU(2)_R$ instead. In this case $P_{(4)}^{(R)}=6e^{-\sigma} ||\varphi||^2$ and $V'=3/(2{\rm Im}z_1)$ which are again directly proportional to each other.

\subsubsection{A New Basis\label{anewbasis}}
The GMZ rotation we discussed in the last subsection resolves the block-diagonality problem of translational symmetries but there are a few more steps to take in order to find a symplectic section that will allow us to find de Sitter vacua. First, it is convenient to work in a symplectic section that satisfies the constraint
\begin{equation}
 X^A \eta_{AB} X^B = F_A \eta^{AB} F_B = 0\label{so2ninv}
\end{equation}
for $\eta_{AB} = {\rm diag}(++-...-)$\footnote{In general, the order of the $+$ and $-$ entries depend on the type of gauging but their numbers are fixed.} so that the $SO(2,n-1)$ invariance is evident. Note that we restrict our analysis to the generic Jordan family (\ref{skmanin4}). Other types of scalar manifolds will be discussed in section \ref{magical4d}.

 Under infinitesimal translations $z^M\rightarrow z^M +b^M$, $\Omega_0$ transforms as in (\ref{omatrf}). We noted that $(X^0, F_I, X^M)$ transform among themselves, as do $(F_0, X^I, F_M)$. This time we are exchanging some of $X^I$ with $F_I$ keeping in mind that we are constrained by (\ref{so2ninv}). Exchanging all of $X^I$ with $F_I$ will not leave this equation invariant. Therefore we decompose the index $I$ as $I=(1,i)$ and swap $X^1$ with $F_1$ and keep the other $X^i$ and $F_i$ intact. By looking at (\ref{omatrf}) we see that one must have \begin{equation} b^M C_{iMN} X^N=0 \label{cimn0}\end{equation} in order to keep the translations block diagonal. This is indeed satisfied for all types of gaugings of the generic Jordan family isometries.

As we discussed earlier, the bare graviphoton in 4 dimensions is a linear combination of the vectors $A^0_\mu$ and $A^1_\mu$ which are obtained by reduction from 5 dimensions. By taking a linear combination of $X^0$ and $F_1$ ($F_0$ and $X^1)$ for $\tilde{X}^0$ ($\tilde{F}_0$) we isolate the bare graviphoton as $\tilde{A}^0_\mu$. The new symplectic section $\tilde{\Omega}$ is given by the rotation of $\Omega_0$ by 
\begin{equation}
 \tilde{\mathcal{S}}=\left(
\begin{array}{cccccccc}
 \frac{1}{\sqrt{2}}&0&0&0&0&\frac{1}{\sqrt{2}}&0&0\\[3pt]
0&0&0&\tilde{\Lambda}^M_{1N}&0&0&0&0\\
0&0&\delta^j_{\phantom{j}i}&0&0&0&0&0\\
\frac{1}{\sqrt{2}}&0&0&0&0&-\frac{1}{\sqrt{2}}&0&0\\
0&-\frac{1}{\sqrt{2}}&0&0&\frac{1}{\sqrt{2}}&0&0&0\\
0&0&0&0&0&0&0&\tilde{\Lambda}_{1M}^{\phantom{1M}N}\\
0&0&0&0&0&0&\delta_j^{\phantom{j}i}&0\\
0&\frac{1}{\sqrt{2}}&0&0&\frac{1}{\sqrt{2}}&0&0&0
\end{array}
\right).\label{orcansS}
\end{equation}
The rescaling $\tilde{\Lambda}^M_{IN}\equiv \sqrt{2} \Lambda^M_{IN} = \frac{2}{\sqrt{3}} \Omega^{MP} C_{INP}$ is done for future convenience.
It is easy to verify that the matrix $S$ is symplectic. More explicitly, we have

\begin{equation}\begin{array}{rcl}
 \tilde{\Omega} &=& \left(
\begin{array}{c}
 \tilde{X}^0\\
 \tilde{X}^M\\
 \tilde{X}^j\\
 \tilde{X}^1\\
 \tilde{F}_0\\
 \tilde{F}_M\\
 \tilde{F}_j\\
 \tilde{F}_1
\end{array}
\right)= \tilde{\mathcal{S}} \Omega_0 = \tilde{\mathcal{S}} \left(
\begin{array}{c}
  X^0\\ X^1\\X^i\\ X^N\\ F_0\\ F_1\\F_i\\F_N
\end{array}
\right)
\end{array}\nonumber\qquad\qquad\qquad\qquad\qquad\qquad\quad
\end{equation}

\begin{equation}\begin{array}{rcl}
 &=& \left(
\begin{array}{c}
  \frac{1}{\sqrt{2}}-\frac{1}{\sqrt{6}} (C_{1JK} z^J z^K + C_{1MN} z^M z^N)\\ \tilde{\Lambda}^M_{1N} z^N\\z^i\\
\frac{1}{\sqrt{2}}+\frac{1}{\sqrt{6}} (C_{1JK} z^J z^K + C_{1MN} z^M z^N) \\ 
-\frac{1}{\sqrt{2}} z^1 + \frac{1}{3\sqrt{6}} (C_{IJK} z^I z^J z^K + 3 C_{IMN} z^I z^M z^N)\\ 
-\frac{2}{\sqrt{3}} \tilde{\Lambda}_{1M}^{\phantom{1M}P} C_{PNI} z^N z^I\\
-\frac{1}{\sqrt{3}} C_{iJK} z^J z^K \\
\frac{1}{\sqrt{2}} z^1 + \frac{1}{3\sqrt{6}} (C_{IJK} z^I z^J z^K + 3 C_{IMN} z^I z^M z^N)
\end{array}
\right).\label{omegaBdefined}
\end{array}\end{equation}
Here $0$ is now the graviphoton index. The combined infinitesimal $z^M\rightarrow z^M + b^M$ translation and infinitesimal $K_{(5)}$ transformation with parameter $\alpha^I$ is generated by the symplectic matrix
\begin{equation}
 \tilde{\mathcal{O}}\equiv \tilde{\mathcal{S}} \mathcal{O} \tilde{\mathcal{S}}^{-1} = \unity_{2n+2} +\left(
\begin{array}{cc}
 \tilde{B} & \tilde{C}\\
\tilde{C}^T & -\tilde{B}^T
\end{array}
\right),
\end{equation}
with
\begin{equation}\tilde{B}=\left(
 \begin{array}{cccc}
  0 & \frac{1}{\sqrt{2}} \tilde{\Lambda}_{1M}^{\phantom{1M}P} B_{P1} &0&0\\
\frac{1}{\sqrt{2}} \tilde{\Lambda}_{1N}^M  b^N & \alpha^I \Lambda_{IN}^M &0& \frac{1}{\sqrt{2}} \tilde{\Lambda}_{1N}^M  b^N\\
0&0&\alpha^I f_{Ij}^k&0\\
0&-\frac{1}{\sqrt{2}} \tilde{\Lambda}_{1M}^{\phantom{1M}P} B_{P1}&0&0
 \end{array}\right),\nonumber
\end{equation}
\begin{equation}
\tilde{C} =\frac{1}{\sqrt{2}}\left(
\begin{array}{cccc}
 0&0& -\alpha^I f_{Ij}^1 & 0\\
0&0&0&0\\
-\alpha^I f_{I1}^j&0&0&\alpha^I f_{I1}^j\\
0&0&\alpha^I f_{Ij}^1&0
\end{array}
\right)
\end{equation}
where $B_{IM}:=-\frac{2}{\sqrt{3}} C_{IMN} b^N$. In order to represent the combined translations and $K_{(5)}$ transformations by block diagonal matrices one must have an algebra with $f_{Ij}^1=f_{I1}^j=0$. Here the index $1$ corresponds to the five dimensional graviphoton, which can only be a gauge field if the gauge group is Abelian because it is a singlet under the action of five dimensional isometry group $SO(\tilde{n}-1,1)$. Therefore this condition is automatically satisfied and hence $\tilde{C}=0$.
Next, by setting $\tilde{B}^C_{\phantom{C}B}=\alpha^A f^C_{AB}$ one can find
\begin{equation}
 f^j_{ik},\quad f^M_{IN}= \Lambda^M_{IN},\quad f^0_{MN}=-f^1_{MN}=-\frac{1}{\sqrt{3}} \Lambda_{1M}^{\phantom{1M}P} C_{1PN} ,\quad f^M_{N0}=f^M_{N1} =-{\Lambda}^M_{1N}\label{str_const_w_cc}
\end{equation}
as non-vanishing components, as well as $\alpha^M = - b^M$.

\subsection{de Sitter Vacua\label{dsvacin4}}
We will now demonstrate how to obtain stable de Sitter vacua by starting with the holomorphic section (\ref{omegaBdefined}). The model to be considered is $4D,\, \mathcal{N}=2$ supergravity coupled to $n=4$ vector multiplets with gauge group $K_{(4)}=SO(2,1)\times U(1)_R$. This model can be trivially extended to arbitrary $n$ as we will discuss at the end of this section. Note that this type of gauging was first used in \cite{FTP02,FTP03} to obtain de Sitter vacua where the authors preferred to use Calabi-Vesentini coordinates to parametrize the complex scalars. The mapping between our notation and theirs can be found in appendix \ref{app_trf}. 

\subsubsection{Potential \texorpdfstring{$P_{(4)}^{(T)}$}{P(4)(T)} from Global Isometry Gauging}
The global isometry group $G_{(4)}$ for the model with 4 vector multiplets is $SU(1,1)\times SO(2,3)$. A potential is introduced by gauging the subgroup $SO(2,1)\subset SO(2,3)$.
\begin{equation}
P_{(4)}^{(T)} = e^{K} (\check{X}^{A}\bar{K}_{A}^{\ensuremath{\tilde{I}}}) g_{\bar{\ensuremath{\tilde{I}}}\ensuremath{\tilde{J}}}(\bar{\check{X}}^{B}
 K_{B}^{\ensuremath{\tilde{J}}}).\label{pt_VP} 
\end{equation}
The structure constants $f^A_{BC}$ of the $SO(2,1)$ algebra (\ref{so21_alg}) read
\begin{equation}
 f^3_{02}=f^2_{03}=-f^3_{20}=-f^2_{30}=1,\qquad f^0_{32}=-f^0_{23}=1.
\end{equation}
The gauge fields are the ``timelike'' $\tilde{A}_\mu^0, \tilde{A}_\mu^2$ and the ``spacelike'' $\tilde{A}_\mu^3$ with respect to $SO(2,3)$ with signature $(++---)$; and the Killing vectors determined by (\ref{prekill4}) and (\ref{kill4}) are given by
\begin{equation}
 \vec{K}_0 = \left(\begin{array}{c}
                    0\\ - w_3\\ -w_2\\0
                   \end{array}
 \right),
\vec{K}_2 = \left(\begin{array}{c}
                    0\\ - \frac{1}{\sqrt{2}}w_2 w_3\\ \frac{1}{2\sqrt{2}}(2-w_2^2-w_3^2+w_4^3)\\ -\frac{1}{\sqrt{2}} w_3 w_4
                   \end{array}
 \right),\nonumber
\end{equation}
\begin{equation}
\vec{K}_3 = \left(\begin{array}{c}
                    0\\ \frac{1}{2\sqrt{2}}(2+w_2^2+w_3^2+w_4^3)\\ \frac{1}{\sqrt{2}} w_2 w_3\\ \frac{1}{\sqrt{2}} w_2 w_4
                   \end{array}
 \right)\label{so21killings}
\end{equation}
where we defined $w_{\tilde{I}}\equiv z^{\tilde{I}}$. The full potential term is given in (\ref{ptso21app}). It simplifies significantly when evaluated at Re$(w_i) = 0$ 
\begin{equation}
 P_{(4)}^{(T)}\arrowvert_{{\rm Re} (w_i) = 0} = \frac{\left(\text{Im}w_2^2-\text{Im}w_3^2\right) \left(2+ ||\text{Im}w||^2\right){}^2 }{16\,
   \text{Im}w_1\, ||\text{Im}w||^4}
\end{equation}
with $||\text{Im}w||^2 \equiv \left(\text{Im}w_2^2-\text{Im}w_3^2-\text{Im}w_4^2\right)$. Note also that this potential term satisfies 
\begin{equation}\frac{\textstyle\partial P_{(4)}^{(T)}}{\textstyle\partial {\rm Re} (w_i)}\arrowvert_{{\rm Re} (w_i) = 0} =0.
 \end{equation}

\subsubsection{\texorpdfstring{$U(1)_R$}{U(1)R} Potential}
We are considering a theory with $n=4$ vector multiplets, and the vector field that gauges the $U(1)_R$-symmetry is $\tilde{A}^1_\mu$. Hence we choose the moment map to be
\begin{equation}
 \mathcal{P}^x_A = \delta^{x3} \delta_{A1}.
\end{equation}
Then the $U(1)_R$ potential term is given by 
\begin{equation}
 P_{(4)}^{(R)}=  U^{11} - 3 \bar{\tilde{L}}^1 \tilde{L}^1 \label{pot_ULL}
\end{equation}
with the following definitions
\begin{equation}\begin{array}{rcl}
 \tilde{L}^A &\equiv& S\, L^A\\
U^{AB} \, &\equiv& \,  f^A_{\tilde{I}} \, f^B_{\bar{\tilde{J}}} \,
g^{\tilde{I} \bar{\tilde{J}}} =
-{\frac{1}{2}} \, \left ( {\rm Im}{\cal N} \right )^{-1 \vert
AB} \, -\, \bar{\tilde{L}}^A \tilde{L}^B ,
\end{array}\end{equation}
$A,B = (0,2,3,4,1)$.

\subsubsection{No dRW-rotation}
For simplicity let's assume no de Roo-Wagemans rotation. One can show that \cite{ABCDFFM96}
\begin{equation}
 U^{AB} - 3 \bar{\tilde{L}}^A \tilde{L}^B = -\frac{\eta^{AB}}{2 \text{Im} w_1}\label{pruab}
\end{equation}
with $\eta^{AB}:=\text{diag}(++---)$. Then the potential (\ref{pot_ULL}) is
\begin{equation}
 P_{(4)}^{(R)}= \frac{1}{2 {\rm{Im}} w_1 } \sim  \frac{1}{e^{\sigma}h^1 } \sim  e^{-\sigma}||\varphi||^2 \label{ftppot}
\end{equation}
where an overall positive multiplier is neglected. We note that this potential is proportional to the last term of (\ref{pr14}) and because of the diagonality of (\ref{pruab}), one cannot get a term proportional to the first term by using a linear combination of vectors as the gauge field. One way to interpret this is: Due to the symplectic rotation (\ref{orcansS}), the five dimensional gauge field ${A}_{\hat{\mu}}^1$ is decomposed in two parts. One part contributes to the four dimensional gauge vector $\tilde{A}^1_\mu$ and the other to the four dimensional bare graviphoton $\tilde{A}^0_\mu$. It is this second part of ${A}_{\hat{\mu}}^1$ that leads to the first term of (\ref{pr14}), which does not contribute to the four dimensional gauge field in this particular choice of the holomorphic section.

\subsubsection{dRW-rotation \label{secdrw}}
The de Roo-Wagemans matrix (\ref{dRWmatrix}) rotates the symplectic section (\ref{omegaBdefined}, with $n=4$) to
\begin{equation}
 \left(
\begin{array}{c}
 \frac{1}{2\sqrt{2}} \left(2-||w||^2\right) \\
 w_2 \\
 w_3 \\
 w_4 \\
 \frac{1}{2\sqrt{2}} \left(2+||w||^2\right) \left(\cos \theta + w_1 \sin \theta \right) \\
 -\frac{1}{2\sqrt{2}} w_1\left(2-||w||^2\right)  \\
 -w_1 w_2 \\
 w_1 w_3 \\
 w_1 w_4 \\
 -\frac{1}{2\sqrt{2}} \left(2+||w||^2\right) \left(\sin \theta -w_1 \cos \theta  \right)
\end{array}
\right),\label{my-symp-sec}
\end{equation}
where $||w||^2\equiv [w_2^2 - w_3^2 - w_4^2]$. Using Mathematica we evaluated the potential as
\begin{equation}
 P_{(4)}^{(R)}= \frac{|{\rm{cos}}\theta+w_1 {\rm{sin}}\theta|^2}{2 {\rm Im} w_1} .
\end{equation}
This potential agrees with \cite{FTP02} by applying the coordinate transformations outlined in appendix \ref{app_trf}.

\subsubsection{Critical Points}
The total potential of the current model with $n=4$ vector multiplets and $K_{(4)}=SO(2,1)\times U(1)_R$ gauge group evaluated at Re$(w_i) = 0$ is given by
\begin{equation}\begin{array}{rcl}
 P_{(4)}\arrowvert_{{\rm Re} (w_i) = 0} &=& (P_{(4)}^{(T)} + \lambda P_{(4)}^{(R)})\arrowvert_{{\rm Re} (w_i) = 0} \\[10pt] &=&\frac{1}{2 \text{Im}w_1} (\frac{\left(\text{Im}w_2^2-\text{Im}w_3^2\right) \left(2+||\text{Im}w||^2\right){}^2}{8
    ||\text{Im}w||{}^4}+\lambda |{\rm{cos}}\theta+w_1 {\rm{sin}}\theta|^2).\end{array}
\end{equation}
The critical points of this potential have coordinates which obey
\begin{equation} 
  w_1=-\cot\theta+\frac{i \csc\theta}{\sqrt{\lambda}},\qquad
(\text{Im}w_2)^2 -(\text{Im}w_3)^2 =2,\qquad \text{Re}w_i = 0, \qquad
\text{Im}w_4=0\label{cxcoorsforcrit}
\end{equation}
and the potential evaluated at these points is
\begin{equation}
  P_{(4)}\arrowvert_{\phi^c}=\sqrt{\lambda } \sin \theta  = \frac{1}{{\rm Im} w_1^c}
\end{equation}
which is positive definite in the physically relevant region\footnote{The imaginary part of $w_1$ is proportional to $1/||\varphi||^2$ which has to be positive definite in order to have positive kinetic terms in the Lagrangian. See section \ref{GJF} for a more thorough discussion.} ($0<\theta<\pi$). Writing (\ref{cxcoorsforcrit}) in terms of real scalar fields we obtain the conditions
\begin{equation}
 A^i_c=\varphi^4_c=0,\quad A^1_c=-\sqrt{3}\cot\theta,\quad e^{3\sigma_c}=\frac{6\sqrt{6}\csc\theta}{\sqrt{\lambda}},\quad [(\varphi^2_c)^2-(\varphi^3_c)^2]=6\, e^{-2\sigma_c}.
\end{equation}
We see that for a given $\theta$, the values of all the scalars, including the dilaton $\sigma$, at the critical point are fixed. The only exception is that the term $[(\varphi^2_c)^2-(\varphi^3_c)^2]$ is fixed whereas $\varphi^2_c$ and $\varphi^3_c$ are not, individually. Observe that this was also the case in five dimensions when the gauge group was $K_{(5)}=SO(1,1)\times U(1)_R$ (c.f. subsection \ref{so11u1rhypers}).

The stability of this family of critical points can be studied by calculating the eigenvalues of the Hessian of the potential evaluated at the extremum. When this is normalized by the inverse of the metric (\ref{gibarj})
\begin{equation}
 g^{\tilde{I}\bar{\tilde{J}}}\arrowvert_{\phi^c} = \left(
\begin{array}{cccc}
 4\, {\text Im} w_1^2 & 0 & 0 & 0 \\
 0 & 4\, ({\text Im} w_2^2-1) & 4\, {\text Im} w_2\, {\text Im} w_3 & 0 \\
 0 & 4\, {\text Im} w_2\, {\text Im} w_3 & 4\, ({\text Im} w_2^2-1) & 0 \\
 0 & 0 & 0 & 4
\end{array}
\right),
\end{equation}
it gives the mass matrix of the scalar fields
\begin{equation}
 \frac{\partial_{\tilde{I}} \partial^{\tilde{J}} P_{(4)}}{P_{(4)}}\arrowvert_{\phi^c} =\left(
\begin{array}{cccc}
 2 & 0 & 0 & 0 \\
 0 & \frac{{\text Im} w_2^2}{2} & \frac{1}{2}\, {\text Im} w_2\, {\text Im} w_3 & 0 \\
 0 & -\frac{1}{2}\, {\text Im} w_2\, {\text Im} w_3 & -\frac{{\text Im} w_3^2}{2}  & 0 \\
 0 & 0 & 0 & 1
\end{array}
\right)
\end{equation}
with ``complex'' eigenvalues $(2, 1, 1, 0)$\footnote{The reason why we called these ``complex'' eigenvalues is based on the fact that the derivatives $\partial_{\tilde{I}}$ are with respect to the complex scalars $z^{\tilde{I}}$. The same mass matrix can be obtained by taking the derivatives with respect to $\bar{z}^{\tilde{\bar{I}}}$.}. Thus the family critical points corresponds to stable de Sitter vacua.

One can extend this result to a theory coupled to an arbitrary number $n>2$ of vector multiplets by trivially extending the holomorphic section and the value of the potential at the extremum will not change. The mass matrix will contain $n-3$ diagonal entries with the value $1$ and the values of the extra scalars at the extremum will be zero.

\subsection{The Five Dimensional Connection}
Dimensionally reducing $5D, \mathcal{N}=2$ YMESGT with isometry gauging group $K_{(5)}$ yields a $4D, \mathcal{N}=2$ YMESGT with an isometry gauging group $K_{(4)}=K_{(5)}\ltimes\mathcal{H}^{n_T +1}$ \cite{GMZ05b} where $n_T$ is the number of tensor multiplets coupled to the theory and $\mathcal{H}^{n_T +1}$ is the Heisenberg group generated by translations and the central charge. This Heisenberg group factor exist only if tensors are coupled to the theory. 

The model discussed in the last section with gauge group $K_{(4)}=SO(2,1)\times U(1)_R$ has stable de Sitter vacua. Unfortunately, it cannot be obtained from five dimensions directly. One can immediately think of gauging a subgroup $K_{(5)}=SO(1,2)$ of the global isometry group for one of the three families (\ref{3families}) in five dimensions. For the generic Jordan Family, the resulting theory after dimensional reduction still has the gauge group $K_{(4)}=SO(1,2)$. This type of gauging does not yield a scalar potential in five dimensions because tensors are absent from the theory, but it does in four dimensions due to the last term of (\ref{POT4__}). For the magical Jordan family there will be tensors transforming under $SO(1,2)$ hence the gauge group in four dimensions is $SO(1,2)\ltimes\mathcal{H}^{n_T +1}$; and for the generic non-Jordan family $SO(1,2)$ is not gaugable because one cannot find vector fields that transform under the adjoint representation of $SO(1,2)$ to use as the gauge fields. The first two of these families allow for four dimensional theories with an $SO(1,2)$ factor in the gauge group but this is not the same $SO(2,1)$ gauge group factor we discussed in the last section. The former one is a subgroup of $SO(1,2)\times SO(1,n-3) \subset SO(2,n-1)$ and has one timelike and two spacelike dimensions whereas the latter is a subgroup of $SO(2,1)\times SO(n-2) \subset SO(2,n-1)$ and has two timelike and one spacelike dimensions. Therefore the model with $SO(2,1)$ gauge group factor we discussed in the last section does not originate from five dimensions.\footnote{However this does not rule out the possibility that the $SO(1,2)$ gauging may result in non-Minkowski ground states in four dimensions. See subsection \ref{so12sec} for this type of gauging.}

Nevertheless, this is not the end of the story. In five dimensions, de Sitter vacua were found for the $SO(1,1)\times U(1)_R$ gauging and in four dimensions they were found for the $SO(2,1)\times U(1)_R$ gauging. In this section, we will show that under an appropriate group contaction of $SO(2,1)$ one can find a theory, which can be obtained from the five dimensional $SO(1,1)\times U(1)_R$ theory under another appropriate group contraction, and that has a potential that allows stable de Sitter ground states.

\subsubsection{Contracting the Algebra}
A geometrical interpretation for the contraction can be given by introducing the $n$-di\-men\-sion\-al inhomogeneous coordinates $u_a\,(a=0, 2, 3, ...\, , n)$ that parametrize a hyperboloid embedded in $n$-dimensional space by $u_a \eta^{ab} u_b = R^2$ where $\eta_{ab}={\rm diag}(++-...-)$ and $R$ is the radius of curvature. The scalars $v_k\, (k=2,3,...\, ,n)$ parametrize an $(n-1)$-dimensional hypersurface. This hypersurface is mapped onto the hyperboloid embedded in $n$-dimensional space by the stereographical projection
\begin{equation} 
 \begin{array}{rcl}
  u_0&=&{\displaystyle \frac{R^2-||v||^2 }{R^2+||v||^2}}\,R,\\[10pt]
u_k &=&{\displaystyle \frac{2 R^2\, v_k}{R^2+||v||^2}}
 \end{array}
\end{equation}
where $||v||^2=[v_2^2 - v_3^2 - ... - v_n^2]$. The inverse mapping is
\begin{equation}
 v_k = \frac{R u_k}{R+u_0}.
\end{equation}
For $n=4$, the $SO(2,1)$ symmetry on the $4$-dimensional hyperboloid is generated by the Killing vectors, which in terms of homogeneous hypersurface coordinates are formulated by
\begin{equation}
 \vec{K}_0 = \left(\begin{array}{c}
                    0\\ - w_3\\ -w_2\\0
                   \end{array}
 \right),
\vec{K}_2 = \left(\begin{array}{c}
                    0\\[10pt]{\displaystyle - \frac{w_2 w_3}{ R}}\\[10pt] {\displaystyle \frac{R^2-w_2^2-w_3^2+w_4^2}{2 R}}\\[10pt]{\displaystyle -\frac{w_3 w_4}{ R}}
                   \end{array}
 \right),
\vec{K}_3 = \left(\begin{array}{c}
                    0\\[10pt]{\displaystyle \frac{R^2+w_2^2+w_3^2+w_4^2}{2 R}}\\[10pt]{\displaystyle \frac{w_2 w_3}{ R}}\\[10pt]{\displaystyle \frac{w_2 w_4}{R}}
                   \end{array}
 \right).
\end{equation}
Note that if the real $v_i$ are extended to the complex $w_i$ and $R=\sqrt{2}$, these are the same Killing vectors we evaluated in (\ref{so21killings}). By taking the large $R$ limit, the hyperboloid is locally flattened and the group is contracted \cite{IW53, Gil74} to
$SO(1,1)\ltimes \mathbb{R}^{(1,1)}$. Let us observe this by defining the new generators as
\begin{equation}
 \vec{K}_0'\equiv \vec{K}_0,\qquad \vec{K}_2'\equiv \frac{{2} \vec{K}_2}{R},\qquad \vec{K}_3'\equiv \frac{{2} \vec{K}_3}{R}\label{contr_K}
\end{equation}
and evaluating the Lie brackets
\begin{equation}
 [\vec{K}_0',\vec{K}_2']=\vec{K}_3',\qquad  [\vec{K}_0',\vec{K}_3']=\vec{K}_2',\qquad [\vec{K}_2',\vec{K}_3']=-\frac{4}{R^2} \vec{K}_0'.\label{contr_alg}
\end{equation}
By taking the limit $R\rightarrow \infty$, the last of these Lie brackets vanishes and we see that the new Killing vectors generate the Lie algebra of the Poincare group in two dimensions which is the semi-direct product of ``Lorentz boosts'' $SO(1,1)$ with ``translations'' $\mathbb{R}^{(1,1)}$.

Meanwhile, for the five dimensional gauge group $K_{(5)}=SO(1,1)$ the structure constants (\ref{str_const_w_cc}) determine the following algebra in four dimensions:
\begin{equation}
 \begin{array}{cclcrcl}
\displaystyle [\frac{T_0 - T_1}{\sqrt{2}} , T_2] &=&0,&\quad&\displaystyle [\frac{T_0 + T_1}{\sqrt{2}} , T_3]=T_2, \\[10pt]
{[}\displaystyle\frac{T_0 - T_1}{\sqrt{2}} , T_3{]} &=&0,&\quad& \displaystyle[\frac{T_0 + T_1}{\sqrt{2}} , T_2]=T_3,\\[10pt]
{[}T_2, T_3{]} &=&\displaystyle \frac{T_0 - T_1}{\sqrt{2}}.
 \end{array}\label{so11wcc}
\end{equation}
They define the Lie algebra of a central extension of the Lie algebra $SO(1,1) \sdsum {\mathbb R}^{(1,1)}$ , with central charge corresponding to the generator $\frac{1}{\sqrt{2}}(T_0 - T_1)$. Here ``$\sdsum$'' denotes ``semi-direct sum''. $\frac{1}{\sqrt{2}}(T_0 + T_1)$ rotates $T_2$ and $T_3$ into each other and corresponds to the bare graviphoton in 5 dimensions which acted as the $SO(1,1)$ gauge field. Note that this result parallels completely the situation in the subsection \ref{gmzrot} (c.f. \cite{GMZ05b}).

By defining the new generators
\begin{equation}
\left(
 \begin{array}{c}
  W_0\\W_1\\W_2\\W_3
 \end{array}
\right)=\left(
\begin{array}{cccc}
 1&1&0&0\\
 -\beta&\beta&0&0\\
0&0&\beta&0\\
0&0&0&\beta
\end{array}
\right)\left(
 \begin{array}{c}
  (T_0-T_1)/\sqrt{2}\\(T_0+T_1)/\sqrt{2}\\T_2\\T_3
 \end{array}
\right)
\end{equation}
one can rewrite the algebra as
\begin{equation}
 \begin{array}{rcl}
  [W_2,W_3]&=&\frac{1}{2}\left(\beta^2 W_0 - \beta W_1\right)\\
{[}W_0,W_2]&=&W_3\\
{[}W_0,W_3]&=&W_2\\
{[}W_1,W_2]&=&\beta W_3\\
{[}W_1,W_3]&=&\beta W_2.\end{array}
\end{equation}
In the limit $\beta\rightarrow 0$ the transformation matrix above becomes noninvertible, but this is expected since information is generically lost during group contractions, and the algebra reduces to $SO(1,1) \sdsum {\mathbb R}^{(1,1)}$ without central charge. This is the same algebra as (\ref{contr_alg}) in the large $R$ region. Thus the two different limits of the two different theories overlap. Now, we will calculate the extrema of the scalar potential they will generate.
\begin{comment}
\begin{equation}
 \begin{array}{rcl}
    [W_2,W_3]&=&0\\
{[}W_0,W_2]&=&W_3\\
{[}W_0,W_3]&=&W_2\\
{[}W_1,W_2]&=&0\\
{[}W_1,W_3]&=&0.
 \end{array}\label{contracted_algebra}
\end{equation}
\end{comment}
\subsubsection{Potential by \texorpdfstring{$(SO(1,1)\ltimes \mathbb{R}^{(1,1)}) \times U(1)_R$}{(SO(1,1) s R(1,1)) x U(1)R} Gauging}
Using the Killing vectors (\ref{contr_K}) in the large $R$ limit, the potential (\ref{pt_VP}) is calculated as in (\ref{ptso11app}). When evaluated at ${\rm Re} (w_k) = 0,\, (k=2,3,...,n)$ it takes the form
\begin{equation}
 P_{(4)}^{(T)}\arrowvert_{{\rm Re} (w_k) = 0} = \frac{\left(\text{Im}w_2^2-\text{Im}w_3^2\right) \left(4+||\text{Im}w||^2\right){}^2 }{64\,
   \text{Im}w_1\, ||\text{Im}w||^4}\label{ptso11wocc}
\end{equation}
where $||\text{Im}w||^2 \equiv \left(\text{Im}w_2^2-\text{Im}w_3^2-...-\text{Im}w_n^2\right)$. This potential term satisfies
\begin{equation}\frac{\textstyle\partial P_{(4)}^{(T)}}{\textstyle\partial {\rm Re} (w_k)}\arrowvert_{{\rm Re} (w_k) = 0} =0.\label{wk_drv_vanishes}
 \end{equation}

The dRW-rotation is done prior to the gauging. One must choose the $U(1)_R$ gauge field $\tilde{A}_\mu^b$ among $\tilde{A}_\mu^i\,(i=4,...,n)$ \footnote{Choosing $\tilde{A}_\mu^1$ as the  $U(1)_R$ gauge field as we did in subsection \ref{secdrw} would result rotating $\tilde{X}^1$ and $\tilde{F}_1$ into each other. But in this case, the presence of tensors makes it impossible to keep the translations block diagonal.} and dRW-rotate $\tilde{X}^b$ and $\tilde{F}_b$ into each other. The dRW-matrix is given by
\begin{equation}
 \mathcal{R} = \left(\begin{array}{ccccc}
                      \unity_{n-1}&0&0&0&0\\
		      0&\cos\theta&0&\sin\theta&0\\
		      0&0&\unity_n&0&0\\
		      0&\sin\theta&0&\cos\theta&0\\
		      0&0&0&0&1
                     \end{array}\label{dRWmatrix-so11}
\right)
\end{equation}
where we chose $b=n$. Note that with this type of gauging one must have $n>3$ vector multiplets (c.f. $n>2$ for the $SO(2,1)\times U(1)_R$ gauging after the dRW-rotation (\ref{dRWmatrix})). The calculation of the $U(1)_R$ potential is similar to the last case but it has the same expression
\begin{equation}
P_{(4)}^{(R)}= \frac{|{\rm{cos}}\theta+w_1 {\rm{sin}}\theta|^2}{2 {\rm Im} w_1}.\label{prafterrw}
\end{equation}
The critical points of the total potential $P_{(4)} = P_{(4)}^{(T)} +\lambda P_{(4)}^{(R)}$ are given by
\begin{equation} 
  w_1^c=-\cot\theta+\frac{i \csc\theta}{\sqrt{2 \lambda}},\qquad
(\text{Im}w_2^c)^2 -(\text{Im}w_3^c)^2 =4,\qquad \text{Re}w_k^c = 0, \qquad
\text{Im}w_i^c=0
\end{equation}
and the value of the potential evaluated at these points is
\begin{equation}
  P_{(4)}\arrowvert_{\phi^c}=\sqrt{\frac{\lambda }{2}} \sin \theta  = \frac{1}{{2\, \rm Im} w_1^c}.
\end{equation}
Writing these in terms of real scalars, we again see that for a given $\theta$, the values of all the scalars, including the dilaton $\sigma$, at the critical point are fixed. The only exception is that the term $[(\varphi^2_c)^2-(\varphi^3_c)^2]$ is fixed whereas $\varphi^2_c$ and $\varphi^3_c$ are not, individually.

The mass matrix for this potential evaluated at the family of critical points is
\begin{equation}
 \frac{\partial_{\tilde{I}} \partial^{\tilde{J}} P_{(4)}}{P_{(4)}}\arrowvert_{\phi^c} =\left(
\begin{array}{cccc}
 2 & 0 & 0 & 0 \\
 0 & \frac{{\text Im} w_2^2}{4} & \frac{1}{4}\, {\text Im} w_2\, {\text Im} w_3 & 0 \\
 0 & -\frac{1}{4}\, {\text Im} w_2\, {\text Im} w_3 & -\frac{{\text Im} w_3^2}{4}  & 0 \\
 0 & 0 & 0 & 1
\end{array}
\right)
\end{equation}
which has eigenvalues $(2,1,1,0)$ and hence the extrema correspond to stable de Sitter vacua. The zero eigenvalue is due to the remaining $SO(1,1)$ symmetry and means that there is a flat direction along the extrema.\\[5pt]

%\underline{\textbf{The effect of group contraction to the potential}}
\paragraph{The effect of group contraction to the potential:}

Without the contraction outlined in the last subsection, i.e. using the structure constants of the algebra (\ref{so11wcc}), the potential $P_{(4)}^{(T)}$ evaluated at ${\rm Re} (w_i) = 0$ is given by
\begin{equation}
\frac{\text{Im}w_2^2-\text{Im}w_3^2}{2\,
   \text{Im}w_1\, ||\text{Im}w||^4}.\label{ptso11wcc}
\end{equation}
Subtracting this expression from (\ref{ptso11wocc}) will give the contribution of the group contraction to the scalar potential as
\begin{equation}
 \frac{\left(\text{Im}w_2^2-\text{Im}w_3^2\right) P_+ P_- }{64\,
   \text{Im}w_1\, ||\text{Im}w||^4},
\end{equation}
where $P_{\pm}=||\text{Im}w||^2 +4(1\pm\sqrt{2})$. This term is positive definite in the neighborhood of the extrema, where $||\text{Im}w||^2 \sim 4$. A quick calculation shows that the $P_{(4)}^{(T)}$ potential (\ref{ptso11wcc}), together with the $P_{(4)}^{(R)}$ potential (\ref{prafterrw}) does not have any critical points.
\begin{comment}\footnote{Note that there is a slight change in the calculation of $P_{(4)}^{(R)}$. When the central charge is involved, $\tilde{X}^1$ and $\tilde{F}_1$ contribute to the $P_{(4)}^{(T)}$. In order to make the dRW-rotation commute with the gauging one must choose another gauge field, for instance $\tilde{A}^4_\mu$, for $U(1)_R$ gauging. But one can show that dRW-rotating  $\tilde{X}^4$ and $\tilde{F}_4$ into each other will result in the same potential term (\ref{prafterrw}) where we chose $\tilde{A}^1_\mu$ as the gauge field.}\end{comment}

\subsection{More Examples}
\subsubsection{\texorpdfstring{$(SO(1,1)\ltimes \mathbb{R}^{(1,1)}) \times SU(2)_R$}{(SO(1,1) s R(1,1)) x SU(2)R} gauging}
In order to do such a gauging along with a dRW-rotation one must have $n>5$ vector multiplets.  $P^{(T)}$ is as given in (\ref{ptso11wocc}). 

For the $SU(2)_R$ gauging, the moment map is as defined in (\ref{mmsu2}) and the gauge fields are chosen to be $\tilde{A}_\mu^b\, (b=n-2,n-1,n)$. $\tilde{X}^b$ and $\tilde{F}_b$ are rotated into each other via the dRW-matrix
\begin{equation}
 \mathcal{R}=\left(\begin{array}{ccccc}
                    \unity_{n-3}&0&0&0&0\\
		    0&\cos\theta \unity_3 & 0 & \sin\theta \unity_3&0\\
		    0&0&\unity_{n-2}&0&0\\
		    0&-\sin\theta \unity_3 & 0 & \cos\theta \unity_3&0\\
		    0&0&0&0&1
                   \end{array}
\right).
\end{equation}
The resulting $SU(2)_R$ potential is given by
\begin{equation}\begin{array}{rcl}
P_{(4)}^{(R)}&=& \left(U^{(AB)} -3 \bar{\tilde{L}}^{(A} \tilde{L}^{B)}\right) \mathcal{P}^x_A \mathcal{P}^x_B\\[10pt]
&=&\displaystyle\sum^n_{y=n-2} \left(U^{\left(yy\right)} - 3 \bar{\tilde{L}}^{(y} \tilde{L}^{y)} \right)\\[15pt]
&=& \displaystyle\frac{3\,|{\rm{cos}}\theta+w_1 {\rm{sin}}\theta|^2}{2 {\rm Im} w_1}
\end{array}\end{equation}
which differs from (\ref{prafterrw}) only by a factor $3$. Each $SU(2)_R$ generator gives the same contribution as the $U(1)_R$ generator in the Abelian case. The total potential is
\begin{equation}
\begin{array}{rcl}
 P_{(4)}\arrowvert_{{\rm Re} (w_k) = 0}&=&(P_{(4)}^{(T)}+\lambda P_{(4)}^{(R)})_{{\rm Re} (w_k) = 0}\\[15pt]
&=&\displaystyle\frac{\left(\text{Im}w_2^2-\text{Im}w_3^2\right) \left(4+||\text{Im}w||^2\right){}^2 }{64\,
   \text{Im}w_1\, ||\text{Im}w||^4}+ \displaystyle\frac{3\lambda\,|{\rm{cos}}\theta+w_1 {\rm{sin}}\theta|^2}{2 {\rm Im} w_1}.
\end{array}
\end{equation}
The critical points of the total potential $P_{(4)} = P_{(4)}^{(T)} +\lambda P_{(4)}^{(R)}$ are given by
\begin{equation} 
  w_1^c=-\cot\theta+\frac{i  \csc\theta}{\sqrt{6 \lambda}},\qquad
(\text{Im}w_2^c)^2 -(\text{Im}w_3^c)^2 =4,\qquad \text{Re}w_k^c = 0, \qquad
\text{Im}w_i^c=0
\end{equation}
and the value of the potential evaluated at these points is
\begin{equation}
  P_{(4)}\arrowvert_{\phi^c}=\sqrt{\frac{3\lambda }{2}} \sin \theta  = \frac{1}{{2\, \rm Im} w_1^c}.
\end{equation}
Writing these in terms of real scalars, we again see that for a given $\theta$, the values of all the scalars, including the dilaton $\sigma$, at the critical point are fixed. The only exception is that the term $[(\varphi^2_c)^2-(\varphi^3_c)^2]$ is fixed whereas $\varphi^2_c$ and $\varphi^3_c$ are not, individually.

The mass matrix for this potential evaluated at the family of critical points is
\begin{equation}
 \frac{\partial_{\tilde{I}} \partial^{\tilde{J}} P_{(4)}}{P_{(4)}}\arrowvert_{\phi^c} =\left(
\begin{array}{cccc}
 2 & 0 & 0 & 0 \\
 0 & \frac{{\text Im} w_2^2}{4} & \frac{1}{4}\, {\text Im} w_2\, {\text Im} w_3 & 0 \\
 0 & -\frac{1}{4}\, {\text Im} w_2\, {\text Im} w_3 & -\frac{{\text Im} w_3^2}{4}  & 0 \\
 0 & 0 & 0 & \unity_{n-3}
\end{array}
\right)
\end{equation}
which has eigenvalues $(2,\underbrace{1,...\, ,1}_{(n-2)\,{\rm times}},0)$ and hence the extrema correspond to stable de Sitter vacua with a flat direction due to the remaining $SO(1,1)$ symmtery.

\subsubsection{\texorpdfstring{$(SO(1,1)\ltimes \mathbb{R}^{(1,1)}) \times U(1)_R$}{(SO(1,1) s R(1,1)) x U(1)R} gauging with hypers}
The authors of \cite{FTP02} studied a model with 5 vector multiplets and 4 hypermultiplets and the scalars of the hypermultiplets spanned the hyperbolic space $\frac{SO(4,2)}{SO(4)\times SO(2)}$ and the gauge group was $SO(2,1)\times SU(2)$. Here we shall consider the coupling of a single hypermultiplet to supergravity and arbitrary number $n$ of vector multiplets. We use the same symmetric space $\mathcal{M}_Q = \frac{SU(2,1)}{SU(2)\times U(1)}$ formalism for the scalar manifold of a single hypermultiplet that we also studied on five dimensions in section \ref{sec2}. The scalars that span this space are $q^X=(V,\theta,\tau,\sigma)$. Gauging $(SO(1,1)\ltimes \mathbb{R}^{(1,1)}) \times U(1)_R$ gives three contributions to the scalar potential.

$P_{(4)}^{(T)}$ is not affected by the hyper coupling so we take it as given in (\ref{ptso11wocc}). Meanwhile, gauging hyper isometries introduces the potential term (\ref{phafterred}) which is written in the canonical form as\cite{ABCDFFM96, WKV00}
\begin{equation}
 P_{(4)}^{(H)} = 4\, e^\mathcal{K} (K^X_A \bar{\tilde{X}}^A) g_{XY} (K^Y_B \tilde{X}^B)
\end{equation}
with $K_A^X = V_A\, Y^a\, T_a^X\quad (a=1,2,3)$, where $V_A$ determine the linear combination of vectors to use as the $U(1)_R$ gauge field. $Y^a$, on the other hand, determine the linear combination of the hyper-isometries $T_a^X$ that are gauged. $T_a^X$, the Killing vectors that generate the symmetries of the isometry group $SU(2,1)$ are given in appendix \ref{appendixhypers}. At the base point $q^c=(V=1, \, \theta=\tau=\sigma=0)$ of the hyperscalar manifold, where the hyperspace metric $g_{XY}$ (\ref{hypermetric}) becomes diagonal, this potential satisfies
\begin{equation}
{P}_{(4)}^{(H)}\arrowvert_{q^c}=\frac{\partial {P}_{(4)}^{(H)}}{\partial w_{\tilde{I}}}\arrowvert_{q^c}=\frac{\partial {P}_{(4)}^{(H)}}{\partial q}\arrowvert_{q^c}= 0
\end{equation}
because of the vanishing Killing vectors at that point. The third contribution is the $U(1)_R$ potential
\begin{equation}
 P_{(4)}^{(R)}=(U^{(AB)} - 3 \bar{\check{L}}^{(A} \check{L}^{B)}) \vec{\mathcal{P}}_A \vec{\mathcal{P}}_B
\end{equation}
where the momentum map is is written in terms of the Killing prepotentials as $ \vec{\mathcal{P}}_A = V_A\, Y^a\, \vec{P}_a$. We choose $\tilde{A}_\mu^n$ as the $U(1)_R$ gauge field and set $V_n=1$. The dRW-rotation matrix that mixes the electric and the magnetic components of the holomorphic section is given in (\ref{dRWmatrix-so11}). 

The total potential 
\begin{equation}
\begin{array}{rcl}
 P_{(4)}\arrowvert_{{\rm Re} (w_i) = 0,\, q^c}&=&\left[P_{(4)}^{(T)}+\lambda (\,P_{(4)}^{(R)}+P_{(4)}^{(H)}\,)\right]_{{\rm Re} (w_k) = 0,\, q^c}\\[15pt]
&=&\displaystyle\frac{\left(\text{Im}w_2^2-\text{Im}w_3^2\right) \left(4+||\text{Im}w||^2\right){}^2 }{64\,
   \text{Im}w_1\, ||\text{Im}w||^4}+ \displaystyle\frac{\lambda\,|{\rm{cos}}\theta+w_1 {\rm{sin}}\theta|^2 (Y^a Y^a)}{8 {\rm Im} w_1}
\end{array}\nonumber
\end{equation}
has extrema at
\begin{eqnarray} 
  \phi^c=\bigg\{ w_1^c=-\cot\theta+\frac{i \sqrt{2} \csc\theta}{\sqrt{\lambda (Y^a Y^a)}},\qquad
(\text{Im}w_2^c)^2 -(\text{Im}w_3^c)^2 =4,\nonumber\\[12pt]  \text{Re}w_k^c = 0, \qquad
\text{Im}w_i^c=0,\qquad V^c=1,\qquad \theta^c=\tau^c=\sigma^c=0\bigg\},\nonumber
\end{eqnarray}
where it takes the value
\begin{equation}
  P_{(4)}\arrowvert_{\phi^c}=\frac{\sqrt{\lambda (Y^a Y^a)}}{2\sqrt{2}} \sin \theta  = \frac{1}{{2\, \rm Im} w_1^c}.\nonumber
\end{equation}
The values of all the scalars at the critical point are fixed, except $w_2$ and $w_3$, which satisfy $(\text{Im}w_2^c)^2 -(\text{Im}w_3^c)^2 =4$. This remaining $SO(1,1)$ symmetry leads to a flat direction along the extrema.
Joining the scalar indices $\zeta=(\tilde{I}, X)$, the expression for the mass matrix is written as
\begin{equation}
 \frac{\partial_{\zeta} \partial^{\xi} P_{(4)}}{P_{(4)}}\arrowvert_{\phi^c} =\left(
\begin{array}{cccccccc}
 2 &0&0&0 \\
 0& \frac{{\text Im} w_2^2}{4} & \frac{1}{4}\,\scriptstyle {\text Im} w_2\, {\text Im} w_3 &0 &\ldots\\
0& -\frac{1}{4}\,\scriptstyle {\text Im} w_2\, {\text Im} w_3 & -\frac{{\text Im} w_3^2}{4} &0 \\
 0&0 &0  & \unity_{n-3}\\
&\vdots&&&\frac{1}{4}\\
&&&&&\qquad 1\\
&&&&&&\qquad\frac{1}{2}\\
&&&&&&&\qquad\frac{1}{8}
\end{array}
\right),\nonumber
\end{equation}
where the last 4 entries belong to the hypers. This matrix has all non-negative eigenvalues $(2,\underbrace{\textstyle 1,...\, ,1}_{(n-2)\,{\rm times}},$ $0,\frac{1}{4},1,\frac{1}{2},\frac{1}{8})$, where again the last 4 entries are the masses of the hyperscalars, and hence the extrema correspond to stable de Sitter vacua.

\subsubsection{Yet Another Holomorphic Section?}
Applying a symplectic transformation
\begin{equation}\begin{array}{rcl}
\tilde{X}^A&\rightarrow& \tilde{F}_A\\
\tilde{F}_A&\rightarrow& - \tilde{X}^A
\end{array}\end{equation}
on the holomorphic section (\ref{omegaBdefined}, with $n=4$), acting on it with the dRW-matrix (\ref{dRWmatrix-so11}) and gauging $K_{(4)}=(SO(1,1)\ltimes \mathbb{R}^{(1,1)} ) \times U(1)_R$ lead to the scalar potential
\begin{equation}\begin{array}{rcl}
 P_{(4)}\arrowvert_{{\rm Re} (w_k) = 0}&=&\left[P_{(4)}^{(T)}+\lambda P_{(4)}^{(R)}\right]_{{\rm Re} (w_k) = 0} \label{pot_XtoF}\\[10pt]
&=&\frac{\left(\text{Im}w_2^2-\text{Im}w_3^2\right) \left(4+||\text{Im}w||^2\right){}^2 (\text{Re}w_1^2 + \text{Im}w_1^2)}{64\,
   \text{Im}w_1\, ||\text{Im}w||^4} + \lambda \frac{|{\rm{sin}}\theta-w_1 {\rm{cos}}\theta|^2}{2 {\rm Im} w_1},
\end{array}\end{equation}
which has critical points at
\begin{equation}
 w_1^c = \frac{\sqrt{2\lambda} \sin\theta}{1+2\lambda \cos^2 \theta}\left(\sqrt{2\lambda} \cos\theta + i\right), \qquad ({\rm Im}w^c_2)^2 - ({\rm Im}w_3^c)^2 =4,\qquad {\rm Re}w^c_k ={\rm Im}w^c_i = 0. \nonumber
\end{equation}
At these family of critical points, the potential takes the value of
\begin{equation}
  P_{(4)}\arrowvert_{\phi^c}=\sqrt{\frac{\lambda }{2}} \sin \theta 
\end{equation}
and the mass matrix $\frac{\partial_{\tilde{I}} \partial^{\tilde{J}} P_{(4)}}{P_{(4)}}$ has eigenvalues $(2,1,1,0)$ and hence this corresponds to stable dS vacua with a flat direction for $0<\theta<\pi$.

Observe that, when $\theta=\pi /2$, apart from the $X^0\&F_1$ and $X^1\&F_0$ mixing in the transformation (\ref{orcansS}), this corresponds to the GMZ holomorphic section we introduced in subsection \ref{gmzrot}. Although this seems like just a specific case, it will play an important role in finding stable de Sitter vacua when we study general homogeneous scalar manifolds below.

\subsubsection{de Sitter Vacua from GMZ Holomorphic Section}
The procedure of obtaining de Sitter ground states using the GMZ holomorphic section, which is obtained by acting on $\Omega_0$ by the transformation matrix (\ref{GMZ_Sdef}), involves a similar group contraction. Consider five dimensional $SO(1,1)$ gauged YMESGT coupled to $\tilde{n}=3$ vector multiplets that has $A_{\hat{\mu}}^1$ as the gauge field. The vectors $A_{\hat{\mu}}^2$ and $A_{\hat{\mu}}^3$ are charged under the gauge group and need to be dualized to tensors. After the dimensional reduction this becomes a four dimensional theory coupled to $n=4$ vector multiplets with a gauge group $K_{(4)}=SO(1,1)\ltimes \mathbb{R}^{(1,1)}$ with central charge\cite{GMZ05b}. The structure constants are
\begin{equation}
 f_{23}^0 = -\sqrt{2},\qquad f_{13}^2=f_{12}^3=\frac{1}{\sqrt{2}}
\end{equation}
which are antisymmetric in the lower indices. With these structure constants one can calculate the Killing vectors (\ref{kill4}) that generate $K_{(4)}$ as
\begin{equation}
 \vec{K}_0=0,\qquad \vec{K}_1=\left(\begin{array}{c}
                                     0\\w_3/\sqrt{2}\\w_2/\sqrt{2}\\0
                                    \end{array}
\right),\qquad \vec{K}_2=\left(\begin{array}{c}
                                     0\\-1\\0\\0
                                    \end{array}
\right),\qquad \vec{K}_1=\left(\begin{array}{c}
                                     0\\0\\-1\\0
                                    \end{array}
\right).
\end{equation}
The contraction will be done by going to a basis with the following rotation of the Killing vectors:
\begin{equation}
 \vec{K}_0'=\vec{K}_0 - \vec{K}_1,\qquad \vec{K}_1'=\vec{K}_0 + \vec{K}_1,\qquad \vec{K}_2'=\vec{K}_2 ,\qquad \vec{K}_3'=\vec{K}_3.
\end{equation}
It is straightforward to show that the new Killing vectors generate $SO(1,1)\ltimes \mathbb{R}^{(1,1)}$ without central charge. After some calculation we found that the potential $P^{(T)}$, defined in (\ref{pt_VP}), is indeed equal to the $P^{(T)}$ given in (\ref{pot_XtoF}). 

In addition to $K_{(4)}$, one can also gauge the $U(1)_R$ symmetry. Choosing the gauge field as $A_\mu^4$, this will result in a potential term $P_{(4)}^{(R)}=1/(2$ Im$w_1)$ that is the $P_{(4)}^{(R)}$ given in (\ref{pot_XtoF}) when $\theta=\pi /2$. The calculation in the last subsection shows that the total potential $P_{(4)}=P_{(4)}^{(T)} +\lambda P_{(4)}^{(R)}$ has de Sitter minima with a flat direction. 

\subsubsection{\texorpdfstring{$SO(1,2)$}{SO(1,2)} Gauging from 5 Dimensions \label{so12sec}}
One can start with a gauged YMESGT in five dimensions with a isometry gauging group $K_{(5)}=SO(1,2)$. For the generic Jordan Family, the only charged vector fields are the gauge fields $A_{\hat{\mu}}^2, A_{\hat{\mu}}^3$ and $A_{\hat{\mu}}^4$ which transform under the adjoint representation of this $SO(1,2)$. There are no tensor fields and no scalar potential is introduced. 

After the dimensional reduction, the gauge group is still $SO(1,2)$ but this is a different subgroup of the global isometry group in four dimensions than what we gauged in section \ref{dsvacin4}. The former one is a subgroup of $SO(1,1)\times SO(1,n-2) \subset SO(2,n-1)$ and has one timelike and two spacelike dimensions whereas the latter is a subgroup of $SO(2,1)\times SO(n-2) \subset SO(2,n-1)$ and has two timelike and one spacelike dimensions. In contrast to the case before the dimensional reduction, gauging $SO(1,2)$ results in a scalar potential in four dimensions due to the second term in (\ref{potin4}). Taking the structure constants as $f^2_{34}=-f^3_{42}=-f^4_{23}=1$ this potential is evaluated to be
\begin{equation}
 P_{(4)}^{(T)}=\frac{Q_{23} + Q_{24} - Q_{34}}{2 \text{Im}w_1 ||\text{Im}w||^4}
\end{equation}
with $Q_{kl} = \left(w_k \bar{w}_l - w_l \bar{w}_k\right)^2$. Unfortunately, this potential does not admit any ground states other than Minkowskian. One can gauge $U(1)_R$ (in four dimensions) in addition to the $SO(1,2)$ symmetry which adds the term (\ref{prafterrw}) to the potential. But it is easy to verify that the total potential does not have any critical points in this case.

At this point, perhaps it is worth rementioning that the four dimensional theories that have different holomorphic sections as their starting points, which are related by just a symplectic transformation, describe different physics. For the generic Jordan family, an $K_{(5)}=SO(1,2)\times U(1)_R$ gauged YMESGT has Minkowski and anti-de Sitter ground states in five dimensions. The Minkowski ground states survive in four dimensions if one works with the GMZ holomorphic section, due to a term in the $U(1)_R$ potential that doesn't exist in the potential that is derived from our holomorphic section. We stressed this issue below equation (\ref{ftppot}).

\subsection{Beyond Generic Jordan Family \label{magical4d}}
For the holomorphic section we obtained by the rotation (\ref{orcansS}), satisfying the equation (\ref{cimn0}) was crucial to keep the translations block diagonal. This equation is trivially satisfied for the generic Jordan family ($C_{iMN}=0$) but for other types of scalar manifolds, such as the magical Jordan family, it does not hold in general. This problem can be evaded by dRW-rotating all of $\tilde{X}^i$ and $\tilde{F}_i$ by $\pi/2$ radians. The entire symplectic transformation matrix, including the dRW-rotation with $\theta=\pi/2$, 
\begin{equation}
 \check{\mathcal{S}}=\left(
\begin{array}{cccccccc}
0&\frac{1}{\sqrt{2}}&0&0&-\frac{1}{\sqrt{2}}&0&0&0\\
0&0&0&0&0&0&0&D^{MN}\\
0&0&\delta^j_{\phantom{j}i}&0&0&0&0&0\\
0&-\frac{1}{\sqrt{2}}&0&0&-\frac{1}{\sqrt{2}}&0&0&0\\
 \frac{1}{\sqrt{2}}&0&0&0&0&\frac{1}{\sqrt{2}}&0&0\\
0&0&0&D_{MN}&0&0&0&0\\
0&0&0&0&0&0&\delta_j^{\phantom{j}i}&0\\
\frac{1}{\sqrt{2}}&0&0&0&0&-\frac{1}{\sqrt{2}}&0&0
\end{array}
\right)\label{orcansnewerS}
\end{equation}
acts on $\Omega_0$ in the following way
\begin{equation}\begin{array}{rcl}
 \check{\Omega} &=& \left(
\begin{array}{c}
 \check{X}^0\\
 \check{X}^M\\
 \check{X}^j\\
 \check{X}^1\\
 \check{F}_0\\
 \check{F}_M\\
 \check{F}_j\\
 \check{F}_1
\end{array}
\right)= \check{\mathcal{S}} \Omega_0 = \check{\mathcal{S}} \left(
\begin{array}{c}
  X^0\\ X^1\\X^i\\ X^N\\ F_0\\ F_1\\F_i\\F_N
\end{array}
\right).
\end{array}\nonumber
\end{equation}
Here $D_{MN}=-\sqrt{2}\, \Omega_{MN}$ and $D_{MN} D^{NP} = \delta^P_M$ and again, we decomposed the index $I$ as $I=(1,i)$.

Furthermore, in order to gauge $\check{K}\equiv K_{(4)} \times U(1)_R=(SO(1,1)\ltimes \mathbb{R}^{(1,1)}) \times U(1)_R$ which was the four dimensional gauge group for the theories with de Sitter solutions that originate from five dimensions, the isometry group needs to contain a subgroup $SO(2,r-1)$ with $r\geq 3$.

So far, we studied \textit{symmetric} space scalar manifolds only. Now we relax this restriction and look for \textit{homogeneous} (but not necessarily symmetric) space scalar manifolds that  have $SO(2,r-1),\, r\geq 3$ as a subsector. We have to re-analyze how the holomorphic section transforms under the translations $z^M \rightarrow z^M + b^M$ because $C_{IJM}$ does not necessarily vanish in homogeneous spaces:
\begin{equation}
 \check{\Omega} \rightarrow \check{\Omega}
+\left(\begin{array}{c}
        \frac{1}{\sqrt{2}} b^M D_{MN} \check{X}^N\\
	\sqrt{\frac{2}{3}} D^{MN} b^P \left\{C_{1NP} (\check{X}^1 - \check{X}^0) +\sqrt{2} C_{iNP} \check{X}^i \right\}\\
	0\\
	\frac{1}{\sqrt{2}} b^M D_{MN} \check{X}^N\\
	-\sqrt{\frac{2}{3}} b^M \left\{ C_{1MN} D^{NP} \check{F}_P +C_{11M} \frac{\check{X}^0 - \check{X}^1}{\sqrt{2}} - C_{1jM} \check{X}^j\right\}\\
	\frac{1}{\sqrt{2}} D_{MN} b^N (\check{F}_0 +\check{F}_1)\\
	-\sqrt{\frac{2}{3}} b^M \left\{ C_{iMN} D^{NP} \check{F}_P +C_{i1M} \frac{\check{X}^0 - \check{X}^1}{\sqrt{2}} - C_{ijM} \check{X}^j\right\}\\
	-\sqrt{\frac{2}{3}} b^M \left\{ C_{1MN} D^{NP} \check{F}_P +C_{11M} \frac{\check{X}^0 - \check{X}^1}{\sqrt{2}} - C_{1jM} \check{X}^j\right\}
       \end{array}
\right)\nonumber
\end{equation}
Observe that in order to keep the translations block diagonal, i.e. to have $\check{X}^A$ and $\check{F}_A$ transform among themselves,
\begin{equation}
C_{IJM}\stackrel{!}{=}0
\end{equation}
must hold.

de Wit and Van Proeyen classified homogeneous very special manifolds and gave their corresponding cubic polynomials in \cite{dWP91, dWP95}. These spaces are of the form $G/H$ where $G$ is the isometry group and $H$ is its isotropy subgroup. $G$ is not necessarily semi-simple, thus not all the homogeneous spaces have a clear name. In their classification, the homogeneous spaces are denoted as $L(q,P)$. Here, $q$ characterizes the real Clifford algebras ($\mathcal{C}(q+1,0)$) that are in one-to-one correspondence with homogeneous special manifolds. These have signatures $(q+1,1)$ for real (in five dimensions), $(q+2,2)$ for K\"{a}hler (in four dimensions) manifolds, which are related to each other with what is called \textbf{r}-map. The non-negative integer $P$ denotes the multiplicity of the representation of the Clifford algebra. For $q\neq 4m$ ($m$ is a non-negative integer), $P$ is unique. When $q=4m$, there are two inequivalent representations. In this case the homogeneous space is denoted by $L(4m, P, \dot{P})$. Note that $L(4m, P, 0) =L(4m, 0,P)\equiv L(4m,P)$. \begin{table}[tbp]
\label{tbl:symvs}
\begin{center}
\begin{tabular}{|c|c|cc|}
\hline
$L(q,P)$&$n$&real & K\"ahler  \\
\hline\hline
$L(-1,0)$&$2$&$SO(1,1)$&$\left[\frac{SU(1,1)}{U(1)}\right]^2$ \\[10pt] \hline
$L(-1,P)$&$2+P$&$\frac{SO(P+1,1)}{SO(P+1)}$&\\[10pt] \hline
$L(0,P)$&$3+P$&$ SO(1,1)\times \frac{SO(P+1,1)}{SO(P+1)}$&
$\frac{SU(1,1)}{U(1)}\times\frac{SO(P+2,2)}{SO(P+2)\times SO(2)}$\\[10pt] \hline
$L(1,1)$&$6$&
$\frac{S\ell(3,R)}{SO(3)}$&$\frac{Sp(6)}{U(3)
}$\\[10pt]
$L(2,1)$&$9$&
$\frac{S\ell(3,C)}{SU(3)}$&$\frac{SU(3,3)}{SU(3)\times
SU(3)\times U(1)}$\\[10pt]
$L(4,1)$&$15$&
$\frac{SU^*(6)}{Sp(3)}$&$\frac{SO^*(12)}{SU(6)\times
U(1)}$\\[10pt]
$L(8,1)$&$27$&
$\frac{E_6}{F_4}$&$\frac{E_7}{E_6\times
 U(1)}$\\[10pt]
\hline \end{tabular}
\end{center}\caption[Symmetric very special manifolds]{\textit{Symmetric very special manifolds.} $L(-1,P)$, which correspond to the generic non-Jordan Family, are symmetric in five dimensions, but not their images under the \textbf{r}-map. $L(0,P)$ is the generic Jordan family and the last 4 entries are the magical Jordan family manifolds. The number $n$ is the complex dimension of the K\"{a}hler space, which also is the number of vector multiplets in 4 dimensions. Table is adapted from \cite{dWP91}.}   \end{table} Table \ref{tbl:symvs} lists the special cases where $L(q,P)$ are symmetric manifolds. The cubic polynomial that has an invariance group that acts transitively on the special real manifolds can be specified in the general form 
\begin{equation}
 N(h)=C_{\tilde{I}\tilde{J}\tilde{K}} h^{\tilde{I}}h^{\tilde{J}}h^{\tilde{K}}=  3\left\{ \hat{h}^1 (\hat{h}^2)^2 - \hat{h}^1 (\hat{h}^\beta)^2 - \hat{h}^2 (\hat{h}^m)^2 +\gamma_{\beta mn} \hat{h}^\beta \hat{h}^m \hat{h}^n \right\},\label{mgcf}
\end{equation}
where the index $\tilde{I}=1,...,n$ is decomposed into $I=1,2,\beta,m$, with $\beta=3,...\,,(q+3)$ and $m=(q+4), ...\,, (q+3 + (P+\dot{P}) \mathcal{D}_{q+1} = n)$. The dimension $\mathcal{D}_{q+1}$ of the irreducible representation of the Clifford algebra with positive signature in $q+1$ dimensions is given by
\begin{eqnarray}
 \mathcal{D}_{q+1} = 1\,\, {\rm for }\,\, q=-1,0,\qquad \mathcal{D}_{q+1} = 2\,\, {\rm for }\,\, q=1,\qquad \mathcal{D}_{q+1} = 4\,\, {\rm for }\,\, q=2,\nonumber\\
 \mathcal{D}_{q+1} = 8\,\, {\rm for }\,\, q=3,4,\qquad \mathcal{D}_{q+1} = 16\,\, {\rm for }\,\, q=5,6,7,8,\qquad \mathcal{D}_{q+8} = 16 \mathcal{D}_{q}.\nonumber
\end{eqnarray}

The constraint $r\geq 3$ translates into $q\geq 0$. Hence we immediately see that $\check{K}$ is not gaugable for generic non-Jordan family $L(-1,P)$. Let us investigate the cases ($q= 0$) and ($q>0$) separately.\\[10pt]
\underline{\textbf{Case 1:} ($q=0$)}\\
If either of $P$ or $\dot{P}$ vanishes the homogeneous space corresponds to the symmetric generic Jordan family, which we have studied already. For non-vanishing $P$ and $\dot{P}$ one can write the cubic polynomial as
\begin{equation}
N(h) = 3 \left\{ h^1 \left[(h^2)^2 - (h^3)^2 - (h^x)^2 \right] - (h^2 - h^3) (h^{\dot{x}})^2\right\}
\end{equation}
after the reparametrization
\begin{equation}
 \hat{h}^1 = h^2 + h^3,\quad \hat{h}^2 = \frac{h^1 + h^2 - h^3}{2},\quad \hat{h}^3 = \frac{-h^1 + h^2 - h^3}{2},\quad \hat{h}^x=h^x,\quad \hat{h}^{\dot{x}}=h^{\dot{x}}\nonumber
\end{equation}
where the index $m$ is decomposed into $P$ indices $x$ and $\dot{P}$ indices $\dot{x}$. The fields $h^2$ and $h^3$ are charged under the gauge group $K_{(4)}$ and the corresponding vector fields $A_\mu^2$ and $A_\mu^3$ need to be dualized to tensor fields. Hence the index $\tilde{I}=(I,M)$ is split as follows: $I=1,x,\dot{x};\,\, M=2,3$. But then $C_{\dot{x}\dot{y}M} \neq 0$ and hence the translations will not remain block diagonal, i.e. $K_{(4)}$ is not gaugable in the standard way.\\[18pt]
\underline{\textbf{Case 2:} ($q> 0$)}\\
All of these spaces $L(q>0,P,\dot{P})$, which also include the symmetric magical Jordan family for $(q=1,2,4,8;\, P=1)$, contain $K_{(4)}=SO(1,1)\ltimes \mathbb{R}^{(1,1)}$ subsectors. Consider the cubic form in the most general form as given in (\ref{mgcf}). Choosing $A_\mu^1$ as the gauge field one can find a $K_{(4)}$-generator such that $\hat{h}^2$ and $\hat{h}^3$ rotate into each other keeping $(\hat{h}^2)^2 - (\hat{h}^3)^2$ fixed. Because they are charged under the gauge group the corresponding vector fields need to be dualized to tensor fields. The rest of $\hat{h}^\beta$ are $K_{(4)}$-singlets and a linear combination of the corresponding vector fields can be used as the $U(1)_R$ (or even $SU(2)_R$ if $q\geq 3$) gauge field(s). $h^m$ form $(P+\dot{P}) \mathcal{D}_{q+1}/2$ doublets under $K_{(4)}$ and their corresponding vector fields are dualized to tensor fields. All the conditions are satisfied and we conclude that the homogeneous spaces of the type $L(q>0,P)$ admit stable de Sitter vacua when $\check{K}$ is gauged.

\section{Discussions\label{sec4}}
Stable de Sitter vacua of $4D, \mathcal{N}=2$ YMESGT's were found in \cite{FTP02, FTP03}. The main goal of this paper was to relate these four dimensional theories to the theories in five dimensions with various gaugings. The authors of these papers asserted that three ingredients are necessary to obtain de Sitter vacua:
\begin{itemize}
 \item non-compact gauge groups,
 \item Fayet-Iliopoulos (FI) terms,
 \item de Roo-Wagemans (dRW) rotation.
\end{itemize}
The non-compact gauge group they used in the three models they studied is $SO(2,1)$. We showed that this is \textit{not} the only gauge group that admits a potential that one needs to obtain de Sitter vacua. One can indeed contract this group to $SO(1,1)\ltimes \mathbb{R}^{(1,1)}$ and de Sitter vacua is preserved under this contraction. We need to emphasize that whereas the $SO(2,1)$ gauged theories do not directly descend from five dimensions, their contracted counterparts do. FI terms are available for gauge groups that have $U(1)$ or $SU(2)$ factors. The variation of such terms in the Lagrangian is a total derivative and they yield supersymmetric terms in the action. \cite{BDKP04,Pro04} point out that adding FI terms to the Lagrangian is indeed equivalent to gauging $R$-symmetry. In the three models they studied; Fre, Trigiante and van Proeyen considered $\mathcal{N}=2$ supergravity with a complex scalar manifold of the form $ \mathcal{M}_{VS}^4 = \mathcal{ST}[2,n-1]=\frac{SU(1,1)}{U(1)}\times\frac{SO(2,n-1)}{SO(2)\times SO(n-1)}$, parametrized by Calabi-Vesentini coordinates. These correspond to symmetric generic Jordan family which describes the geometry of a real manifold of the form $ \mathcal{M}_{VS}^5=\frac{SO(\tilde{n}-1,1) \times SO(1,1)}{SO(\tilde{n}-1)}$ in five dimensions. Here $n$ and $\tilde{n}$ denote the number of vector multiplets coupled to supergravity in four and five dimensions, respectively and they are related by $n = \tilde{n} +1$. For such theories one has a certain amount of freedom to choose a holomorphic (symplectic) section upon dimensional reduction. This freedom is parametrized by dRW-angles $\theta$. Different choices of $\theta$ yield different gauged models with different physics. We use the notation of \cite{GST83a,GMZ05b} to parametrize the complex manifold instead of Calabi-Vesentini coordinates for two main reasons. First, in the former parametrization the five dimensional connection is as clear as it could be, as the complex scalar fields are obtained directly from dimensional reduction and second, generalizing the results to homogeneous manifolds is significantly easier. The mapping between two parametrizations can be found in appendix \ref{app_trf}.

As we stressed earlier, stable de Sitter vacua exist in five dimensional $SO(1,1)\times R_s$ gauged YMESGT's where the $R_s$ denotes a subgroup of the full $R$-symmetry group $SU(2)_R$ \cite{GZ01, CS05, Oge06}. These theories descend to four dimensional theories that have the gauge group $(SO(1,1)\ltimes \mathbb{R}^{(1,1)})\times R_s$ with a central charge \cite{GMZ05b}. The procedure in establishing de Sitter ground states in four dimensions from these theories include finding an appropriate holomorphic section by means of a dRW-rotation, and contracting the gauge group. The contraction rotates some of the group generators into each other, eliminates the central charge and gives a positive definite contribution to the potential. Without this contribution the potential does not have any ground states and that makes the group contraction essential. We showed that these theories can also be obtained from four dimensional $SO(2,1) \times R_s$ gauged YMESGT's, which were considered in \cite{FTP02, FTP03} to have stable de Sitter vacua, by means of a different contraction.

In analogy to five dimensions, the theories with generic Jordan family scalar manifolds have stable de Sitter vacua for $(SO(1,1)\ltimes \mathbb{R}^{(1,1)})\times R_s$ gaugings. $R_s$ can be either $U(1)_R$ or $SU(2)_R$. In either case, the de Sitter minima in four dimensions we found has a flat direction. Recall that this was also the case in five dimensions before the dimensional reduction. In addition to vector/tensor multiplets, one can couple a universal hypermultiplet and simultaneously gauge $U(1)$ or $SU(2)$ symmetry of its quaternionic hyperscalar manifold. We showed that, again in analogy to five dimensions, this type of extra gauging preserves the nature of the de Sitter ground states. The theories with non-compact hyper isometry gauging, which lead to stable de Sitter ground states in five dimensions, still need to be checked in four dimensions to complete the analogy. This topic is not covered in this paper and we leave it for future investigation.

Same results can be achieved by starting either with the G\"{u}naydin-McRey\-nolds-Za\-ger\-mann (GMZ) symplectic section \cite{GMZ05b} or with the symplectic section we introduced in (\ref{omegaBdefined}) which has a closer connection to the Calabi-Vesentini basis used in \cite{FTP02, FTP03}. While in either case a gauge group contraction that rotates some of the generators into each other and eliminates the central charge is essential, it should be noted that one does not need an extra dRW-rotation for the GMZ symplectic section, because it is already ``dRW-rotated'' by $\theta=\pi/2$ radians with respect to our symplectic section. 

\begin{comment}\subsection{Extension to Homogeneous Spaces}\end{comment}
In four dimensions, general homogeneous (but not necessarily symmetric) scalar manifolds $L(q,P)$ admit de Sitter vacua provided that they contain a $(SO(1,1)\ltimes \mathbb{R}^{(1,1)})\times R_s$ subsector. These spaces are limited to $L(q\geq 0,P)$. For the symmetric generic Jordan family spaces $L(0,P)$, one has the freedom to choose the dRW-angle from $0<\theta<\pi$. This choice affects the values of the scalar fields and the value of the potential at the de Sitter minima. For the spaces of type $L(q>0,P)$, on the other hand, the value of the dRW-angle has to be fixed to $\theta=\pi/2$ because otherwise the translational variations of the holomorphic section do not become block diagonal and one cannot gauge the theory in the standard way. Observe that the GMZ symplectic section carries this rotation to begin with.

The spaces of the type $L(q\geq 0,P)$ have de Sitter minima but one can analyze them for other ground states. However, the analysis of extrema of the homogeneous spaces in their full generality is involved and requires a separate study.

\begin{comment}\subsection{Embedding in String Theory?}\end{comment}
Having found the recipe that starts with five dimensional $SO(1,1)\times U(1)_R$ gauged $\mathcal{N}=2$ YMESGT and ends with stable de Sitter vacua in four dimensions, one can ask the question: Is it possible to embed the theory into a fundamental superstring theory or M-theory? There are several directions one can take to answer this question. Compactifications of Type IIA and type IIB superstring theories on Calabi-Yau threefolds yield ungauged supergravity theories in four dimensions. Using the same method, it was shown in \cite{CCDF95} that $5D,\, \mathcal{N}=2$ MESGT coupled to hypers can be obtained by compactifying 11 dimensional supergravity. In particular, the Hodge number $h_{(1,1)}$ of the threefold corresponds to the number of vector fields (including graviphoton) in the resulting $5D,\, \mathcal{N}=2$ MESGT, whereas $h_{(2,1)} +1$ corresponds to the number of hypermultiplets. Type IIA or Type IIB supergravity in ten dimensions compactified on $T^6$ results in $N=8$ supergravity in four dimensions. Similarly, $5D,\,\mathcal{N}=8$ supergravity can be obtained by compactifying eleven dimensional supergravity on $T^6$. By orbifolding (modding out by discrete groups) the four dimensional theory, Sen and Vafa considered examples of models with broken supersymmetries \cite{SV95}. In one of the several models the scalar manifold belongs to the generic Jordan family. As was pointed out in \cite{Gun06}, another model they considered is the $J_3^{\mathbb{H}}$ of the magical Jordan family. These results can be extended to the $11D$-to-$5D$ compactifications. However these are ungauged theories. Whether one can obtain gauged versions of these theories by turning on fluxes is an open problem to be investigated.

Meanwhile, after solving the stabilization problem of compactification of internal dimensions \cite{GKP01, KKLT03}, it was possible to find de Sitter vacua from string theory. This moduli stabilization fixes the runaway behavior of the axion-dilaton fields, which was also a problem we encountered upon dimensional reduction in the beginning of section \ref{sec3}. In the original KKLT scenario, the moduli stabilization brings the minimum of the scalar potential to a finite negative value. Then the addition of an anti-$D3$ brane lifts this minimum to a state with positive vacuum energy. In our construction, on the other hand, a similar effect was established through dRW-rotation, and gauging the non-compact $SO(1,1)$ subgroup of the global isometry group of the scalar manifold simultaneously with a subgroup of the $R$-symmetry group. Finding a relation between our and a KKLT-like scenario is an interesting problem and we leave this for a future study.

\paragraph{Acknowledgements:} I am grateful to my academic advisor Murat G\"unaydin for his suggestions and guidance throughout this work. I would also like to thank Mario Trigiante for his help with clearing out confusions regarding FI-term computations in section \ref{sec3}. This work was supported in part by the National Science Foundation under grant number PHY-0555605. Any opinions, findings and conclusions or recommendations expressed in this material are those of the author and do not necessarily reflect the views of the National Science Foundation.
\newpage
{\LARGE{\textbf{Appendices}}}
\appendix
\section[``Very Special Geometry'', the Lagrangians and the Potentials]{``Very Special Geometry'', the Lagrangians in 5 and 4 Dimensions and the Derivation of the  Potential Terms\label{appendixveryspecial}}
%\numberwithin{equation}{chapter}
%\renewcommand{\theequation}{A.\arabic{equation}}
%\setcounter{equation}{0}
%\newpage
%\*{Appendix A: The ``Very Special Geometry''}
The bosonic sector of the $5D, \mathcal{N}=2$ gauged Yang-Mills-Einstein supergravity\footnote{For the full Lagrangian, see \cite{GZ00, CD00}} coupled to tensor- and hypermultiplets is described by the Lagrangian (with metric signature $(-++++)$) \cite{GZ99, EGZ01, CD00}
\begin{equation}\begin{array}{rcl}
\hat{e}^{-1}\mathcal{L}^{(5)}&=&\displaystyle-\frac{1}{2}\hat{R}-\frac{1}{4}\stackrel{o}{a}_{\tilde{I}\tilde{J}}\mathcal{H}^{\tilde{I}}_{{\hat{\mu}}{\hat{\nu}}}\mathcal{H}^{\tilde{J}{\hat{\mu}}{\hat{\nu}}}-\frac{1}{2}g_{XY}\mathcal{D}_{\hat{\mu}} q^X \mathcal{D}^{\hat{\mu}} q^Y\\[7pt]
&&-\displaystyle\frac{1}{2}g_{\tilde{x}\tilde{y}} \mathcal{D}_{\hat{\mu}} \varphi^{\tilde{x}} \mathcal{D}^{\hat{\mu}} \varphi^{\tilde{y}} +\frac{\hat{e}^{-1}}{6\sqrt{6}} C_{IJK} \epsilon^{{\hat{\mu}}{\hat{\nu}}{\hat{\rho}}{\hat{\sigma}}{\hat{\tau}}} F^I_{{\hat{\mu}}{\hat{\nu}}} F^J_{{\hat{\rho}}{\hat{\sigma}}} A^K_{\hat{\tau}}\\[7pt]
&&\displaystyle+\frac{\hat{e}^{-1}}{4 g} \epsilon^{{\hat{\mu}}{\hat{\nu}}{\hat{\rho}}{\hat{\sigma}}{\hat{\tau}}} \Omega_{MN} B^M_{{\hat{\mu}}{\hat{\nu}}} \mathcal{D}_{\hat{\rho}} B^N_{{\hat{\sigma}}{\hat{\tau}}} -P_{(5)}(\varphi, q).\label{Lin5}
\end{array}\end{equation}
Here, non-Abelian field strengths $\mathcal{F}^I_{{\hat{\mu}}{\hat{\nu}}}\equiv  F^I_{{\hat{\mu}}{\hat{\nu}}} + g f^I_{JK} A^J_{\hat{\mu}} A^K_{\hat{\nu}}\quad (I=1,2,...,n_V+1)$ of the gauge group $K_{(5)}$ and the self-dual tensor fields $B^M_{{\hat{\mu}}{\hat{\nu}}}\quad (M=1,2,...,2n_T)$ are grouped together to define the tensorial quantity $\mathcal{H}^{\tilde{I}}_{{\hat{\mu}}{\hat{\nu}}} \equiv (\mathcal{F}^I_{{\hat{\mu}}{\hat{\nu}}},B^M_{{\hat{\mu}}{\hat{\nu}}})$ with $\tilde{I}=1,2,...,n_V+2n_T+1$. The potential term $P_{(5)}(\varphi, q)$ is given by 
\begin{equation}
 P_{(5)}(\varphi, q) = g^2 (P_{(5)}^{(T)}(\varphi) + \lambda P_{(5)}^{(R)}(\varphi, q)+\kappa P_{(5)}^{(H)}(q))
\end{equation}
where
\begin{equation}
 \begin{array}{rcl}
  P_{(5)}^{(T)} &=& 2 W_{\tilde{x}} W^{\tilde{x}},\\[4pt]
P_{(5)}^{(R)} &=& -4 \vec{P}\cdot \vec{P} +2 \vec{P}^{\tilde{x}}\cdot \vec{P}_{\tilde{x}},\\[4pt]
P_{(5)}^{(H)} &=& 2 \mathcal{N}_{X} \mathcal{N}^{X}
 \end{array}
\end{equation}
and $\lambda=g_R^2/g^2$ , $\kappa=g_H^2/g^2$. The quantities given in the above expression are defined as
\begin{equation}
 \begin{array}{rcl}
 W_{\tilde{x}}&\equiv&\displaystyle-\frac{\sqrt{6}}{8} \Omega^{MN} h_{M\tilde{x}} h_N = \frac{\sqrt{6}}{4} h^I K_I^{\tilde{x}},\\[7pt]
\vec{P}&\equiv&h^I \vec{P}_I,\\[7pt]
\vec{P}_{\tilde{x}}&\equiv&h^I_{\tilde{x}} \vec{P}_I,\\[7pt]
\mathcal{N}^{X}&\equiv&\displaystyle\frac{\sqrt{6}}{4}h^I K_I^{X},\label{wapadef}
\end{array}
\end{equation}
where $K_I^{\tilde{x}}$ and $K_I^{X}$ are Killing vectors acting on the scalar and the hyperscalar parts of the total scalar manifold $\mathcal{M}^5_{scalar} = \mathcal{M}^5_{VS} \otimes \mathcal{M}_{Q}$; $\vec{P}_I$ are the Killing prepotentials which will be defined below; $\Omega^{MN}$ is the inverse of $\Omega_{MN}$, which is the  constant invariant anti-symmetric tensor of the gauge group $K_{(5)}$; and $h^I$ and $h^I_{\tilde{x}}$ are elements of the very special manifold $\mathcal{M}^5_{VS}$ described by the hypersurface 
\begin{equation}
 N(h)=C_{\tilde{I}\tilde{J}\tilde{K}} h^{\tilde{I}} h^{\tilde{J}} h^{\tilde{K}} = 1,\qquad \tilde{I},\tilde{J},\tilde{K} = 1,...,\tilde{n}+1
\end{equation}
of the $\tilde{n}+1$ dimensional space $M=\{h^{\tilde{I}} \in \mathbb{R}^{\tilde{n}+1}|N(h)=C_{\tilde{I}\tilde{J}\tilde{K}} h^{\tilde{I}} h^{\tilde{J}} h^{\tilde{K}}>0\}$ with metric 
\begin{equation}
 a_{IJ} = -\frac{1}{3}\partial_I \partial_J \text{ln} N(h).
\end{equation}
The terms $P_{(5)}^{(T)}$ and $P_{(5)}^{(H)}$ are semi-positive definite in the physically relevant region, whereas $P_{(5)}^{(R)}$ can have both signs. $\mathcal{M}^5_{VS}$ is determined completely by the totally symmetric tensor $C_{\tilde{I}\tilde{J}\tilde{K}}$. The scalar field metric on this hypersurface is the induced metric from the embedding space, which is given by
\begin{equation}
 g_{\tilde{x}\tilde{y}}=\frac{3}{2} a_{\tilde{I}\tilde{J}} h^{\tilde{I}}_{,\tilde{x}} h^{\tilde{J}}_{,\tilde{y}} \arrowvert_{N=1}=-3C_{\tilde{I}\tilde{J}\tilde{K}} h^{\tilde{I}} h^{\tilde{J}}_{,\tilde{x}} h^{\tilde{K}}_{,\tilde{y}}\arrowvert_{N=1}\label{appgxy}
\end{equation}
where ``$,\tilde{x}$'' denotes a derivative with respect to $\varphi^{\tilde{x}}$. The definitions
\begin{equation}\begin{array}{rcl}
\stackrel{o}{a}_{\tilde{I}\tilde{J}} &\equiv& a_{\tilde{I}\tilde{J}} \arrowvert_{N=1}=-2C_{\tilde{I}\tilde{J}\tilde{K}} h^{\tilde{K}}+ 3 h_{\tilde{I}} h_{\tilde{J}},\\
h_{\tilde{I}}&\equiv& C_{\tilde{I}\tilde{J}\tilde{K}} h^{\tilde{J}} h^{\tilde{K}} = \stackrel{o}{a}_{\tilde{I}\tilde{J}} h^{\tilde{J}},\\
h^{\tilde{I}}_{\tilde{x}}&\equiv&-\sqrt{\frac{3}{2}} h^{\tilde{I}}_{,\tilde{x}},\\
h_{\tilde{I}\tilde{x}}&\equiv&\stackrel{o}{a}_{\tilde{I}\tilde{J}} h^{\tilde{J}}_{\tilde{x}}=\sqrt{\frac{3}{2}}h_{\tilde{I},\tilde{x}}
\end{array}\label{appdefs}\end{equation}
help us write the algebraic constraints of the very special geometry
\begin{equation}
 \begin{array}{rcl}
h^{\tilde{I}} h_{\tilde{I}}&=&1,\\
h^{\tilde{I}}_{\tilde{x}} h_{\tilde{I}}&=&h_{\tilde{I}\tilde{x}} h^{\tilde{I}}=0,\\
h^{\tilde{I}}_{\tilde{x}} h^{\tilde{J}}_{\tilde{y}} \stackrel{o}{a}_{\tilde{I}\tilde{J}} &=& g_{\tilde{x}\tilde{y}}.
\end{array}\label{appalco}
\end{equation}
There are also differential constraints to be satisfied:
\begin{equation}
 \begin{array}{rcl}
h_{\tilde{I} \tilde{x};\tilde{y}}&=&\sqrt{\frac{2}{3}}\left(g_{\tilde{x}\tilde{y}} h_{\tilde{I}} +T_{\tilde{x}\tilde{y}\tilde{z}} h^{\tilde{z}}_{\tilde{I}}\right),\\
h^{\tilde{I}}_{ \tilde{x};\tilde{y}}&=&-\sqrt{\frac{2}{3}}\left(g_{\tilde{x}\tilde{y}}h^{\tilde{I}}+T_{\tilde{x}\tilde{y}\tilde{z}} h^{\tilde{I}\tilde{z}}\right)
\end{array}
\end{equation}
where ``$;$'' is the covariant derivative using the Christoffel connection calculated from the metric $g_{\tilde{x}\tilde{y}}$ and
\begin{equation}
 T_{\tilde{x}\tilde{y}\tilde{z}}\equiv C_{\tilde{I}\tilde{J}\tilde{K}} h^{\tilde{I}}_{ \tilde{x}} h^{\tilde{J}}_{ \tilde{y}} h^{\tilde{K}}_{ \tilde{z}}.
\end{equation}
Using (\ref{appgxy}),(\ref{appdefs}) and (\ref{appalco}) one can derive
\begin{eqnarray}
 \stackrel{o}{a}_{\tilde{I}\tilde{J}}&=&h_{\tilde{I}} h_{\tilde{J}} + h_{\tilde{I}}^{ \tilde{x}} h_{\tilde{J} \tilde{x}},\\
h_{\tilde{I}}^{ \tilde{x}} h_{\tilde{J} \tilde{x}}&=& -2 C_{\tilde{I}\tilde{J}\tilde{K}} h^{\tilde{K}} + 2 h_{\tilde{I}} h_{\tilde{J}}.
\end{eqnarray}
The indices $\tilde{I},\tilde{J},\tilde{K}$ are raised and lowered by  $\stackrel{o}{a}_{\tilde{I}\tilde{J}}$ and its inverse $\stackrel{o}{a}^{\tilde{I}\tilde{J}}$. $P_{(5)}^{(T)}$ can now be written in a more compact form
\begin{equation}
 \begin{array}{rcl}
  P_{(5)}^{(T)}&=& \frac{3}{8} \Omega^{MN} \Omega^{PR} C_{MRI} h_N h_P h^I\\
&=&\frac{3\sqrt{6}}{16} \Lambda_I^{MN} h_M h_N h^I.
 \end{array}
\end{equation}
with $\Lambda^M_{IN}$ being the transformation matrices of the tensor fields under the gauge group $K_{(5)}$
\begin{equation}
 \Lambda_I^{MN} = \Lambda^M_{IP} \Omega^{PN}=\frac{2}{\sqrt{6}} \Omega^{MR} C_{IRP} \Omega^{PN}.
\end{equation}
 Gauging the $R$-symmetry introduces the potential term $P_{(5)}^{(R)}=-4 \vec{P}\cdot \vec{P} +2 \vec{P}^{\tilde{x}}\cdot \vec{P}_{\tilde{x}}$, where $\vec{P}=h^I \vec{P}_I$ and $\vec{P}_{\tilde{x}}=h^I_{\tilde{x}}\vec{P}_I$ are vectors that transform under the $R$-symmetry group that is being gauged. For the $SU(2)_R$ gauging one can take
\begin{equation}
 \vec{P}_I = \vec{e}_I\nonumber
\end{equation}
where $\vec{e}_I$ satisfy $\vec{e}_i\times \vec{e}_j=d_{ij}^{\phantom{ij}{k}} \vec{e}_k$ and $\vec{e}_i\cdot  \vec{e}_j=\delta_{ij}$ when $i,j,k$ are the $SU(2)_R$ adjoint indices ($d_{ij}^{\phantom{ij}{k}}$ are the $SU(2)$ structure constants); and $\vec{e}_I=0$ otherwise. With this convention and the use of (\ref{appdefs}) and (\ref{appalco}) the potential term simplifies to
\begin{equation}
P_{(5)}^{(R)}=-4 C^{ij\tilde{K}}\delta_{ij} h_{\tilde{K}}.
\end{equation}
If the $U(1)_R$ subgroup of $SU(2)_R$ is being gauged one can take
\begin{equation}
 \vec{P}_I = V_I \vec{e},\nonumber
\end{equation}
where $\vec{e}$ is an arbitrary vector in the $SU(2)$ space and $V_I$ are some constants that define the linear combination of the vector fields $A^I_{\hat{\mu}}$ that is used as the $U(1)_R$ gauge field
\begin{equation}
A_{\hat{\mu}} [U(1)_R]=V_I A^I_{\hat{\mu}}.\nonumber
\end{equation}
The potential term then can be written as
\begin{equation}
P_{(5)}^{(R)}=-4C^{IJ\tilde{K}}V_I V_J h_{\tilde{K}}.
\end{equation}
If tensors are coupled to the theory the $V_I$ have to be constrained by
\begin{equation}
V_I f^I_{JK}=0\nonumber
\end{equation}
with $f^I_{JK}$ being the structure constants of $K_{(5)}$. When the target manifold $\mathcal{M}_{VS}$ is associated with a Jordan algebra, the following equality holds componentwise
\begin{equation}
C^{\tilde{I}\tilde{J}\tilde{K}}=C_{\tilde{I}\tilde{J}\tilde{K}}= {\rm const}.\nonumber
\end{equation}

After the dimensional reduction from 5 to 4, the Lagrangian (\ref{Lin5}) becomes \cite{GMZ05b}
\begin{eqnarray}
 e^{-1}\mathcal{L}^{(4)} &=&-\frac{1}{2}R -\frac{3}{4}
 {\stackrel{\circ}{a}}_{\ensuremath{\tilde{I}}\ensuremath{\tilde{J}}}(\mathcal{D}_{\mu}\tilde{h}^{\ensuremath{\tilde{I}}})(\mathcal{D}^{\mu}\tilde{h}^{\ensuremath{\tilde{J}}})
-\frac{1}{2}e^{-2\sigma}{\stackrel{\circ}{a}}_{IJ}(\mathcal{D}_{\mu} A^{I})(\mathcal{D}^{\mu}A^{J}) \nonumber\\
&&-\frac{1}{2}e^{-2\sigma} g_{XY} (\mathcal{D}_{\mu} q^X) (\mathcal{D}^{\mu} q^Y)-e^{-2\sigma}{\stackrel{\circ}{a}}_{IM}(\mathcal{D}_{\mu}A^{I})B^{\mu M}
-\frac{1}{2}e^{-2\sigma}{\stackrel{\circ}{a}}_{MN} B_{\mu}^{M}B^{\mu N} \nonumber\\
&&+\frac{e^{-1}}{g} \epsilon^{\mu\nu\rho\sigma}\Omega_{MN}B_{\mu\nu}^{M}
(\partial_{\rho} B_{\sigma}^{N}+ g A_{\rho}^{I}\Lambda_{IP}^{N}B_{\sigma}^{P})\nonumber\\
& &  +\frac{e^{-1}}{g} \epsilon^{\mu\nu\rho\sigma}\Omega_{MN}W_{\mu\nu}B_{\rho}^{M}B_{\sigma}^{N} +\frac{e^{-1}}{2\sqrt{6}} C_{MNI}\epsilon^{\mu\nu\rho\sigma}B_{\mu\nu}^{M}
B_{\rho\sigma}^{N} A^{I}\nonumber\\
& &  -\frac{1}{4}
e^{\sigma}{\stackrel{\circ}{a}}_{MN}B_{\mu\nu}^{M}B^{N\mu\nu}
-\frac{1}{2}e^{\sigma}{\stackrel{\circ}{a}}_{IM}(\mathcal{F}_{\mu\nu}^{I}+2W_{\mu\nu}A^{I})B^{M\mu\nu}\nonumber\\
& &  -\frac{1}{4}e^{\sigma}{\stackrel{\circ}{a}}_{IJ}(\mathcal{F}_{\mu\nu}^{I}+2W_{\mu\nu}A^{I}
 )( \mathcal{F}^{J\mu\nu}+2W^{\mu\nu}A^{J})  -\frac{1}{2}e^{3\sigma}W_{\mu\nu}W^{\mu\nu} \nonumber \\
   & &   +\frac{e^{-1}}{2\sqrt{6}} C_{IJK}\epsilon^{\mu\nu\rho\sigma} \Big\{
\mathcal{F}_{\mu\nu}^{I}\mathcal{F}_{\rho\sigma}^{J}A^{K} + 2 \mathcal{F}_{\mu\nu}^{I}W_{\rho\sigma}A^{J}A^{K} +\frac{4}{3}W_{\mu\nu}W_{\rho\sigma} A^{I}A^{J}A^{K} \Big\} \nonumber\\
       & & -g^2 P_{(4)},  \label{Lin4}
\end{eqnarray}
where\begin{eqnarray}
\tilde{h}^{\ensuremath{\tilde{I}}}& \equiv&  e^\sigma h^{\ensuremath{\tilde{I}}},\\
\mathcal{D}_{\mu}A^{I} & \equiv  & \partial_{\mu} A^{I} +g A_{\mu}^{J}f_{JK}^{I}A^{K},\\
\mathcal{F}_{\mu\nu}^{I}  &  \equiv  &    2\partial_{[\mu}A_{\nu]}^{I} + gf_{JK}^{I}A_{\mu}^{J}A_{\nu}^{K} ,\\
\mathcal{D}_{\mu}\tilde{h}^{\ensuremath{\tilde{I}}}& \equiv & \partial_{\mu} \tilde{h}^{\ensuremath{\tilde{I}}} + g A_{\mu}^{I}M_{I\ensuremath{\tilde{K}}}^{\ensuremath{\tilde{I}}}\tilde{h}^{\ensuremath{\tilde{K}}},\\
\mathcal{D}_{\mu} q^X & \equiv  & \partial_{\mu} q^X + g_H A_{\mu}^{I} K^X_I,
\end{eqnarray}
and the total scalar potential, $P_{(4)}$, is given by
\begin{equation}
 P_{(4)}=P_{(4)}^{(T)} + \frac{g_H^2}{g^2} P_{(4)}^{(H)},
\end{equation}
where
\begin{equation}
P_{(4)}^{(T)}\equiv e^{-\sigma}P_{(5)}^{(T)} +\frac{3}{4} e^{-3\sigma}{\stackrel{\circ}{a}}_{\ensuremath{\tilde{I}}\ensuremath{\tilde{J}}} (A^{I}M_{(I)\ensuremath{\tilde{K}}}^{\ensuremath{\tilde{I}}} h^{\ensuremath{\tilde{K}}})
(A^{J}M_{(J)\tilde{L}}^{\ensuremath{\tilde{J}}}h^{\tilde{L}})\label{potin4}
\end{equation}
and
\begin{equation}
 P_{(4)}^{(H)}\equiv e^{-\sigma}P_{(5)}^{(H)} + \frac{1}{2} e^{-3\sigma} (A^I K_I^X) g_{XY} (A^J K_J^Y),\label{phafterred}
\end{equation}
which would get an additional term of the form
\begin{equation}
 \frac{g_R^2}{g^2} P_{(4)}^{(R)}\equiv \frac{g_R^2}{g^2} e^{-\sigma} P_{(5)}^{(R)}(h^{\tilde{I}})\label{prin4}
\end{equation}
if the $R$-symmetry is being gauged. The transformation matrices $M_{(I)\tilde{K}}^{\ensuremath{\tilde{J}}}$ that correspond to the gauge group $K_{(5)}$ are decomposed as follows
\begin{equation}
 M_{(I)\tilde{K}}^{\ensuremath{\tilde{J}}}=\left(
\begin{array}{cc}
 f^J_{IK} & 0 \\      0 & \Lambda_{IM}^N
\end{array}
\right).\label{mijk}
\end{equation}
$f^J_{IK}$ are always antisymmetric in the lower two indices.
%\cleardoublepage
%\thispagestyle{empty}
\section{Killing Vectors of the Hyper-isometry\label{appendixhypers}}
%\renewcommand{\theequation}{B.\arabic{equation}}
%\setcounter{equation}{0}
%\numberwithin{equation}{chapter}
%\section*{Appendix B: Killing vectors of the hyper-isometry}
The eight Killing vectors $k^X_\alpha$ that generate isometry group $SU(2,1)$ of the hyperscalar manifold are given by \cite{CDKP01}
\begin{equation}\begin{array}{l}
\vec{k}_1=\left( \begin{array}{c}0\\1\\0\\0\end{array}\right),\quad
\vec{k}_2=\left( \begin{array}{c}0\\2\theta\\0\\1\end{array}\right),\quad
\vec{k}_3=\left( \begin{array}{c}0\\-2\tau\\1\\0\end{array}\right),\quad
\vec{k}_4=\left( \begin{array}{c}0\\0\\-\tau\\\theta\end{array}\right),\\
\vec{k}_5=\left( \begin{array}{c}V\\\sigma\\\theta/2\\\tau/2\end{array}\right),\quad
\vec{k}_6=\left( \begin{array}{c}2V\sigma\\\sigma^2-(V+\theta^2+\tau^2)^2\\\sigma\theta-\tau(V+\theta^2+\tau^2)\\\sigma\tau+\theta(V+\theta^2+\tau^2)\end{array}\right),\\
\vec{k}_7=\left( \begin{array}{c}-2V\theta\\-\sigma\theta+V\tau+\tau(\theta^2+\tau^2)\\\frac{1}{2}(V-\theta^2+3\tau^2)\\-2\theta\tau-\sigma/2\end{array}\right),\quad
\vec{k}_8=\left( \begin{array}{c}-2V\tau\\-\sigma\tau-V\theta-\theta(\theta^2+\tau^2)\\-2\theta\tau+\sigma/2\\\frac{1}{2}(V+3\theta^2-\tau^2)\end{array}\right).\label{tis}
\end{array}\end{equation}
The corresponding prepotentials are
\begin{equation}\begin{array}{l}
\vec{p}_1=\left( \begin{array}{c}0\\0\\-\frac{1}{4V}\end{array}\right),\quad
\vec{p}_2=\left( \begin{array}{c}-\frac{1}{\sqrt{V}}\\0\\-\frac{\theta}{V}\end{array}\right),\quad
\vec{p}_3=\left( \begin{array}{c}0\\\frac{1}{\sqrt{V}}\\\frac{\tau}{V}\end{array}\right),\quad
\vec{p}_4=\left( \begin{array}{c}-\frac{\theta}{\sqrt{V}}\\-\frac{\tau}{\sqrt{V}}\\\frac{1}{2}-\frac{\theta^2+\tau^2}{2V}\end{array}\right),\\
\vec{p}_5=\left( \begin{array}{c}-\frac{\tau}{2\sqrt{V}}\\\frac{\theta}{2\sqrt{V}}\\-\frac{\sigma}{4V}\end{array}\right),\quad
\vec{p}_6=\left( \begin{array}{c}-\frac{1}{\sqrt{V}}[\sigma\tau+\theta(-V+\theta^2+\tau^2)]\\\frac{1}{\sqrt{V}}[\sigma\theta-\tau(-V+\theta^2+\tau^2)]\\-\frac{V}{4}-\frac{1}{4V}[\sigma^2+(\theta^2+\tau^2)^2]+\frac{3}{2}(\theta^2+\tau^2)\end{array}\right),\\
\vec{p}_7=\left( \begin{array}{c}\frac{4\theta\tau+\sigma}{2\sqrt{V}}\\\frac{3\tau^2-\theta^2}{2\sqrt{V}}-\frac{\sqrt{V}}{2}\\-\frac{3}{2}\tau+\frac{1}{2V}[\sigma\theta+\tau(\theta^2+\tau^2)]\end{array}\right),\quad
\vec{p}_8=\left( \begin{array}{c}-\frac{3\theta^2-\tau^2}{2\sqrt{V}}+\frac{\sqrt{V}}{2}\\\frac{\sigma-4\theta\tau}{2\sqrt{V}}\\\frac{3}{2}\theta+\frac{1}{2V}[\sigma\tau-\theta(\theta^2+\tau^2)]\end{array}\right).
\end{array}\end{equation}

It is easier to see that the Killing vectors close to the $SU(2,1)$ algebra if they are recasted in the following combinations
\begin{equation}
\begin{array}{rl}
SU(2)&\quad \left\{ \begin{array}{rcl}T_1&=&\frac{1}{4}(k_2-2k_8),\\T_2&=&\frac{1}{4}(k_3-2k_7),\\T_3&=&\frac{1}{4}(k_1+k_6-3k_4),\end{array}\right. \\[25pt]
U(1)&\quad \left\{ \begin{array}{rcl} T_8&=&\frac{\sqrt{3}}{4}(k_4+k_1+k_6),\end{array}\right.
\end{array}\quad
\frac{SU(2,1)}{U(2)} \quad \left\{ \begin{array}{rcl}T_4&=&k_5,\\T_5&=&-\frac{1}{2}(k_1-k_6),\\T_6&=&-\frac{1}{4}(k_3+2k_7),\\T_7&=&-\frac{1}{4}(k_2+2k_8).\end{array}\right.
\label{Tvex}
\end{equation}
This basis is chosen for convenience such that the generators $T_1, T_2, T_3$ and $T_8$ are the isotropy group of the point $(V, \sigma, \theta, \tau)=(1,0,0,0)$. The metric hyperscalar manifold becomes diagonal at this point. In all the theories that have hyper coupling, we will take this basis point $q^C$ for a possible candidate of the hyper-coordinates of a critical point. The Killing vectors $K_I^X$ are then given by $V_I^\alpha k^X_\alpha$ and the corresponding prepotentials $\vec{P}_I$ are $V_I^\alpha \vec{p}_\alpha$, where $V_I^\alpha$ are constants that determine which isometries are being gauged and what linear combination of vector fields being used. In particular, 
\[
K_I^X=\left\{ \begin{array}{ll}
         T_1^X, T_2^X, T_3^X&\text{for } SU(2) \text{ gauging},\\
         V_I W^k T_k^X,\quad k=1,2,3,8& \text{for } U(1) \text{ gauging},\\
         V_I W^k T_k^X,\quad k=4,5,6,7& \text{for } SO(1,1) \text{ gauging},
        \end{array}
\right.
\]
where $V_I$ and $W^k$ are constants depending on the model.

\section{Transformations Between Two Parametrizations \label{app_trf}}
%\numberwithin{equation}{chapter}
For $\mathcal{N}=2$ supergravity coupled to $n$ vector multiplets and no tensors, the symplectic section (\ref{om0}) takes the following form\cite{GMZ05b}
\begin{equation}
 {\Omega}_0=\left(
\begin{array}{c}
1 \\ z_1\\ z_2\\z_a \\ \frac{1}{2} z_1 ||z||^2 \\ - \frac{1}{2} ||z||^2\\
- z_1 z_2 \\  z_1 z_a   \\
\end{array}\right).
\end{equation}
Here $||z||^2 = [(z_2)^2 - (z_3)^2 - ... - (z_{n})^2]$ and $a=3,...,n$.
FTP\cite{FTP02} use Calabi-Vesentini coordinates for which ($X^\Lambda, F_\Lambda=\eta_{\Lambda\Sigma} S X^\Sigma\ ;
\ X^\Lambda X^\Sigma \eta_{\Lambda\Sigma}=0\ ,\eta_{\Lambda\Sigma}={\rm diag}(+,+,-,\ldots,-)$) holds. More explicitly \cite{FTP02},
\begin{equation}
 {\Omega}_{CV}=\left(\begin{array}{c}
 X^\Lambda\\ F_\Sigma               
               \end{array}
\right)
=
\left(
\begin{array}{c}
 \frac{1}{2}(1+||y||^2)\\ \frac{1}{2}i(1-||y||^2)\\y_1\\y_{a-1} \\ \frac{1}{2}S(1+||y||^2)\\ \frac{1}{2}iS(1-||y||^2)\\-S y_1\\-S y_{a-1} 
\end{array}
\right)\label{holsec}
\end{equation}
where $||y||^2 = y_1^2 + ... + y_{n-1}^2$. The transformations between the two notations are given by
\begin{equation}
 \begin{array}{rcl}
  \frac{1}{2}(1+||y||^2)&=& \frac{1}{2\sqrt{2}} (2- ||z||^2)\\
\frac{1}{2}i(1-||y||^2)&=& z_2\\
y_{a-2}&=& z_a\\
y_{n-1} &=& \frac{1}{2\sqrt{2}} (2+ ||z||^2)\\
S&=&-z_1.
 \end{array}\label{ztoytrf}
\end{equation}
The matrix for the symplectic rotation $C\, {\Omega}_{CV} = {\Omega}_0$ is given by
\begin{equation}
C=\left(
 \begin{array}{cccccccc}
  \frac{1}{\sqrt{2}}&0&\frac{1}{\sqrt{2}}&0&0&0\\
0&0&0&-\frac{1}{\sqrt{2}}&0&\frac{1}{\sqrt{2}}\\
0&\unity_{n-1}&0&0&0&0\\
0&0&0&\frac{1}{\sqrt{2}}&0&\frac{1}{\sqrt{2}}\\
\frac{1}{\sqrt{2}}&0&-\frac{1}{\sqrt{2}}&0&0&0\\
0&0&0&0&\unity_{n-1}&0
 \end{array}
\right)
\end{equation}
It is easy to see that the symplectic section (\ref{holsec}) together with the coordinate transformations (\ref{ztoytrf}) is a particular case of (\ref{omegaBdefined}) and also that $C = S^{-1}$.
\section{Various Potential Terms}
%\numberwithin{equation}{chapter}
The $P^{(T)}$ potential terms given here are calculated for $\mathcal{N}=2,\,4D$ YMESGT coupled to $n=4$ vector/tensor multiplets.

Gauging $K_{(4)}=SO(2,1)$ symmetry results in the following potential
\begin{eqnarray}
 \scriptstyle&& \scriptstyle-i\,\,\big[ (\bar{w}_2^2-\bar{w}_3^2) w_2^4+2 \bar{w}_2 (-\bar{w}_2^2+\bar{w}_3^2+\bar{w}_4^2+2)
   w_2^3+(\bar{w}_2^4-2 (w_3^2-\bar{w}_3 w_3+w_4^2 +\bar{w}_3^2+\bar{w}_4^2+4) \bar{w}_2^2+2 w_3^2 \bar{w}_3^2+2
   w_4^2 \bar{w}_3^2\nonumber\\
&& \scriptstyle\quad+(\bar{w}_3^2+\bar{w}_4^2){}^2+4 (\bar{w}_4^2+1)-2 w_3 \bar{w}_3
   (\bar{w}_3^2+\bar{w}_4^2+2)) w_2^2+2 \bar{w}_2 ((\bar{w}_2^2-\bar{w}_3^2-\bar{w}_4^2-2)
   w_3^2+8 \bar{w}_3 w_3\nonumber\\
&& \scriptstyle\quad+(w_4^2+2) (\bar{w}_2^2-\bar{w}_3^2-\bar{w}_4^2-2)) w_2-w_3^2 \bar{w}_2^4-4
   w_3^2-(\bar{w}_3 (w_3^2-\bar{w}_3 w_3+w_4^2+2)-w_3 \bar{w}_4^2) (\bar{w}_3
   w_3^2-(\bar{w}_3^2+\bar{w}_4^2+4) w_3\nonumber\\
&& \scriptstyle\quad+(w_4^2+2) \bar{w}_3)+\bar{w}_2^2 (w_3^4-2 \bar{w}_3
   w_3^3+2 (w_4^2+\bar{w}_3^2+\bar{w}_4^2) w_3^2-2 (w_4^2+2) \bar{w}_3
   w_3+(w_4^2+2){}^2)\big]\nonumber\\
&& \scriptstyle\quad\big/\big[{2 (w_1-\bar{w}_1)
   \left((w_2-\bar{w}_2){}^2-(w_3-\bar{w}_3){}^2-(w_4-\bar{w}_4){}^2\right){}^2}\big].\label{ptso21app}
\end{eqnarray}

Gauging $K_{(4)}=SO(1,1)\ltimes \mathbb{R}^{(1,1)}$ (no central charge) symmetry, on the other hand, results in the following potential
\begin{eqnarray}
 \scriptstyle && \scriptstyle-i\,\,\big [-2 (-(w_2-\bar{w}_2){}^2-(w_3-\bar{w}_3){}^2+(w_4-\bar{w}_4){}^2)
   (-\bar{w}_2^2+\bar{w}_3^2+\bar{w}_4^2+2) w_2^2+8 (w_2-\bar{w}_2) (w_3-\bar{w}_3) \bar{w}_3
   w_2\nonumber\\
&& \scriptstyle\quad+4 \bar{w}_2 (-(w_2-\bar{w}_2){}^2-(w_3-\bar{w}_3){}^2+(w_4-\bar{w}_4){}^2)
   w_2-8 w_3 (w_2-\bar{w}_2) (w_3-\bar{w}_3) (-\bar{w}_2^2+\bar{w}_3^2+\bar{w}_4^2+2) w_2\nonumber\\
&& \scriptstyle\quad+2 (-w_2^2+w_3^2+w_4^2+2) (w_2-\bar{w}_2) (w_3-\bar{w}_3) \bar{w}_3
   (-\bar{w}_2^2+\bar{w}_3^2+\bar{w}_4^2+2) w_2\nonumber\\
&& \scriptstyle\quad+(-w_2^2+w_3^2+w_4^2+2) \bar{w}_2
   (-(w_2-\bar{w}_2){}^2-(w_3-\bar{w}_3){}^2+(w_4-\bar{w}_4){}^2)
   (-\bar{w}_2^2+\bar{w}_3^2+\bar{w}_4^2+2) w_2\nonumber\\
&& \scriptstyle\quad+8 w_3 (w_2-\bar{w}_2) \bar{w}_2
   (w_3-\bar{w}_3)-8 (-w_2^2+w_3^2+w_4^2+2) (w_2-\bar{w}_2) \bar{w}_2
   (w_3-\bar{w}_3) \bar{w}_3\nonumber\\
&& \scriptstyle\quad-2 (-w_2^2+w_3^2+w_4^2+2) \bar{w}_3^2
   (-(w_2-\bar{w}_2){}^2-(w_3-\bar{w}_3){}^2-(w_4-\bar{w}_4){}^2)\nonumber\\
&& \scriptstyle\quad+4 w_3 \bar{w}_3
   (-(w_2-\bar{w}_2){}^2-(w_3-\bar{w}_3){}^2-(w_4-\bar{w}_4){}^2)\nonumber\\
&& \scriptstyle\quad-2
   (-w_2^2+w_3^2+w_4^2+2) \bar{w}_2^2
   (-(w_2-\bar{w}_2){}^2-(w_3-\bar{w}_3){}^2+(w_4-\bar{w}_4){}^2)\nonumber\\
&& \scriptstyle\quad+2 w_3
   (-w_2^2+w_3^2+w_4^2+2) (w_2-\bar{w}_2) \bar{w}_2 (w_3-\bar{w}_3)
   (-\bar{w}_2^2+\bar{w}_3^2+\bar{w}_4^2+2)\nonumber\\
&& \scriptstyle\quad-2 w_3^2
   (-(w_2-\bar{w}_2){}^2-(w_3-\bar{w}_3){}^2-(w_4-\bar{w}_4){}^2)
   (-\bar{w}_2^2+\bar{w}_3^2+\bar{w}_4^2+2)\nonumber\\
&& \scriptstyle\quad+w_3 (-w_2^2+w_3^2+w_4^2+2) \bar{w}_3
   (-(w_2-\bar{w}_2){}^2-(w_3-\bar{w}_3){}^2-(w_4-\bar{w}_4){}^2)
   (-\bar{w}_2^2+\bar{w}_3^2+\bar{w}_4^2+2)\big]\nonumber\\
&& \scriptstyle\quad\big/\big[2 (w_1-\bar{w}_1)
   \left((w_2-\bar{w}_2){}^2-(w_3-\bar{w}_3){}^2-(w_4-\bar{w}_4){}^2\right){}^3\big].\label{ptso11app}
\end{eqnarray}
\newpage

   \bibliographystyle{unsrtorc}
   \addcontentsline{toc}{section}{Bibliography}
   \bibliography{Biblio-Database}

\begin{thebibliography}{100}

\bibitem{FNF76}
Daniel~Z. Freedman, P.~{van Nieuwenhuizen}, and S.~Ferrara.
\newblock {Progress Toward a Theory of Supergravity}.
\newblock {\em Phys. Rev.}, D13:3214--3218, 1976.

\bibitem{DZ76}
S.~Deser and B.~Zumino.
\newblock {Consistent Supergravity}.
\newblock {\em Phys. Lett.}, B62:335, 1976.

\bibitem{Nie81}
P.~{Van Nieuwenhuizen}.
\newblock {Supergravity}.
\newblock {\em Phys. Rept.}, 68:189--398, 1981.

\bibitem{SS89}
A.~Salam and E.~Sezgin.
\newblock {\em Supergravities in Diverse Dimensions. Vol. 1, 2}.
\newblock World Scientific, 1989.
\newblock For a review and references see the reprint volumes.

\bibitem{Lin07}
Andrei Linde.
\newblock {Inflationary Cosmology}.
\newblock {\em Lect. Notes Phys.}, 738:1--54, 2008.
\newblock \href {http://arxiv.org/abs/0705.0164} {\path{arXiv:0705.0164}}.

\bibitem{Lin90}
A.~Linde.
\newblock {\em Particle Physics and Inflationary Cosmology}.
\newblock Harwood Academic Publishers, Switzerland, 1990.

\bibitem{LL00}
A.~R. Liddle and D.~H. Lyth.
\newblock {\em Cosmological Inflation and Large-Scale Structure}.
\newblock Cambridge Univ Press, 2000.

\bibitem{Per97}
S.~Perlmutter et~al.
\newblock {Discovery of a Supernova Explosion at Half the Age of the Universe
  and its Cosmological Implications}.
\newblock {\em Nature}, 391:51--54, 1998.
\newblock \href {http://arxiv.org/abs/astro-ph/9712212}
  {\path{arXiv:astro-ph/9712212}}.

\bibitem{Rie98}
Adam~G. Riess et~al.
\newblock {Observational Evidence from Supernovae for an Accelerating Universe
  and a Cosmological Constant}.
\newblock {\em Astron. J.}, 116:1009--1038, 1998.
\newblock \href {http://arxiv.org/abs/astro-ph/9805201}
  {\path{arXiv:astro-ph/9805201}}.

\bibitem{Wei87}
Nathan Weiss.
\newblock {Possible Origins of a Small Nonzero Cosmological Constant}.
\newblock {\em Phys. Lett.}, B197:42, 1987.

\bibitem{Wet87}
C.~Wetterich.
\newblock {Cosmology and the Fate of Dilatation Symmetry}.
\newblock {\em Nucl. Phys.}, B302:668, 1988.

\bibitem{RP87}
Bharat Ratra and P.~J.~E. Peebles.
\newblock {Cosmological Consequences of a Rolling Homogeneous Scalar Field}.
\newblock {\em Phys. Rev.}, D37:3406, 1988.

\bibitem{KLPS01}
R.~Kallosh, A.~D. Linde, S.~Prokushkin, and M.~Shmakova.
\newblock Gauged supergravities, {de Sitter} space and cosmology.
\newblock {\em Phys. Rev. D}, 65:105016, 2002.
\newblock \href {http://arxiv.org/abs/hep-th/0110089}
  {\path{arXiv:hep-th/0110089}}.

\bibitem{FTP02}
P.~Fre, M.~Trigiante, and A.~{Van Proeyen}.
\newblock Stable {de Sitter} vacua from {N = 2} supergravity.
\newblock {\em Class. Quant. Grav.}, 19:4167, 2002.
\newblock \href {http://arxiv.org/abs/hep-th/0205119}
  {\path{arXiv:hep-th/0205119}}.

\bibitem{FTP03}
P.~Fre, M.~Trigiante, and A.~{Van Proeyen}.
\newblock {N = 2} supergravity models with stable {de Sitter} vacua.
\newblock {\em Class. Quant. Grav.}, 20:S487, 2003.
\newblock \href {http://arxiv.org/abs/hep-th/0301024}
  {\path{arXiv:hep-th/0301024}}.

\bibitem{BM03}
Klaus Behrndt and Swapna Mahapatra.
\newblock {{de Sitter} vacua from {N = 2} gauged supergravity}.
\newblock {\em JHEP}, 01:068, 2004.
\newblock \href {http://arxiv.org/abs/hep-th/0312063}
  {\path{arXiv:hep-th/0312063}}.

\bibitem{KKLT03}
S.~Kachru, R.~Kallosh, A.~D. Linde, and S.~P. Trivedi.
\newblock {de Sitter} vacua in string theory.
\newblock {\em Phys. Rev. D}, 68:046005, 2003.
\newblock \href {http://arxiv.org/abs/hep-th/0301240}
  {\path{arXiv:hep-th/0301240}}.

\bibitem{BHM02}
P.~Berglund, T.~Hubsch, and D.~Minic.
\newblock {de Sitter} spacetimes from warped compactifications of iib string
  theory.
\newblock {\em Phys. Lett. B.}, 534:147, 2002.
\newblock \href {http://arxiv.org/abs/hep-th/0112079}
  {\path{arXiv:hep-th/0112079}}.

\bibitem{MSS02}
A.~Maloney, E.~Silverstein, and A.~Strominger.
\newblock {de Sitter} space in noncritical string theory.
\newblock \href {http://arxiv.org/abs/hep-th/0205316}
  {\path{arXiv:hep-th/0205316}}.

\bibitem{BKQ03}
C.~P. Burgess, R.~Kallosh, and F.~Quevedo.
\newblock {de Sitter} string vacua from supersymmetric d-terms.
\newblock {\em JHEP}, 0310:056, 2003.
\newblock \href {http://arxiv.org/abs/hep-th/0309187}
  {\path{arXiv:hep-th/0309187}}.

\bibitem{BCK04}
M.~Becker, G.~Curio, and A.~Krause.
\newblock {de Sitter} vacua from heterotic {M-theory}.
\newblock {\em Nucl. Phys. B}, 693:223, 2004.
\newblock \href {http://arxiv.org/abs/hep-th/0403027}
  {\path{arXiv:hep-th/0403027}}.

\bibitem{Buc04}
E.~I. Buchbinder.
\newblock Raising anti {de Sitter} vacua to {de Sitter} vacua in heterotic
  {M-theory}.
\newblock {\em Phys. Rev. D}, 70:066008, 2004.
\newblock \href {http://arxiv.org/abs/hep-th/0406101}
  {\path{arXiv:hep-th/0406101}}.

\bibitem{Gra05}
M.~Grana.
\newblock Flux compactifications in string theory: A comprehensive review.
\newblock {\em Phys. Rept.}, 423:91, 2006.
\newblock For a review and further references see.
\newblock \href {http://arxiv.org/abs/hep-th/0509003}
  {\path{arXiv:hep-th/0509003}}.

\bibitem{VZ06}
Giovanni Villadoro and Fabio Zwirner.
\newblock {D terms from D-branes, gauge invariance and moduli stabilization in
  flux compactifications}.
\newblock {\em JHEP}, 03:087, 2006.
\newblock \href {http://arxiv.org/abs/hep-th/0602120}
  {\path{arXiv:hep-th/0602120}}.

\bibitem{ISS06}
Kenneth Intriligator, Nathan Seiberg, and David Shih.
\newblock {Dynamical SUSY breaking in meta-stable vacua}.
\newblock {\em JHEP}, 04:021, 2006.
\newblock \href {http://arxiv.org/abs/hep-th/0602239}
  {\path{arXiv:hep-th/0602239}}.

\bibitem{Sil07}
Eva Silverstein.
\newblock {Simple de Sitter Solutions}.
\newblock 2007.
\newblock \href {http://arxiv.org/abs/0712.1196} {\path{arXiv:0712.1196}}.

\bibitem{Cov08}
Laura Covi et~al.
\newblock {de Sitter vacua in no-scale supergravities and {Calabi-Yau} string
  models}.
\newblock 2008.
\newblock \href {http://arxiv.org/abs/0804.1073} {\path{arXiv:0804.1073}}.

\bibitem{Hul01a}
C.~M. Hull.
\newblock {de Sitter} space in supergravity and m theory.
\newblock {\em JHEP}, 0111:012, 2001.
\newblock \href {http://arxiv.org/abs/hep-th/0109213}
  {\path{arXiv:hep-th/0109213}}.

\bibitem{Hul01b}
C.~M. Hull.
\newblock Domain wall and {de Sitter} solutions of gauged supergravity.
\newblock {\em JHEP}, 0111:061, 2001.
\newblock \href {http://arxiv.org/abs/hep-th/0110048}
  {\path{arXiv:hep-th/0110048}}.

\bibitem{GH01}
G.~W. Gibbons and C.~M. Hull.
\newblock {de Sitter} space from warped supergravity solutions.
\newblock \href {http://arxiv.org/abs/hep-th/0111072}
  {\path{arXiv:hep-th/0111072}}.

\bibitem{dRWP03}
M.~{de Roo}, D.~B. Westra, and S.~Panda.
\newblock {de Sitter} solutions in {N = 4} matter coupled supergravity.
\newblock {\em JHEP}, 0302:003, 2003.
\newblock \href {http://arxiv.org/abs/hep-th/0212216}
  {\path{arXiv:hep-th/0212216}}.

\bibitem{dRWPT03}
M.~{de Roo}, D.~B. Westra, S.~Panda, and M.~Trigiante.
\newblock Potential and mass-matrix in gauged {N = 4} supergravity.
\newblock {\em JHEP}, 0311:022, 2003.
\newblock \href {http://arxiv.org/abs/hep-th/0310187}
  {\path{arXiv:hep-th/0310187}}.

\bibitem{KP04}
Renata Kallosh and Sergey Prokushkin.
\newblock {Supercosmology}.
\newblock 2004.
\newblock \href {http://arxiv.org/abs/hep-th/0403060}
  {\path{arXiv:hep-th/0403060}}.

\bibitem{RRW06}
Jan Rosseel, Thomas {Van Riet}, and Dennis~B. Westra.
\newblock {Scaling cosmologies of {N = 8} gauged supergravity}.
\newblock {\em Class. Quant. Grav.}, 24:2139--2152, 2007.
\newblock \href {http://arxiv.org/abs/hep-th/0610143}
  {\path{arXiv:hep-th/0610143}}.

\bibitem{GST84}
M.~Gunaydin, G.~Sierra, and P.~K. Townsend.
\newblock {Gauging The D = 5 {Maxwell-Einstein} Supergravity Theories: More On
  Jordan Algebras}.
\newblock {\em Nucl. Phys. B}, 253:573, 1985.

\bibitem{GRW85}
M.~Gunaydin, L.~J. Romans, and N.~P. Warner.
\newblock Gauged {N = 8} supergravity in five-dimensions.
\newblock {\em Phys. Lett. B.}, 154:268, 1985.

\bibitem{GRW86}
M.~Gunaydin, L.~J. Romans, and N.~P. Warner.
\newblock Compact and noncompact gauged supergravity theories in
  five-dimensions.
\newblock {\em Phys. Lett. B.}, 272:298, 1986.

\bibitem{PPvN85}
M.~Pernici, K.~Pilch, and P.~{van Nieuwenhuizen}.
\newblock Gauged {N = 8} d = 5 supergravity.
\newblock {\em Nucl. Phys. B}, 259:460, 1985.

\bibitem{Mal98}
J.~M. Maldacena.
\newblock The {large N} limit of superconformal field theories and
  supergravity.
\newblock {\em Adv. Theor. Math. Phys.}, 2:231, 1998.
\newblock \href {http://arxiv.org/abs/hep-th/9711200}
  {\path{arXiv:hep-th/9711200}}.

\bibitem{GKP98}
S.~S. Gubser, I.~R. Klebanov, and A.~M. Polyakov.
\newblock Gauge theory correlators from non-critical string theory.
\newblock {\em Phys. Lett. B.}, 428:105, 1998.
\newblock \href {http://arxiv.org/abs/hep-th/9802109}
  {\path{arXiv:hep-th/9802109}}.

\bibitem{Wit98}
E.~Witten.
\newblock {Anti-de Sitter} space and holography.
\newblock {\em Adv. Theor. Math. Phys.}, 2:253, 1998.
\newblock \href {http://arxiv.org/abs/hep-th/9802150}
  {\path{arXiv:hep-th/9802150}}.

\bibitem{AGMOO99}
O.~Aharony, S.~S. Gubser, J.~M. Maldacena, H.~Ooguri, and Y.~Oz.
\newblock {Large N} field theories, string theory and gravity.
\newblock {\em Phys. Rept.}, 323:183, 2000.
\newblock For an extensive list of references on AdS/CFT dualities see the
  review paper.
\newblock \href {http://arxiv.org/abs/hep-th/9905111}
  {\path{arXiv:hep-th/9905111}}.

\bibitem{RS99a}
L.~Randall and R.~Sundrum.
\newblock A large mass hierarchy from a small extra dimension.
\newblock {\em Phys. Rev. Lett.}, 83:3370, 1999.
\newblock \href {http://arxiv.org/abs/hep-ph/9905221}
  {\path{arXiv:hep-ph/9905221}}.

\bibitem{RS99b}
L.~Randall and R.~Sundrum.
\newblock An alternative to compactification.
\newblock {\em Phys. Rev. Lett.}, 83:4690, 1999.
\newblock \href {http://arxiv.org/abs/hep-th/9906064}
  {\path{arXiv:hep-th/9906064}}.

\bibitem{KL00}
R.~Kallosh and A.~D. Linde.
\newblock Supersymmetry and the brane world.
\newblock {\em JHEP}, 0002:005, 2000.
\newblock \href {http://arxiv.org/abs/hep-th/0001071}
  {\path{arXiv:hep-th/0001071}}.

\bibitem{CCDF95}
A.~C. Cadavid, A.~Ceresole, R.~D'Auria, and S.~Ferrara.
\newblock Eleven-dimensional supergravity compactified on {Calabi-Yau}
  threefolds.
\newblock {\em Phys. Lett. B.}, 357:76, 1995.
\newblock \href {http://arxiv.org/abs/hep-th/9506144}
  {\path{arXiv:hep-th/9506144}}.

\bibitem{FKS95}
S.~Ferrara, R.~Kallosh, and A.~Strominger.
\newblock {N = 2} extremal black holes.
\newblock {\em Phys. Rev. D}, 52:5412, 1995.
\newblock \href {http://arxiv.org/abs/hep-th/9508072}
  {\path{arXiv:hep-th/9508072}}.

\bibitem{FK96}
S.~Ferrara and R.~Kallosh.
\newblock Supersymmetry and attractors.
\newblock {\em Phys. Rev. D}, 54:1514, 1996.
\newblock \href {http://arxiv.org/abs/hep-th/9602136}
  {\path{arXiv:hep-th/9602136}}.

\bibitem{FGK97}
S.~Ferrara, G.~W. Gibbons, and R.~Kallosh.
\newblock Black holes and critical points in moduli space.
\newblock {\em Nucl. Phys. B}, 500:75, 1997.
\newblock \href {http://arxiv.org/abs/hep-th/9702103}
  {\path{arXiv:hep-th/9702103}}.

\bibitem{GM85}
M.~Gunaydin and N.~Marcus.
\newblock The spectrum of the {$S^5$} compactification of the chiral {N = 2}, d
  = 10 supergravity and the unitary supermultiplets of {U(2, 2/4)}.
\newblock {\em Class. Quant. Grav.}, 2:L11, 1985.

\bibitem{KRvN85}
H.~J. Kim, L.~J. Romans, and P.~{van Nieuwenhuizen}.
\newblock The mass spectrum of chiral {N = 2} d = 10 supergravity on {$S^5$}.
\newblock {\em Phys. Rev. D}, 32:389, 1985.

\bibitem{NV00}
H.~Nastase and D.~Vaman.
\newblock On the nonlinear kk reductions on spheres of supergravity theories.
\newblock {\em Nucl. Phys. B}, 583:211, 2000.
\newblock \href {http://arxiv.org/abs/hep-th/0002028}
  {\path{arXiv:hep-th/0002028}}.

\bibitem{CLPST00}
M.~Cvetic, H.~Lu, C.~N. Pope, A.~Sadrzadeh, and T.~A. Tran.
\newblock Consistent {SO(6)} reduction of type iib supergravity on {S(5)}.
\newblock {\em Nucl. Phys. B}, 586:275, 2000.
\newblock \href {http://arxiv.org/abs/hep-th/0003103}
  {\path{arXiv:hep-th/0003103}}.

\bibitem{Tra01}
T.~A. Tran.
\newblock {\em Gauged supergravities from spherical reductions}.
\newblock PhD thesis, 2001.
\newblock For more examples on the subject in general, see the PhD thesis of
  the author.
\newblock \href {http://arxiv.org/abs/hep-th/0109092}
  {\path{arXiv:hep-th/0109092}}.

\bibitem{HW95}
Petr Horava and Edward Witten.
\newblock {Heterotic and type I string dynamics from eleven dimensions}.
\newblock {\em Nucl. Phys.}, B460:506--524, 1996.
\newblock \href {http://arxiv.org/abs/hep-th/9510209}
  {\path{arXiv:hep-th/9510209}}.

\bibitem{HW96}
P.~Horava and E.~Witten.
\newblock Eleven-dimensional supergravity on a manifold with boundary.
\newblock {\em Nucl. Phys. B}, 475:94, 1996.
\newblock \href {http://arxiv.org/abs/hep-th/9603142}
  {\path{arXiv:hep-th/9603142}}.

\bibitem{LOSW99a}
A.~Lukas, B.~A. Ovrut, K.~S. Stelle, and D.~Waldram.
\newblock The universe as a domain wall.
\newblock {\em Phys. Rev. D}, 59:086001, 1999.
\newblock \href {http://arxiv.org/abs/hep-th/9803235}
  {\path{arXiv:hep-th/9803235}}.

\bibitem{LOSW99b}
A.~Lukas, B.~A. Ovrut, K.~S. Stelle, and D.~Waldram.
\newblock Heterotic {M-theory} in five dimensions.
\newblock {\em Nucl. Phys. B}, 552:246, 1999.
\newblock \href {http://arxiv.org/abs/hep-th/9806051}
  {\path{arXiv:hep-th/9806051}}.

\bibitem{ELPP98}
J.~R. Ellis, Z.~Lalak, S.~Polorski, and W.~Pokorski.
\newblock Five-dimensional aspects of {M-theory} dynamics and supersymmetry.
\newblock {\em Nucl. Phys. B}, 540:149, 1999.
\newblock \href {http://arxiv.org/abs/hep-ph/9805377}
  {\path{arXiv:hep-ph/9805377}}.

\bibitem{GZ99}
M.~Gunaydin and M.~Zagermann.
\newblock The gauging of five-dimensional, {N = 2} {Maxwell-Einstein}
  supergravity theories coupled to tensor multiplets.
\newblock {\em Nucl. Phys. B}, 572:131, 2000.
\newblock \href {http://arxiv.org/abs/hep-th/9912027}
  {\path{arXiv:hep-th/9912027}}.

\bibitem{dWP92}
Bernard {de Wit} and Antoine {Van Proeyen}.
\newblock {Broken sigma model isometries in very special geometry}.
\newblock {\em Phys. Lett.}, B293:94--99, 1992.
\newblock \href {http://arxiv.org/abs/hep-th/9207091}
  {\path{arXiv:hep-th/9207091}}.

\bibitem{Cre80}
E.~Cremmer.
\newblock {\em Supergravities In 5 Dimensions}.
\newblock Cambridge Univ Press, 1981.

\bibitem{CN80}
A.~H. Chamseddine.
\newblock Coupling the {SO(2)} supergravity through dimensional reduction.
\newblock {\em Phys. Lett. B.}, 96:89, 1980.

\bibitem{GST83b}
M.~Gunaydin, G.~Sierra, and P.~K. Townsend.
\newblock {The Geometry Of {N = 2} {Maxwell-Einstein} Supergravity And Jordan
  Algebras}.
\newblock {\em Nucl. Phys. B}, 242:244, 1984.

\bibitem{CD00}
A.~Ceresole and G.~Dall'Agata.
\newblock General matter coupled {N = 2}, d = 5 gauged supergravity.
\newblock {\em Nucl. Phys. B}, 585:143, 2000.
\newblock \href {http://arxiv.org/abs/hep-th/0004111}
  {\path{arXiv:hep-th/0004111}}.

\bibitem{GZ00}
M.~Gunaydin and M.~Zagermann.
\newblock The vacua of 5d, {N = 2} gauged {Yang-Mills/Einstein}/tensor
  supergravity: Abelian case.
\newblock {\em Phys. Rev. D}, 62:044028, 2000.
\newblock \href {http://arxiv.org/abs/hep-th/0002228}
  {\path{arXiv:hep-th/0002228}}.

\bibitem{GZ01}
M.~Gunaydin and M.~Zagermann.
\newblock {Gauging the full R-symmetry group in five-dimensional, {N = 2}
  {Yang-Mills/Einstein}/tensor supergravity}.
\newblock {\em Phys. Rev. D}, 63:064023, 2001.
\newblock \href {http://arxiv.org/abs/hep-th/0004117}
  {\path{arXiv:hep-th/0004117}}.

\bibitem{CS05}
B.~Cosemans and G.~Smet.
\newblock Stable {de Sitter} vacua in {N = 2}, d = 5 supergravity.
\newblock {\em Class. Quant. Grav.}, 22:2359, 2005.
\newblock \href {http://arxiv.org/abs/hep-th/0502202}
  {\path{arXiv:hep-th/0502202}}.

\bibitem{Oge06}
O.~Ogetbil.
\newblock {A general study of ground states of {N = 2} supergravity theories
  with symmetric scalar manifolds in 5 dimensions}.
\newblock {\em Phys. Rev. D}, 75:065033, 2007.
\newblock \href {http://arxiv.org/abs/hep-th/0612145}
  {\path{arXiv:hep-th/0612145}}.

\bibitem{GMZ05b}
M.~Gunaydin, S.~McReynolds, and M.~Zagermann.
\newblock {The R-map and the coupling of {N = 2} tensor multiplets in 5 and 4
  dimensions}.
\newblock {\em JHEP}, 0601:168, 2006.
\newblock \href {http://arxiv.org/abs/hep-th/0511025}
  {\path{arXiv:hep-th/0511025}}.

\bibitem{Oge08a}
O.~Ogetbil.
\newblock {\em de Sitter Vacua and $\mathcal{N} = 2$ Supergravity}.
\newblock PhD thesis, Penn State University, August 2008.

\bibitem{GST86}
M.~Gunaydin, G.~Sierra, and P.~K. Townsend.
\newblock More on d = 5 {Maxwell-Einstein} supergravity: Symmetric spaces and
  kinks.
\newblock {\em Class. Quant. Grav.}, 3:763, 1986.

\bibitem{CDKP01}
A.~Ceresole, G.~Dall'Agata, R.~Kallosh, and A.~{Van Proeyen}.
\newblock Hypermultiplets, domain walls and supersymmetric attractors.
\newblock {\em Phys. Rev. D}, 64:104006, 2001.
\newblock \href {http://arxiv.org/abs/hep-th/0104056}
  {\path{arXiv:hep-th/0104056}}.

\bibitem{GST83a}
M.~Gunaydin, G.~Sierra, and P.~K. Townsend.
\newblock {Exceptional Supergravity Theories and the MAGIC Square}.
\newblock {\em Phys. Lett.}, B133:72, 1983.

\bibitem{GMZ05a}
M.~Gunaydin, S.~McReynolds, and M.~Zagermann.
\newblock {Unified {N = 2} {Maxwell-Einstein} and {Yang-Mills-Einstein}
  supergravity theories in four dimensions}.
\newblock {\em JHEP}, 09:026, 2005.
\newblock \href {http://arxiv.org/abs/hep-th/0507227}
  {\path{arXiv:hep-th/0507227}}.

\bibitem{FS89}
S.~Ferrara and A.~Strominger.
\newblock {{N = 2} Space-time Supersymmetry and {Calabi-Yau} Moduli Space}.
\newblock 1989.
\newblock Presented at Strings '89 Workshop, College Station, TX, Mar 13-18.

\bibitem{Str90}
A.~Strominger.
\newblock {Special Geometry}.
\newblock {\em Commun. Math. Phys.}, 133:163--180, 1990.

\bibitem{CDF90}
L.~Castellani, R.~D'Auria, and S.~Ferrara.
\newblock {Special Kahler Geometry: An Intrinsic Formulation from {N = 2}
  Space-time Supersymmetry}.
\newblock {\em Phys. Lett.}, B241:57, 1990.

\bibitem{CDGP90}
P.~Candelas, X.~C. {De la Ossa}, P.~S. Green, and L.~Parkes.
\newblock {An Exactly soluble superconformal theory from a mirror pair of
  {Calabi-Yau} manifolds}.
\newblock {\em Phys. Lett.}, B258:118--126, 1991.

\bibitem{dWP84}
B.~{de Wit} and A.~{Van Proeyen}.
\newblock {Potentials and Symmetries of General Gauged {N = 2} Supergravity -
  {Yang-Mills} Models}.
\newblock {\em Nucl. Phys. B}, 245:89, 1984.

\bibitem{Cre84}
E.~Cremmer et~al.
\newblock {Vector Multiplets Coupled to {N = 2} Supergravity: SuperHiggs
  Effect, Flat Potentials and Geometric Structure}.
\newblock {\em Nucl. Phys.}, B250:385, 1985.

\bibitem{dWLP85}
B.~{de Wit}, P.G. Lauwers, and A.~{Van Proeyen}.
\newblock Lagrangians of {N = 2} supergravity - matter systems.
\newblock {\em Nucl. Phys. B}, 255:569, 1985.

\bibitem{dRW85}
M.~{de Roo} and P.~Wagemans.
\newblock {Gauge Matter Coupling in {N = 4} Supergravity}.
\newblock {\em Nucl. Phys.}, B262:644, 1985.

\bibitem{Wag90}
P.~Wagemans.
\newblock {\em {Aspects of {N = 4} supergravity}}.
\newblock PhD thesis, 1990.
\newblock RX-1299 (Groningen).

\bibitem{ABCDFFM96}
L.~Andrianopoli, M.~Bertolini, A.~Ceresole, R.~D'Auria, S.~Ferrara, P.~Fre, and
  T.~Magri.
\newblock {N = 2} supergravity and {N = 2} super {Yang-Mills} theory on general
  scalar manifolds: Symplectic covariance, gaugings and the momentum map.
\newblock {\em J. Geom. Phys.}, 23:111, 1997.
\newblock \href {http://arxiv.org/abs/hep-th/9605032}
  {\path{arXiv:hep-th/9605032}}.

\bibitem{BW83}
J.~Bagger and E.~Witten.
\newblock {Matter Couplings in {N = 2} Supergravity}.
\newblock {\em Nucl. Phys.}, B222:1, 1983.

\bibitem{DFF91}
R.~D'Auria, S.~Ferrara, and P.~Fre.
\newblock {Special and quaternionic isometries: General couplings in {N = 2}
  supergravity and the scalar potential}.
\newblock {\em Nucl. Phys.}, B359:705--740, 1991.

\bibitem{FI74}
P.~Fayet and J.~Iliopoulos.
\newblock {Spontaneously Broken Supergauge Symmetries and Goldstone Spinors}.
\newblock {\em Phys. Lett.}, B51:461--464, 1974.

\bibitem{Wei00}
Steven Weinberg.
\newblock {\em The Quantum Theory of Fields}, volume III Supersymmetry.
\newblock Cambridge Univ Press, 2000.

\bibitem{BDKP04}
P.~Binetruy, G.~Dvali, R.~Kallosh, and A.~{Van Proeyen}.
\newblock {Fayet-Iliopoulos terms in supergravity and cosmology}.
\newblock {\em Class. Quant. Grav.}, 21:3137--3170, 2004.
\newblock \href {http://arxiv.org/abs/hep-th/0402046}
  {\path{arXiv:hep-th/0402046}}.

\bibitem{Pro04}
A.~{Van Proeyen}.
\newblock {Supergravity with Fayet-Iliopoulos terms and R-symmetry}.
\newblock {\em Fortsch. Phys.}, 53:997--1004, 2005.
\newblock \href {http://arxiv.org/abs/hep-th/0410053}
  {\path{arXiv:hep-th/0410053}}.

\bibitem{Mar06}
K.~Maruyoshi.
\newblock {Gauged {N = 2} supergravity and partial breaking of extended
  supersymmetry}.
\newblock 2006.
\newblock \href {http://arxiv.org/abs/hep-th/0607047}
  {\path{arXiv:hep-th/0607047}}.

\bibitem{IW53}
E.~Inonu and Eugene~P. Wigner.
\newblock {On the Contraction of Groups and Their Representations}.
\newblock {\em Proc. Nat. Acad. Sci.}, 39:510--524, 1953.

\bibitem{Gil74}
G.~Gilmore.
\newblock {\em Lie Groups, Lie Algebras, and Some of Their Applications}.
\newblock Wiley-Interscience, 1974.

\bibitem{WKV00}
B.~{de Wit}, Bas Kleijn, and Stefan Vandoren.
\newblock {Superconformal hypermultiplets}.
\newblock {\em Nucl. Phys.}, B568:475--502, 2000.
\newblock \href {http://arxiv.org/abs/hep-th/9909228}
  {\path{arXiv:hep-th/9909228}}.

\bibitem{dWP91}
B.~{de Wit} and A.~{Van Proeyen}.
\newblock {Special geometry, cubic polynomials and homogeneous quaternionic
  spaces}.
\newblock {\em Commun. Math. Phys.}, 149:307--334, 1992.
\newblock \href {http://arxiv.org/abs/hep-th/9112027}
  {\path{arXiv:hep-th/9112027}}.

\bibitem{dWP95}
B.~{de Wit} and A.~{Van Proeyen}.
\newblock {Isometries of special manifolds}.
\newblock 1995.
\newblock in the Proceedings of the meeting on quaternionic structures in
  mathematics and physics, Trieste, September 1994; available on
  http://www.emis.de/proceedings/QSMP94/.
\newblock \href {http://arxiv.org/abs/hep-th/9505097}
  {\path{arXiv:hep-th/9505097}}.

\bibitem{SV95}
Ashoke Sen and Cumrun Vafa.
\newblock {Dual pairs of type II string compactification}.
\newblock {\em Nucl. Phys.}, B455:165--187, 1995.
\newblock \href {http://arxiv.org/abs/hep-th/9508064}
  {\path{arXiv:hep-th/9508064}}.

\bibitem{Gun06}
M.~Gunaydin.
\newblock {From d = 6, {N = 1} to d = 4, {N = 2}, No-scale supergravity models
  and Jordan Algebras}.
\newblock Talk at the conference ``30 years of supergravity'' in Paris, Oct
  2006, 2006.

\bibitem{GKP01}
Steven~B. Giddings, Shamit Kachru, and Joseph Polchinski.
\newblock {Hierarchies from fluxes in string compactifications}.
\newblock {\em Phys. Rev.}, D66:106006, 2002.
\newblock \href {http://arxiv.org/abs/hep-th/0105097}
  {\path{arXiv:hep-th/0105097}}.

\bibitem{EGZ01}
J.~R. Ellis, M.~Gunaydin, and M.~Zagermann.
\newblock Options for gauge groups in five-dimensional supergravity.
\newblock {\em JHEP}, 0111:024, 2001.
\newblock \href {http://arxiv.org/abs/hep-th/0108094}
  {\path{arXiv:hep-th/0108094}}.

\end{thebibliography}
\end{document}